\documentclass[english]{article}
\usepackage[T1]{fontenc}
\usepackage[latin9]{inputenc}
\usepackage{float}
\usepackage{amssymb}
\usepackage{graphicx}
\usepackage[margin=1.5in]{geometry}
\usepackage[authoryear]{natbib}
\RequirePackage{amsthm}
\RequirePackage[cmex10]{amsmath}
\usepackage{algorithm,algorithmic}
\usepackage{pgf,tikz}
\usepackage{ctable} 
\usepackage{subfig}
\usepackage{pdflscape}
\usetikzlibrary{shapes,decorations,arrows,calc,arrows.meta,fit,positioning}
\tikzset{
	-{Latex[length=3.5mm, width=1.1mm]},auto,node distance =1 cm and 1 cm,semithick,
	state/.style ={ellipse, draw, minimum width = 0.7 cm},
	point/.style = {circle, draw, inner sep=0.04cm,fill,node contents={}},
	bidirected/.style={Latex-Latex,dashed},
	el/.style = {inner sep=2pt, align=left, sloped}
}

\numberwithin{figure}{section}

\makeatletter

\providecommand{\tabularnewline}{\\}
\floatstyle{ruled}

\theoremstyle{plain}

\@ifundefined{showcaptionsetup}{}{%
	\PassOptionsToPackage{caption=false}{subfig}}

\makeatother

\usepackage{babel}
\begin{document}
\title{Many Proxy Controls}
\author{Ben Deaner\\(Cowles Foundation, Yale University) \\bendeaner@gmail.com}
\maketitle
\begin{abstract}
	A recent literature considers causal inference using noisy proxies for unobserved confounding factors. The proxies are divided into two sets that are independent conditional on the confounders. One set of proxies are `negative control treatments' and the other are `negative control outcomes'. Existing work applies to low-dimensional settings with a fixed number of proxies and confounders. In this work we consider linear models with many proxy controls and possibly many confounders. A key insight is that if each group of proxies is strictly larger than the number of confounding factors, then a matrix of nuisance parameters has a low-rank structure and a vector of nuisance parameters has a sparse structure. We can exploit the rank-restriction and sparsity to reduce the number of free parameters to be estimated. The number of unobserved confounders is not known a priori but we show that it is identified, and we apply penalization methods to adapt to this quantity. We provide an estimator with a closed-form as well as a doubly-robust estimator that must be evaluated using numerical methods. We provide conditions under which our doubly-robust estimator is uniformly root-$n$ consistent, asymptotically centered normal, and our suggested confidence intervals have asymptotically correct coverage. We provide simulation evidence that our methods achieve better performance than existing approaches in high dimensions, particularly when the number of proxies is substantially larger than the number of confounders.
\end{abstract}

\section{Introduction and Related Literature}

A recent, rapidly growing literature considers the problem of causal inference when a researcher observes only proxies for unobserved confounding factors. For example, one may wish to use test scores to proxy for ability, an unobserved confounder. The proxies are divided into two groups which are independent conditional on the unobserved confounders. One group is a set of negative control treatments: variables that have no direct causal effect on outcomes. The other group of proxies are negative control outcomes: variables that are not directly affected by the treatments.

Compared to standard factor-analytic methods, the proxy control approach has the advantage that the factor structure itself need not be identified. That is, neither the distribution of unobserved factors nor the causal effects of these factors must be identified. Thus proxy control methods may be applied even when the assumptions required for identification of the factor structure do not hold. However, this limits the application of proxy control methods to settings in which the factor structure itself is not of interest.

The proxy control approach is particularly apt for causal inference with high-dimensional data, that is, data that contain many covariates. In these settings one may hope to use the the rich covariates to adjust for confounding. Standard methods treat the covariates as controls, and are valid only if the confounders are non-random once we condition on the covariates. This condition is known as `unconfoundedness', `ignorability', or `selection on observables'. It holds, for example, if all of the confounders are included among the covariates.

By contrast, proxy control methods do not require that the covariates perfectly explain the confounders, only that they are informative about them. Proxy control methods can be valid when the covariates are only noisy proxies for the confounders. Thus proxy control methods allow us to exploit the richness of the covariates without having so assume unconfoundedness. However, to relax this assumption we must place other restrictions on the relationships between the relevant variables.

Nonparametric identification with proxy controls is achieved using the conditional independence assumptions listed in Sub-Figure 1.(a). $Y_i$ is an outcome of interest for an individual $i$ and $X_i$ is a vector of treatments assigned to $i$. $Y_i(\cdot)$ is individual $i$'s treatment response function, or equivalently, $Y_i(x)$ is individual $i$'s potential outcome from the counterfactual treatment level $x$. $W_i$ is a vector of unobserved confounding factors. $D_i$ is a vector of observed characteristics like age and gender. $Z_i$ and $V_i$ are two groups of proxies for $W_i$. $Z_i$ is a vector of negative control treatments and $V_i$ is a vector of negative control outcomes.

Sub-Figure 1.(b) contains two causal diagrams. The diagram on the left, taken from \citet{Tchetgen} differs from that on the right in that it assumes $Z_i$ is a vector of pre-treatment variables. In the diagram on the right, $Z_i$ is a vector of post-treatment variables, as in \citet{Deaner2021}. Each diagram is associated with a nonparametric structural equations model (\citet{Pearl2009}) that implies the conditions in Sub-Figure 1.(a). Note that this is not the only causal structure that implies the conditions in Sub-Figure 1.(a), for example we could allow simultaneous causation between $V_i$ and $W_i$ and between $X_i$ and $Z_i$ (see \citet{Deaner2021}).

1.(b) imposes that $Z_i$ and $V_i$ have no direct causal effect on each other. However, $Z_i$ and $V_i$ may both depend on $W_i$ and $D_i$. The dependence on $W_i$ means that we can understand these variables as noisy proxies for $W_i$. Correspondingly, condition 1.(a).3 implies that, after accounting for $W_i$ and $D_i$, the variables $V_i$ and $Z_i$ are independent.

the conditions in Figure 1 restrict the manner in which the treatments and outcomes are related to the proxies $Z_i$ and $V_i$. These restrictions are asymmetric. 1.(b) imposes that the negative control treatments $Z_i$ can both cause and be caused by the treatment $X_i$.  However, the negative control outcomes $V_i$ cannot cause or be caused by the treatments $X_i$. On the other hand, the diagrams in 1.(b) allow $V_i$ to directly cause the outcomes $Y_i$ whereas $Z_i$ cannot directly cause $Y_i$.

\begin{figure}[h]
\centering
	\caption{Causal Structure of Proxy Controls}
\subfloat[Conditional Independence Restrictions]{

		1.  $Y_i(\cdot)\perp\!\!\!\perp X_i|W_i,D_i$,  \,	2.  $Y_i(\cdot)\perp\!\!\!\perp Z_i|X_i,W_i,D_i$, \,
		3.  $V_i\perp\!\!\!\perp (X_i,Z_i)|W_i,D_i$

}
\centering

\subfloat[Causal Diagrams]{
	\resizebox{!}{90pt}{%
		
		\begin{tikzpicture}
			% nodes %
			
			\node (y) at (1.1,-1.22) [label=below right:$Y_i$,point];
			\node (x) at (-1.1,-1.22) [label=below left:$X_i$,point];
			\node (w) at (0,1.5) [circle, fill=white,label=above:$W_i$,point];
			\node (v) at (1.6,0.46) [label=above right:$V_i$,point];
			\node (z) at (-1.6,0.46) [label=above left:$Z_i$,point];
			\node (d) at (0,-0.1) [label=below:$D_i$,point];

			%	\node (y) at (0.8,-1.1) [label=below right:$Y_i$,point];
			%\node (x) at (-0.8,-1.1) [label=below left:$X_i$,point];
			%\node (w) at (0,1) [circle, fill=white,label=above:$W_i$,point];
			%\node (v) at (1.42,0) [label=above right:$V_i$,point];
			%\node (z) at (-1.42,0) [label=above left:$Z_i$,point];
			%\node (d) at (0,-0.2) [label=below:$D_i$,point];	
			
			\path (d) edge (x);
			\path (d) edge (y);
			\path (d) edge (z);
			\path (d) edge (v);
			\path (d) edge (w);
			
			\path (x) edge (y);
			\path (w)[dashed] edge (x);
			\path (w)[dashed] edge (y);
			\path (w)[dashed] edge (z);
			\path (w)[dashed] edge (v);
			\path (z) edge (x);
			\path (v) edge (y);
		\end{tikzpicture}

		\begin{tikzpicture}
			% nodes %
			
			\node (y) at (1.1,-1.22) [label=below right:$Y_i$,point];
			\node (x) at (-1.1,-1.22) [label=below left:$X_i$,point];
			\node (w) at (0,1.5) [circle, fill=white,label=above:$W_i$,point];
			\node (v) at (1.6,0.46) [label=above right:$V_i$,point];
			\node (z) at (-1.6,0.46) [label=above left:$Z_i$,point];
			\node (d) at (0,-0.1) [label=below:$D_i$,point];

			%	\node (y) at (0.8,-1.1) [label=below right:$Y_i$,point];
			%\node (x) at (-0.8,-1.1) [label=below left:$X_i$,point];
			%\node (w) at (0,1) [circle, fill=white,label=above:$W_i$,point];
			%\node (v) at (1.42,0) [label=above right:$V_i$,point];
			%\node (z) at (-1.42,0) [label=above left:$Z_i$,point];
			%\node (d) at (0,-0.2) [label=below:$D_i$,point];	
			
			\path (d) edge (x);
			\path (d) edge (y);
			\path (d) edge (z);
			\path (d) edge (v);
			\path (d) edge (w);
			
			\path (x) edge (y);
			\path (w)[dashed] edge (x);
			\path (w)[dashed] edge (y);
			\path (w)[dashed] edge (z);
			\path (w)[dashed] edge (v);
			\path (x) edge (z);
			\path (v) edge (y);
		\end{tikzpicture}
	}		
}

\end{figure}

\citet{Miao2018a} show that if the conditional independence restrictions in Sub-Figure 1.(a) hold along with two statistical completeness assumptions and some regularity conditions, then the average structural function is nonparametrically identified. The average structural function is equal to $E[Y_i(x)]$ in potential outcomes notation or $ E[Y_i|do(x)]$ in the notation of \citet{Pearl2009}. \citet{Deaner2021} shows that under the same conditional independence restrictions, related regularity conditions, and weaker completeness conditions, one can identify the conditional average structural function $E[Y_i(x_1)|X_i=x_2]$.\footnote{Identification of the conditional average structural function implies identification of the average structural function. However, the converse does not hold.} Note that \citet{Deaner2021} allows for additional covariates $D_i$ by assuming they are included in both $W_i$, $Z_i$, and $V_i$. Alternatively, one can apply the results in \citet{Miao2018a} and \citet{Deaner2021} within one particular stratum of $D_i$.

\citet{Deaner2021} also provides an alternative characterization of the conditional average structural function and provides sufficient conditions for the conditional independence restrictions in the context of panel data.

Nonparametric estimation in this setting was considered by \citet{Deaner2019}, and later by \citet{Tchetgen}, \citet{Singh2020}, \citet{Cui2020}, and \citet{Kallus2021}.\footnote{
	\citet{Deaner2018} earlier considered nonparametric estimation in an equivalent setting that arises in the context of panel data. \citet{Deaner2019} was expanded to include a method for nonparametric inference in \citet{Deaner2020}.
}
 \citet{Miao2018} consider estimation in parametric models when a `confounding bridge' function is identified.

In this work we consider identification, estimation, and inference in linear models when the set of proxy controls and the set of confounding factors may be high-dimensional. In particular, we allow for the possibility that the number of proxies and confounders grows with the sample size. Existing work applies to low-dimensional settings in which the number of proxies and confounding factors is treated as fixed.

A key insight in this work is that if there are strictly fewer confounders than there are proxies in $V_i$ and $Z_i$, then a matrix of nuisance parameters has a low rank structure. In addition, a vector of nuisance parameters has a sparse structure. We exploit this low-rank structure and sparsity to reduce the number of free parameters to be estimated. This allows for more efficient estimation, particularly when the number of proxies is large. The number of confounders is generally unknown, so we propose a model selection methods that allows us to adapt to this quantity. The model selection methods are based on techniques from the literature on reduced-rank regression and $\ell_1$-penalized regression.

We present three different estimation methods. The first employs a known bound on the number of confounders, the second selects the number of confounders using the data. These first two methods are based on techniques from the literature on reduced-rank regression and have closed-forms. The third estimator is doubly-robust and is an example of a DML2 estimator of the kind analyzed in section 3.2 of \citet{Chernozhukov2018}. \citet{Chernozhukov2018} shows that DML2 estimators are root-$n$ consistent, asymptotically unbiased, and asymptotically Gaussian, under relatively weak conditions on the nuisance parameter estimates. A disadvantage of the doubly-robust estimator (compared to our alternative methods) is that it does not have a closed-form, however the numerical part of the algorithm can be carried out using any standard Lasso procedure.

Linearity also allows us to weaken the identifying assumptions used in the nonparametric case. In particular, we need only assume variables are uncorrelated rather than independent, we can replace statistical completeness with more intuitive full rank conditions, and we avoid the need for regularity conditions like Assumption A3 in \citet{Miao2018}.

The linear proxy control model has a long history, dating back to the work of \citet{Griliches1977} who considers two scalar proxies for a single confounding factor. However, to the best of our knowledge no existing work exploits the dimension reduction when there are fewer confounders than proxies in $V_i$ or $Z_i$, nor does any existing work allow for a growing number of proxies or confounders.

  In sum, our contributions are as follows. We provide a set of identifying assumptions in the linear proxy control model. We present novel estimation methods that allow us to exploit the low rank structure and sparsity in the nuisance parameters when the number of unobserved confounders is less than the number of proxies in $Z_i$ and $V_i$. We develop asymptotic theory for the estimator and an associated inference method, and we provide simulation evidence of the efficacy of our methods.

\section{Model and Identification}

Suppose that for each individual $i$, the potential outcome $Y_i(x)$ from treatment $x$ is linear in $x$ and differs between individuals only by an additive constant:
\[
Y_i(x)=x'\beta_0+U_i
\]

$\beta_0$ in the above is a vector of parameters that is the same for all individuals. $U_i$
is a random scalar that represents heterogeneity in potential outcomes. Note that the model implies treatment effects are the same for all individuals. By the definition of potential outcomes, the realized outcome $Y_i$ satisfies $Y_i=Y_i(X_i)$ and so:
\begin{equation}
Y_i=X_i'\beta_0+U_i\label{eq:1}
\end{equation}

Where $X_i$ is the realized treatment and is a column vector of length $d_X$. Let $W_i$ be a length-$d_W$ column vector of
latent variables and let $V_i$ be a length-$d_V$ vector of proxies. Let $D_i$ be a length-$d_D$ vector of observed control variables whose first entry is $1$. We assume
the following two relationships are linear:
\begin{align}
U_i&=A_0 W_i+L_0 D_i + \varepsilon_i\label{eq:U}\\
V_i&=B_0 W_i+ R_0 D_i + \upsilon_i\label{V}
\end{align}

Where $E[\varepsilon_i (W_i',D_i')]=E[\upsilon_i (W_i',D_i')]=0$. We can always find matrices $A_0$, $B_0$, $L_0$, and $R_0$ that satisfy
these relationships, however we will place assumptions directly on $\varepsilon_i$
and $\upsilon_i$ and so the linear specifications do incur a loss of
generality.

$V_i$ does not directly enter the expression for $U_i$, however this does not rule out the possibility that $V_i$ has a direct causal effect on $Y_i$. We could make this more explicit by including $V_i$ in the expression for $Y_i$ as follows:
\begin{equation}
	Y_i=X_i'\beta_0+F_0 W_i+K_0 D_i +\chi_0 V_i + e_i \label{eq:altU}
	\end{equation}

Where $E[e_i (W_i',D_i')]=0$. If (\ref{V}) holds then the model for $Y_i$ above is equivalent to the combination of (\ref{eq:1}) and (\ref{eq:U}) with $A_0=F_0+\chi_0 B_0$, $L_0=K_0+\chi_0 R_0$, and  $\varepsilon_i = e_i + \chi_0 \upsilon_i$. 

We wish to identify and estimate $\beta_0$ in the model for $Y_i(\cdot)$.
Identification of $\beta_0$ immediately implies identification of
average treatment effects (ATEs):
\[
E[Y_i(x_{1})-Y_i(x_{2})]=(x_{1}-x_{2})'\beta_0
\]

And also the conditional average structural function (CASF):
\[
E[Y_i(x_{1})|X_i=x_{2}]=(x_{1}-x_{2})'\beta_0+E[Y_i|X_i=x_{2}]
\]

Before we state our identifying assumptions let us introduce some additional notation. In particular we define the following objects for each of the variables $H_i=W_i,\, V_i,\, Z_i,\, Y_i,\, X_i$:

\begin{align*}
	\gamma_{H,0}&=E[D_i D_i']^{+}E[D_i H_i']\\
	\omega_{H,0}&=E[(X_i',D_i')'(X_i',D_i')]^{+}E[(X_i',D_i')' H_i']\\
	\tilde{H}_i(\gamma)&=H_i-\gamma'D_i\\
	\bar{H}_i(\omega)&=H_i-\omega'(X_i',D_i')'
\end{align*}

$M^{+}$ denotes the Moore-Penrose pseudo-inverse of a matrix $M$. For notational convenience we sometimes write $\tilde{H}_i=\tilde{H}_i(\gamma_{H,0})$ and $\bar{H}_i=\bar{H}_i(\omega_{H,0})$. So for example,  $\tilde{X}_i$ is the residual from population linear regression of $X_i$ on $D_i$, or in other words, $X_i$ with $D_i$ partialled out. Similarly, $\bar{Z}_i$ is $Z_i$ with both $D_i$ and $X_i$ partialled out.

\theoremstyle{definition}
\newtheorem*{A1.1}{Assumption 1.1 (Model and Exclusion restrictions)}
\begin{A1.1}
	i. (\ref{eq:1}), (\ref{eq:U}), and (\ref{V}) hold. ii. $E[\varepsilon_i X_i]=0$. iii. $E[\varepsilon_i Z_i]=0$, $E[\upsilon_i Z_i]=0$, and $E[\upsilon_i X_i]=0$.
\end{A1.1}
\theoremstyle{definition}

\newtheorem*{A1.2}{Assumption 1.2 ($V_i$ is sufficiently informative about $W_i$)}
\begin{A1.2}
$E[{W}_i \bar{V}_i']$ has full row rank.
\end{A1.2}
\theoremstyle{definition}
\newtheorem*{A1.3}{Assumption 1.3 ($Z_i$ is sufficiently informative about $W_i$)}
\begin{A1.3}
$E[{W}_i \bar{Z}_i']$ has full row rank.
\end{A1.3}
\newtheorem*{A1.4}{Assumption 1.4 (Full support)}
\begin{A1.4} 
	i. $E[\tilde{X}_i \tilde{X}_i']$ is non-singular. ii. $E[\bar{Z}_i \bar{Z}_i']$ and $E[D_i D_i']$ are non-singular.
\end{A1.4}

\theoremstyle{definition}
 Assumptions 1.1.i, 1.1.ii, and 1.1.iii are analogous to conditions 1., 2., and 3., in Sub-Figure 1.(a). Because we impose linearity, we only require zero partial correlations rather than conditional independence.
 
We can formulate an equivalent set of conditions to Assumptions 1.1.ii and 1.1.iii in terms of the model (\ref{eq:altU}). Recall that $\varepsilon_i = e_i + \chi_0 \upsilon_i$. If $E[\upsilon X_i]=0$ and $E[\upsilon Z_i]=0$ then the remaining conditions in Assumptions 1.1.ii and 1.1.iii hold if and only if $E[e_i X_i]=0$ and $E[e_i Z_i]=0$.

Assumption 1.2 requires that once we have accounted for $X_i$ and $D_i$, $V_i$ is a sufficiently informative proxy for the confounders $W_i$. The assumption replaces the statistical completeness condition on $V_i$ required in the nonparametric setting. Note that the assumption is equivalent to the rank condition for identification in linear instrumental variables (IV) estimation: $W_i$ takes the role of the endogenous regressors, $D_i$ and $X_i$ take the role of the exogenous regressors, and $V_i$ acts as a vector of instruments.

Similarly, Assumption 1.3 requires that after accounting for the treatments and observed controls, $Z_i$ is sufficiently informative about the confounders. Again, this is the same condition required for identification in a linear IV model in which $Z_i$ is a vector of instruments for $W_i$, and the variables $X_i$ and $D_i$ are exogenous regressors.

Note that Assumptions 1.2 and 1.3 can only hold if the vectors $Z_i$ and $V_i$ each have weakly larger dimension than $W_i$.

Assumption 1.4.i is a very mild condition that after partialling out the observed controls $D_i$, the treatments are not perfectly colinear. Assumption 1.4.ii is without loss of generality for the identification results because we can ignore any linearly dependent components of $\bar{Z}_i$ and $D_i$.

\theoremstyle{plain}
\newtheorem*{Th1}{Theorem 1}
\begin{Th1}

Under Assumptions 1.1-1.4 $\beta_0$ and $d_W$ are identified. In particular, let $d_W\leq r\leq d_V$, then there exists a vector $\beta$ and matrices $A$, $B$, $C$, and $G$ of respective dimensions $1\times r$, $d_{V}\times r$, $r\times d_{Z}$, and $r\times d_{X}$, with $rank(B)\leq r$ so that:

\begin{equation}
	E\bigg[\bigg(\begin{pmatrix}\tilde{V}_i\\
		\tilde{Y}_i-\beta'\tilde{X}_i
	\end{pmatrix}-\begin{pmatrix}BC & BG\\
	AC & AG
\end{pmatrix}
\begin{pmatrix}\tilde{Z}_i\\
		\tilde{X}_i
	\end{pmatrix}\bigg)(\tilde{Z}_i', \tilde{X}_i')\bigg]=0 \label{sparemom}
\end{equation}

 For any such a solution, $\beta=\beta_0$, and $d_W$ is the smallest value of $r$ such that a solution exists.
\end{Th1}

Theorem 1 provides a characterization of $\beta_0$ and $d_W$ in terms of a set of moment conditions. The moment conditions involve nuisance parameters $A$, $B$, $C$, and $G$, which are not uniquely determined. One solution to the moment conditions is $\beta=\beta_0$, $A=A_0$, $B=B_0$, $C=C_0$, and $G=G_0$, where  $\beta_0$, $A_0$, and $B_0$ are as defined (\ref{eq:1}), (\ref{eq:U}), and (\ref{V}), and $C_0$ and $G_0$ are respectively the population coefficients from linear regression of $\tilde{W}_i$ on $\bar{Z}_i$ and $\tilde{X}_i$.

$\beta_0$ could be estimated directly from the moment conditions in Theorem 1 using the Generalized Method of Moments (GMM) (\citet{Hansen1982}). However, such an estimation method is computationally challenging. It involves minimizing a GMM objective jointly over $\beta$ and matrices  $A$, $B$, $C$, and $G$. Further complicating matters, the moment conditions are non-linear in parameters, the solutions $A$, $B$, $C$, and $G$ are only unique up to non-singular transformations, and (for $r\leq d_V$) the rank constraint on $B$ is non-linear.

As we discuss in Section 3, a more computationally expedient estimator can be attained using an alternative characterization of $\beta_0$ and $d_W$ that is equivalent to that in Theorem 1. The equivalent characterization is captured in Corollary 1 below.

\theoremstyle{plain}
\newtheorem*{C1}{Corollary 1}
\begin{C1}
	Under Assumptions 1.1-1.4 $\beta_0$ and $d_W$ are identified. In particular, if and only if $\beta=\beta_0$ then there exists $M\in\mathbb{R}^{d_V\times (d_Z +d_X)}$ and $\xi\in\mathbb{R}^{d_V}$ so that the moment conditions below hold:
	\begin{align}
		E\big[\big(\tilde{V}_{i}-M(\tilde{Z}_{i}',\tilde{X}_{i}')'\big)(\tilde{Z}_{i}',\tilde{X}_{i}')\big]	&=0 \label{mom1st}\\
		E\big[\big(\tilde{Y}_{i}-\beta'\tilde{X}_{i}-\xi'M(\tilde{Z}_{i}',\tilde{X}_{i}')'\big)(\tilde{Z}_{i}',\tilde{X}_{i}')\big]	&=0 \label{mom2nd}
	\end{align}
  The unique solution $M_0$ to (\ref{mom1st}) has $rank(M_0)=d_W$. Moreover, there exists a $\xi$ with $||\xi||_0\leq d_W$ so that $\beta_0$, $M_0$, and $\xi$ satisfy (\ref{mom2nd}).\footnote{$||v||_0$ is the number of non-zero entries in the vector $v$.}
\end{C1}

$\beta_0$ is the unique value of $\beta$ that satisfies the moment conditions in Corollary 1 for some $M$ and $\xi$. There is a unique matrix $M$ that satisfies the moment conditions, and we refer to this solution as $M_0$, and note that $M_0=B_0(C_0,G_0)$. If $d_W\leq d_V$ then there is not a unique choice of $\xi$ with $||\xi_0||_0\leq d_W$ that satisfies the moment conditions. However, it is useful to refer to one particular solution $\xi_0$ which satisfies the moment conditions, we let $\xi_0$ be the (generically unique) solution with minimal $\ell_1$ norm. Note that if the minimal $\ell_1$ solution is unique then $||\xi_0||_0\leq d_W$. $M_0$ and $\xi_0$ are nuisance parameters and are not of direct interest.

Corollary 1 suggest two different means of adapting to the number of confounding factors $d_W$. Firstly, $d_W$ is the rank of $M_0$. Secondly, there is a solution $\beta,M,\xi$ to the moment conditions where $\xi$ has at most $d_W$ non-zero entries. We exploit these results in Subsection 3.2.

\subsection{Discussion}

To the best of our knowledge, the characterization of $\beta_0$ and the number of confounders $d_W$ using the moment condition in (\ref{sparemom}) is original, and similarly for (\ref{mom1st}) and (\ref{mom2nd}).

The characterization of $\beta_0$ is distinct from the characterization using instrumental variables-type moment conditions in \citet{Miao2018} and implicit in \citet{Griliches1977}. As we discuss below, when our assumptions hold and $r<d_V$, our characterization provides additional restrictions that help identify $\beta_0$.

First let us compare with \citet{Miao2018}. For simplicity let us assume there are no additional controls $D_i$. \citet{Miao2018} assume the existence of a function called a `confounding bridge' which then plays a key role in their analysis. A confounding bridge is a function $b$ with the property that for each  $x$ in the support of $X_i$, with probability $1$:
\[E[Y_i|W_i,X_i=x]=E[b(V_i,x)|W_i,X_i=x]\]
Suppose our Assumptions 1.1-1.4 hold and $\epsilon_i$ and $\upsilon_i$ are mean independent of $W_i$ (rather than just uncorrelated with $W_i$), then our model admits a confounding bridge of the form $b(v,x)= \beta_0'x + A_0(B_0' Q B_0)^{-1} B_0' Q v$, where $Q$ is any non-singular matrix and $A_0$ and $B_0$ are as defined in (\ref{eq:U}) and (\ref{V}).\footnote{Under Assumptions 1.1-1.4 $B_0$  has full column rank and so $B_0' Q B_0$ is non-singular. See the proof of Theorem 1.}

\citet{Miao2018} impose assumptions that imply the confounding bridge is unique and point identified. In our model it may be neither unique nor point identified. In fact, under Assumptions 1.1-1.4  the confounding bridge is generally not unique unless $ d_V = d_W $, otherwise it may depend on the matrix $Q$.\footnote{Under Assumptions 1.1-1.4 $B_0'B_0$ is non-singular. If $B_0'B_0$ is non-singular then $(B_0' Q B_0)^{-1}B_0' Q=(B_0'B_0)^{-1} B_0'$ for all non-singular $Q$ if and only if $B_0$ is a square matrix, i.e., $d_W=d_V$. Strictly speaking, even if $d_W\neq d_V$ the confounding bridge may be unique for certain values of $A_0$ (for example, if $A_0$ is a matrix of zeros).} Even if the confounding bridge is unique, in order to identify the bridge, $Z_i$ must be relevant instruments for $V_i$ after controlling for $X_i$ (see Assumption 5 in \citet{Miao2018}). Again, under our assumptions  this is only possible when $d_V=d_W$.

 Applying \citet{Miao2018} in our model amounts to using GMM to estimate solutions $\beta$ and $\gamma$ to the following moment condition:\footnote{\citet{Miao2018} allow the instruments $(Z_i',X_i')$ to be replaced with any vector of transformations $q(Z_i,X_i)$ with finite variance. However, if $q$ is nonlinear then the resulting moment conditions are valid only when  $\epsilon_i$ and $\upsilon_i$ are mean independent of $W_i$ rather than just uncorrelated with $W_i$.}
\begin{equation}
	E\big[(Y_i -\beta'X_i - \xi V_i)(Z_i',X_i')'\big]=0\label{othmom}
\end{equation}

Under Assumptions 1.1-1.4, the condition above is satisfied when $\beta=\beta_0$ and $\xi=A_0(B_0' Q B_0)^{-1} B_0' Q$ for any non-singular $Q$. When $ d_W < d_V $, the solution $\xi$ is generally not unique, however $\beta_0$ may still be point identified by the moment condition.

(\ref{othmom}) is a standard instrumental variables (IV) moment condition. \citet{Griliches1977} suggests a method for estimation with scalar proxy controls that amounts to performing a standard IV procedure to empirically solve (\ref{othmom}).

Corollary 1 imposes additional structure compared to (\ref{othmom}). Firstly, Corollary 1 shows that there is a $\xi$ with $d_W$ non-zero entries that satisfies the moment condition. Secondly, if (\ref{mom1st}) in Corollary 1 holds, then (\ref{othmom}) is equivalent to (\ref{mom2nd}). Thus Corollary 1 supplements (\ref{othmom}) with an additional set of moment conditions (\ref{mom1st}). (\ref{mom1st}) provides additional identifying power when $d_W<\min\{ d_V,d_Z+d_X \} $ (there are fewer confounding factors than proxies). If $d_W$ is unknown, then (\ref{mom1st}) helps to identify $d_W$, because $d_W$ is rank of the lowest-rank solution $M$ to (\ref{mom1st}). 

Conversely, if $d_W = d_V $ then the additional moment conditions (\ref{mom1st}) do not help identify $\beta_0$ because there is no reduced-rank restriction on the solution $M$ to (\ref{mom1st}). Further, if $d_W = d_V $, then the restriction that $\xi$ have only $d_W$ non-zero entries is trivial.

In sum, the results in Theorem 1 and Corollary 1 provide additional identifying power over existing results whenever there are fewer confounding factors than proxies, and also help identify the number of confounding factors. Our results show precisely how the number of latent factors implies sparsity and low-rank structure within the nuisance parameters.  

\section{Estimation}

For a given $r$, we could estimate $\beta_0$ by GMM from the moment condition (\ref{sparemom}) in Theorem 1. However, the moment condition is non-linear in parameters, the objective is generally non-convex, the rank restriction on $B$ is non-linear when $r\leq d_V$, and the solutions $A$, $B$, $C$, and $D$  are (at most) unique up to a normalization. Thus direct minimization of the GMM objective by standard numerical methods may be computationally infeasible, particularly in high dimensions. This problem also applies when $r$ is chosen using model selection (in which case we select an $r$ that estimates $d_W$).

In order to avoid the computational difficulty of joint GMM estimation using (\ref{sparemom}), we suggest a sequential method of moments estimator (see \citet{Newey1994}) based on the moment conditions in Corollary 1. We use the first set of moment conditions (\ref{mom1st}) to estimate the solution $M_0$. We plug the estimate into the moment conditions (\ref{mom2nd}) and use these to estimate the remaining coefficients, including $\beta_0$.

If $d_W\leq \min\{d_V,d_Z+d_X\}$ then the solution $M_0$ to (\ref{mom1st}) has a reduced-rank structure. Imposing this reduced-rank structure in estimation reduces the number of free nuisance parameters that must be estimated. For estimation of $M_0$, we apply methods from reduced-rank regression. These methods provide closed-form solutions, even when the rank is unknown and must be chosen by model selection. We can also impose a sparse structure on an estimate of $\xi_0$ (recall $\xi=\xi_0$ satisfies the moment conditions in Corollary 1 and $||\xi_0||_0\leq d_W$). To induce sparsity we use $\ell_1$-penalization. In the sub-section on adaptive estimation we propose an estimator with a closed-form that does not induce sparsity in the estimate of $\xi_0$, and also a doubly-robust (but more computationally burdensome) estimator that does impose that $\xi_0$ is sparse.

Let us introduce some notation. Let $n$ be the number of available observations, let $X$ be the matrix with $n$ rows whose $i^{th}$ row is equal to $X_i'$ and similarly for $Y$, $V$, $Z$, and $D$. Recall that the partialled out variables $\tilde{X}_i$, $\tilde{Z}_i$, $\bar{V}_i$, etc., depend on population linear regression parameters. In estimation we must replace these population coefficients with sample analogues. For each $H=V,\, Z,\, Y,\, X$ we define the objects below:
\begin{align*}
	\hat{\gamma}_H&=(D'D)^{+}D'H\\
	\hat{\omega}_H&=\big((X,D)'(X,D)\big)^{+}(X,D)'H\\
	\hat{H}&=H-D\hat{\gamma}_H\\
	\check{H}&=H-(X,D)\hat{\omega}_H
\end{align*} 
We also let $\hat{H}_i$ be the transpose of the $i^{th}$ column vector of $\hat{H}$ and $\check{H}_i$ the transpose of the $i^{th}$ column vector of $\check{H}$. Thus $\hat{\gamma}_X$ is a sample estimate of $\gamma_{X,0}$, $\hat{X}_i$ is an estimate of $\tilde{X}_i$, and similarly for the other variables.

\subsection{Estimation With a Given Rank Restriction}

Corollary 1 states that $M_0$ has rank equal to $d_W$, the number of latent confounding factors. If $r$ upper bounds $d_W$ then $rank(M_0)\leq r$. In this sub-section we consider estimation under this restriction on the rank of $M$ for a fixed choice of $r$. Of particular interest is the case of $d_W$ known, then we can set $r=d_W$..

To estimate $\beta_0$ we first find a rank $\leq r$ matrix $\hat{M}_r$ that approximately solves an empirical analogue of (\ref{mom1st}) in Corollary 1. In particular, we let $\hat{M}_r$ solve the minimization problem below:
\begin{equation}
	\hat{M}_r=\underset{rank(M)\leq r}{\text{argmin}}||\hat{V}-(\hat{Z},\hat{X})M'||_F^2\label{RRR1}
\end{equation}
Where $||\cdot||_F^2$ is the squared  Frobenius norm (the sum of the squared entries of the matrix).

$\hat{M}_r$ has a closed-form solution described in \citet{Reinsel1998} and originally due to \citet{Izenman1975}. To describe the solution, let $\hat{\Omega}=(\hat{Z},\hat{X} )'(\hat{Z},\hat{X} )$ and define a matrix $\hat{Q}$ given by:
\begin{equation}
	\hat{Q}=\hat{V}'(\hat{Z},\hat{X} )
\hat{\Omega}^{+}
\big( \hat{V}'(\hat{Z},\hat{X} )\big)'\label{Q}
\end{equation}

Let $\hat{E}=eigen(\hat{Q})$ be the matrix whose columns are the right eigenvectors of $\hat{Q}$ normalized so that $\hat{E}\hat{E}'$ is the identity and ordered so that the $k^{th}$ column of $\hat{E}$ corresponds to the $k^{th}$ largest eigenvalue. Let $\hat{E}_{[:,1:r]}$ denote the sub-matrix of the first $r$ columns of $\hat{E}$. Then we have:
\begin{equation}
\hat{M}_r=\hat{\Omega}^{+} (\hat{Z},\hat{X})'\hat{V}\hat{E}_{[:,1:r]}\hat{E}_{[:,1:r]}'\label{M}
\end{equation}

Having evaluated $\hat{M}_r$ we solve an empirical analogue of (\ref{mom2nd}) with $M$ in the moment condition replaced by $\hat{M}_r$. In particular, our estimate of $\beta_0$  is the vector $\hat{\beta}_r$ that minimizes the least-squares objective below:
\begin{equation}
\hat{\beta}_r= \underset{\beta\in\mathbb{R}^{d_X}}{\text{argmin}}\underset{\xi\in \mathbb{R}^{d_V}}{\text{min}}||\hat{Y}-\hat{X}\beta-(\hat{Z},\hat{X})\hat{M}_r'\xi ||^2 \label{objknown}
\end{equation}
Where $||\cdot||$ is the Euclidean norm.

Corollary 1 states that there exists $\xi$ that satisfies the moment conditions with $||\xi||_0\leq r$. We could impose this restriction by adding $||\xi||_0\leq r$ as a constraint in the minimization problem (\ref{objknown}). That is, instead of taking the minimum over $\xi\in \mathbb{R}^{d_V}$ we could take the minimum over $\xi\in \mathbb{R}^{d_V}:\,||\xi||_0\leq r$. However, if $\hat{M}_r$ is of rank $r$, then adding this restriction has no effect on the resulting estimator of $\beta_0$. This is because for any vector $\xi$ and rank-$r$ matrix $M$, $M'\xi$ always equals $M' \xi^*$ for some vector $\xi^*$ with only $r$ non-zero components and vice-versa.

The minimizer $\hat{\beta}_r$ in (\ref{objknown}) has a closed-form given below.
\[
\hat{\beta}_r=(I_{d_{X}},0_{d_X\times d_V})\big(\hat{J}_r'(\hat{Z},\hat{X})'(\hat{Z},\hat{X})\hat{J}_r\big)^{+}\hat{J}_r'(\hat{Z},\hat{X})'\hat{Y}
\]
$I_{d_{X}}$ is the $d_X\times d_X$ identity matrix, $0_{d_X\times d_V}$ is a $d_X\times d_V$ matrix of zeros, and $\hat{J}_r=\bigg(\begin{pmatrix}0_{d_{X}\times d_{V}}\\
	I_{d_{X}}
\end{pmatrix},\hat{M}_r\bigg)$. The complete procedure is detailed step-by-step in Algorithm 1 below.

 \begin{algorithm}
 	\caption{Sequential estimate with a fixed $r$}
 	\label{alg1}
 	
 	Returns $\hat{\beta}_r$ the estimate of $\beta_0$.
 	\begin{algorithmic}[1]
 		
 		\FOR{$H=W,\, V,\, Z,\, Y,\, X$}
 		\STATE  $\hat{\gamma}_H\leftarrow (D'D)^{+}D'H$
 		\STATE $\hat{H}\leftarrow H-D\hat{\gamma}_H$
 		\ENDFOR
 		\STATE $\hat{\Omega}\leftarrow(\hat{Z},\hat{X} )'(\hat{Z},\hat{X} )$
 		\STATE $\hat{Q} \leftarrow \hat{V}'(\hat{Z},\hat{X} )
 		\hat{\Omega}^{+}
 		\big( \hat{V}'(\hat{Z},\hat{X} )\big)'
 		$
 		
 		\STATE $\hat{E}\leftarrow eigen(\hat{Q})$
 		\STATE $\hat{M}_r=\hat{\Omega}^{+} (\hat{Z},\hat{X})'\hat{V}\hat{E}_{[:,1:r]}\hat{E}_{[:,1:r]}'$
 		\STATE $\hat{J}_r\leftarrow\bigg(\begin{pmatrix}0_{d_{X}\times d_{V}}\\
 			I_{d_{X}}
 		\end{pmatrix},\hat{M}_r\bigg)$
 		
 		\STATE $\hat{\beta}_r\leftarrow(I_{d_{X}},0_{d_X\times d_V})\big(\hat{J}_r'(\hat{Z},\hat{X})'(\hat{Z},\hat{X})\hat{J}_r\big)^{+}\hat{J}_r'(\hat{Z},\hat{X})'\hat{Y}$
 	\end{algorithmic}
 \end{algorithm}

\subsection{Adaptive Estimation}

Algorithm~\ref{alg1} applies for a specific choice of rank restriction $r$ on the estimate of $M_0$. A smaller value of $r$ results in a greater dimension reduction in the nuisance parameters one must estimate. However, if $r< d_W$ (where $d_W$ is the number of unobserved confounders) then the moment conditions are misspecified, i.e., they have no solution. Ideally, $r$ would be chosen to equal $d_W$ which ensures the number of free parameters is minimized and the moment condition correctly specified. However, $d_W$ is generally unknown. We can choose the rank restriction $r$ by model selection and thus adapt to the unknown quantity $d_W$. These methods also result in an estimate of $d_W$, which may be of interest in itself.

We perform model selection by adding penalty terms in estimation so as to induce a low-rank structure on our estimate of $M_0$ and/or a sparse structure on our estimate of $\xi_0$ (recall $\xi_0$ is a choice of $\xi$ that satisfies the moment conditions and has $d_W$ non-zero entries). If we only apply a penalty in the estimation of $M_0$ and use a particular penalty described below, the resulting estimator has a closed-form. If we penalize in both the estimation of $M_0$ and $\xi_0$ this generally results in an estimator without a closed-form solution, but the required numerical minimization can be performed using a standard Lasso estimator (\cite{Tibshirani1996}). The combination of both penalties results in a doubly-robust estimation method at the cost of additional computation.

In this sub-section we present both an estimator with a closed form that only applies penalization in the estimation of $M_0$, and a doubly-robust estimator without a closed-form that applies penalization in estimation of both $M_0$ and $\xi_0$.

Doubly-robust estimators are insensitive to estimation error in any one of a number of nuisance parameters. They are Neyman Orthogonal/Locally Robust, and generally allow for valid inference using a standard root-$n$ Gaussian approximation under weaker conditions than estimators that are not robust (see for example \cite{Chernozhukov2016} and \cite{Chernozhukov2018}).

Let us first describe the estimator of $M_0$ with an adaptive rank restriction. We replace the first-stage objective (\ref{RRR1}) with a penalized objective which we minimize over all $d_V\times (d_Z+d_X)$ matrices $M$ (rather than those with rank weakly less than $r$). The penalized objective is:
\begin{equation*}
	||\hat{V}-(\hat{Z},\hat{X})M'||_F^2+\lambda_n pen(M)
\end{equation*} 
Where $pen(\cdot)$ is a penalty function and $\lambda_n$ is a scalar penalty parameter that may depend on the data. The penalty function should induce a low-rank structure in the minimizer, for example one could let $pen(M)$ return the nuclear norm of $M$, which is the sum of the singular values of $M$. We focus on the case in which $pen(M)$ simply returns the rank of the matrix $M$. The corresponding objective is given below:
\begin{equation}
	||\hat{V}-(\hat{Z},\hat{X})M'||_F^2+\lambda_n rank(M)\label{RRR2}
\end{equation}

Minimization of the objective above is a penalized reduced-rank regression problem as considered in \citet{Bunea2011}, and admits a closed-form solution. \citet{Bunea2011} show that the solution $\hat{M}$ is a matrix of rank $\hat{r}$, where $\hat{r}$ is the number of eigenvalues of $\hat{Q}$ that exceed $\lambda_n$. Recall that $\hat{Q}$ is defined in (\ref{Q}). $\hat{M}$ is equal to $\hat{M}_r$ defined in (\ref{M}) but with $r$ set to the data-dependent $\hat{r}$. Note that $\hat{r}$ can be understood as an estimator of the number of unobserved confounders $d_W$.

Having obtained the matrix $\hat{M}$ that minimizes the penalized objective, we may proceed as in the previous subsection to attain an estimate $\hat{\beta}$ of $\beta_0$. The full algorithm is given in detail Algorithm~\ref{alg2} below. The first five steps provide the estimates $\hat{r}$ and $\hat{M}$. The procedure does not require any numerical optimization nor simulation.

\begin{algorithm}
	\caption{Sequential estimate with rank selection}
	\label{alg2}
	
	Returns estimates $\hat{\beta}_{\hat{r}}$  of $\beta_0$ and $\hat{r}$ of $d_W$.
	
	\begin{algorithmic}[1]
		
		\FOR{$H=W,\, V,\, Z,\, Y,\, X$}
		\STATE  $\hat{\gamma}_H\leftarrow (D'D)^{+}D'H$
		\STATE $\hat{H}\leftarrow H-D\hat{\gamma}_H$
		\ENDFOR
		\STATE $\hat{\Omega}\leftarrow(\hat{Z},\hat{X} )'(\hat{Z},\hat{X} )$
		\STATE $\hat{Q} \leftarrow \hat{V}'(\hat{Z},\hat{X} )
		\hat{\Omega}^{+}
		\big( \hat{V}'(\hat{Z},\hat{X} )\big)'
		$
		\STATE $\hat{E}\leftarrow eigen(\hat{Q})$
		\STATE $\hat{r}\leftarrow \#\{\text{eigenvalues of $\hat{Q}$ $\geq$ $\lambda_n$}\}$
		\STATE $\hat{M}\leftarrow\hat{\Omega}^{+} (\hat{Z},\hat{X})'\hat{V}\hat{E}_{[:,1:\hat{r}]}\hat{E}_{[:,1:\hat{r}]}'$
		\STATE $\hat{J}\leftarrow\bigg(\begin{pmatrix}0_{d_{X}\times d_{V}}\\
			I_{d_{X}}
		\end{pmatrix},\hat{M}\bigg)$
		\STATE $\hat{\beta}_{\hat{r}}\leftarrow(I_{d_{X}},0_{d_X\times d_V})\big(\hat{J}'(\hat{Z},\hat{X})'(\hat{Z},\hat{X})\hat{J}\big)^{+}\hat{J}'(\hat{Z},\hat{X})'\hat{Y}$

	\end{algorithmic}
\end{algorithm}

In our simulations we select the penalty parameter $\lambda_n$ by cross-validation. More precisely, let $\{\mathcal{I}_j\}_{j=1}^J$ be a partition of the indices $\{1,...,n\}$. Let $\mathcal{I}_{-j}$ denote all the elements of $\{1,...,n\}$ that are not in $\mathcal{I}_j$. For a given value of $\lambda$ of the penalty $\lambda_n$, let $\hat{M}_{j,\lambda}$ be the result from carrying out steps 2 to 9 in Algorithm 2 using only the data with indices in $\mathcal{I}_{-j}$. We choose $\lambda_n$ to be the value of $\lambda$ that minimizes the cross validation objective below:
\[
\sum_{j=1}^J\sum_{i\in \mathcal{I}_j}||\hat{V}_i-\hat{M}_{j,\lambda_n}'(\hat{Z}_i',\hat{X}_i')'||^2
\]

\subsubsection{Doubly-Robust Estimation}

Corollary 1 states that there exists a vector $\xi_0$ with at most $d_W$ non-zero entries so that the moment conditions are satisfied when $\xi=\xi_0$, $\beta=\beta_0$, and $M=M_0$.

The estimator in Algorithm~\ref{alg2} does not impose sparsity on an estimate of  $\xi_0$. We propose a doubly-robust estimator that exploits both the sparsity in $\xi_0$ and the low-rank structure of $M_0$.

Before we describe the procedure as a whole, let us motivate an estimator of $\xi_0$ that induces sparsity. Using (\ref{mom1st}) to substitute out $M$ from (\ref{mom2nd}) and then partialling out $\tilde{X}_i$ from the resulting equation, we get the following:
\begin{equation}
E\big[{\bar{Z}_i}'(\bar{Y}_i-\xi_0'\bar{V}_i)\big]=0 \label{ximoment1}
\end{equation}

Our estimate of $\xi_0$ is indirectly based on the moment condition above. However, it is useful to adapt the moment condition so that $\xi_0$ may take the form of a penalized least-squares estimator. Let $M_{0,[:,1:d_Z]}$ be the sub-matrix of $M_0$ that contains its first $d_Z$ columns. If we substitute the definition of $M_0$ into the condition above and multiply both sides by $M_{0,[:,1:d_Z]}$ then we get the following moment condition:
\begin{equation*}
	E\big[M_{0,[:,1:d_Z]}{\bar{Z}_i}(\bar{Y}_i-\xi_0'M_{0,[:,1:d_Z]}\bar{Z}_i )\big]=0
\end{equation*}

$\xi_0$ satisfies the condition above if and only if it minimizes the following least squares criterion:
\[
E\big[(\bar{Y}_i-\xi_0'M_{0,[:,1:d_Z]}\bar{Z}_i )^2\big]
\]
To estimate $\xi_0$, we minimize a penalized empirical analogue of the criterion above. In particular, our estimate of $\xi_0$ is the vector $\xi$ that minimizes the empirical objective below:
\begin{equation}
	||\check{Y}-\check{Z}\hat{M}_{d_V,[:,1:d_Z]}'\xi||^2_F+\delta_n ||\xi||_1 \label{lassobj}
\end{equation}

Where $||\cdot||_1$ is the $\ell_1$ norm and $\delta_n$ is a penalty parameter. $\hat{M}_{d_V,[:,1:d_Z]}$ is the sub-matrix consisting of the first $d_Z$ columns of $\hat{M}_{d_V}$, which is the estimator of $M_0$ defined in Algorithm 1 for $r=d_V$. Note that the estimate $\hat{M}_{d_V}$ does not have a restricted rank. 

Minimization of (\ref{lassobj}) is an $\ell_1$-penalized least squares problem and can be solved using any standard Lasso algorithm. A number of methods are available for selecting the penalty parameter in Lasso regression. For example, $\delta_n$ could be chosen using cross-validation. In our simulations we normalize each of the regressors and the outcomes so that they have mean zero and unit variance and then perform Lasso with the penalty parameter equal to $d_V /n$.

We now turn to the definition of the doubly-robust moment condition. $M_0$ only enters the doubly-robust moment condition indirectly through a parameter $\mu_0$ defined below:
\[
\mu_0=E[\tilde{X}_i(\tilde{Z}_i',\tilde{X}_i')]M_0'\big(M_{0,[:,1:d_Z]}E[\bar{Z}_i\bar{Z}_i']M_{0,[:,1:d_Z]}'\big)^{+}M_{0,[:,1:d_Z]}
\]

Our estimate $\mu_0$ is a sample analogue of the above. We replace the expectations with sample averages, we replace the partialled out variables with their sample counterparts, and we replace $M_0$ with the adaptive reduced-rank estimate $\hat{M}$ from Algorithm~\ref{alg2}. The estimate $\hat{\mu}$ is given below:
\[
\hat{\mu}=\hat{X}'(\hat{Z},\hat{X})\hat{M}'\big(\hat{M}_{[:,1:d_Z]}\check{Z}'\check{Z}\hat{M}_{[:,1:d_Z]}'\big)^{+}\hat{M}_{0,[:,1:d_Z]}
\]

In the above $\hat{M}_{[:,1:d_Z]}$ is the sub-matrix that contains the first $d_Z$ columns of $\hat{M}$.

We can now define the doubly-robust moment condition. The condition involves a number of nuisance parameters. These nuisance parameters are $\mu_0$, $\xi_0$, and all the nuisance parameters involved in partialling out $D_i$ and $X_i$. The nuisance parameters involved in partialling out are $\gamma_{Y,0}$, $\gamma_{V,0}$, $\gamma_{X,0}$, $\omega_{Z,0}$, $\omega_{Y,0}$, and $\omega_{V,0}$.

The doubly-robust moment condition is as follows:\footnote{Note that $\gamma_{X,0}$ appears twice as an argument in the moment condition, this is not an error. $\gamma_{X,0}$ enters the score function in two different places, and for analytical purposes it is useful to think of these two occurrences as separate parameters.}
\[
E[g_i(\beta_0;\xi_0,\mu_0,\gamma_{Y,0},\gamma_{V,0},\gamma_{X,0},\gamma_{X,0},\omega_{Z,0},\omega_{Y,0},\omega_{V,0})]=0
\]
The score function $g_i(\cdot)$ in the moment condition is defined below:
\begin{align*}
	&g_i(\beta;\xi,\mu,\gamma_Y,\gamma_V,\gamma_{X,1},\gamma_{X,2},\omega_Z,\omega_Y,\omega_V)\\
	=&\tilde{X}_i (\gamma_{X,1}) \big(\tilde{Y}_i (\gamma_Y) -\xi ' \tilde{V}_i (\gamma_V) -\beta' \tilde{X}_i (\gamma_{X,2}) \big)
-\mu \bar{Z}_i (\omega_Z) \big(\bar{Y}_i (\omega_Y) -\xi' \bar{V}_i (\omega_V)\big) 
\end{align*}

The moment condition is doubly-robust because it holds when we replace a single one of its nuisance parameter arguments with any alternative value. For example, for any vector $\xi$ (not just the correct value $\xi_0$), the moment condition holds:
\[
E[g_i(\beta_0;\xi,\mu_0,\gamma_{Y,0},\gamma_{V,0},\gamma_{X,0},\gamma_{X,0},\omega_{Z,0},\omega_{Y,0},\omega_{V,0})]=0
\]

We prove the validity of the moment condition (under Assumptions 1.1-1.4) and establish that it is doubly-robust in Appendix B.

The doubly robust estimate $\hat{\beta}_{DR}$ of $\beta$, sets an empirical expectation of the score function to zero when the nuisance parameters are replaced with first-stage estimates. As we discuss below, it may be advantageous to employ sample-splitting in doubly-robust estimation, but for ease of exposition we begin by defining a doubly robust estimator without sample-splitting.

Recall the definitions of the estimates $\hat{\xi}$, $\hat{\mu}$, $\hat{\gamma}_Y$, $\hat{\gamma}_V$, $\hat{\gamma}_X$, $\hat{\omega}_Z$, $\hat{\omega}_Y$, and $\hat{\omega}_V$. $\hat{\beta}_{DR}$ solves:
\[
\frac{1}{n}\sum_{i=1}^n g_i(\hat{\beta}_{DR};\hat{\xi},\hat{\mu},\hat{\gamma}_Y,\hat{\gamma}_V, \hat{\gamma}_X, \hat{\gamma}_X, \hat{\omega}_Z,\hat{\omega}_Y,\hat{\omega}_V)=0
\]
We can write the solution succinctly as follows:
\[
\hat{\beta}_{DR}=\hat{\Sigma}^{+}\big(\hat{X}' \big(\hat{Y} - \hat{V} \hat{\xi} \big)
-\hat{\mu} \check{Z}' (\check{Y} -\check{V} \hat{\xi})\big)/n
\]
In the above $\hat{\Sigma}=\hat{X}'\hat{X}/n$. Recent work including \cite{Chernozhukov2018} and \cite{Chernozhukov2016} shows that there may be advantages to sample-splitting in doubly robust and locally robust estimators. Below we describe a version of the estimator above that employs sample splitting.

We partition the data into $J$ sub-samples. In particular, let $\{\mathcal{I}_j\}_{j=1}^J$ be a partition of $\{1,...,n\}$ and let $n_j$ be the number of entries in $\mathcal{I}_j$. Thus each index $i=1,...,n$ is a member of precisely one subset $\mathcal{I}_j$ and $\sum_{j=1}^J n_j =n$. We will use the shorthand $\mathcal{I}_{-j}$ to denote all the elements of $\{1,...,n\}$ that are not in $\mathcal{I}_j$ (i.e., the complement of $\mathcal{I}_j$).

For each $j=1,...,J$ the researcher evaluates each of the nuisance parameter estimates using only the observations with indices in $\mathcal{I}_{-j}$, that is, the data outside of the $j^{th}$ subsample. Thus, for each $j$, one evaluates estimates $\hat{\xi}_j$, $\hat{\mu}_j$, $\hat{\gamma}_{Y,j}$, $\hat{\gamma}_{V,j}$, $\hat{\gamma}_{X,j}$, $\hat{\omega}_{Z,j}$, $\hat{\omega}_{Y,j}$, and $\hat{\omega}_{V,j}$. These estimates are calculated using only data outside the $j^{th}$ subsample, but are otherwise identical to $\hat{\xi}$, $\hat{\mu}$, $\hat{\gamma}_{Y}$, $\hat{\gamma}_{V}$, $\hat{\gamma}_{X}$, $\hat{\omega}_{Z}$, $\hat{\omega}_{Y}$, and $\hat{\omega}_{V}$  respectively.

The estimate $\hat{\beta}_{DR}$ with sample-splitting satisfies the formula: 
\begin{equation}
\frac{1}{n}\sum_{j=1}^J\sum_{i\in \mathcal{I}_j} g_i(\hat{\beta}_{DR};\hat{\xi}_j,\hat{\mu}_j,\hat{\gamma}_{Y,j},\hat{\gamma}_{V,j},\hat{\gamma}_{X,j},\hat{\gamma}_{X,j},\hat{\omega}_{Z,j},\hat{\omega}_{Y,j},\hat{\omega}_{V,j})=0 \label{drdef1}
\end{equation}
The solution to the above is as follows:
\[
\hat{\beta}_{DR}=\hat{\Sigma}_X^{+}\frac{1}{n}\sum_{j=1}^J\sum_{i\in \mathcal{I}_j}\big(\hat{X}_{i,j} \big(\hat{Y}_{i,j} - \hat{V}_{i,j}' \hat{\xi}_j \big)
+\hat{\mu}_j \check{Z}_{i,j} (\check{Y}_{i,j} -\check{V}_{i,j}' \hat{\xi}_j)\big)
\]
In the above $\hat{\Sigma}_X = \sum_{j=1}^J\sum_{i\in\mathcal{I}_j}\hat{X}_{j,i}\hat{X}_{j,i}'/n$. The full doubly-robust procedure, with sample-splitting, is detailed in Algorithm~\ref{alg3} in Appendix A.

\subsection{Many Additional Covariates}

In some cases the vector of additional covariates $D_i$ may be high-dimensional, that is, there may be many available additional covariates. In addition, we may believe that only a subset of these covariates are linearly predictive of $V_i$, $Z_i$, $X_i$, and $Y_i$. In this case $\gamma_{H,0}$ and/or $\omega_{H,0}$ may be sparse or approximately sparse for some $H\in \{V,Z,X,Y\}$.

To exploit this sparsity or approximate sparsity, we can replace the linear regression estimators $\hat{\gamma}_{H}$ and $\hat{\omega}_{H}$ with Lasso estimators. Let $\eta_{H,n,\gamma}$ and $\eta_{H,n,\omega}$ be a scalar penalty parameter. For $H=V, \, Z,\, X,\, Y$ we can define the following alternative estimators $\hat{\gamma}_{H}$ and $\hat{\omega}_{H}$:
\begin{align*}
	\hat{\gamma}_{H}&=\underset{\gamma\in\mathbb{R}^{d_D}}{\text{argmin}}\sum_{i=1}^n (H_i-\gamma D_i)^2+\eta_{H,n,\gamma}||\gamma||_1\\	\hat{\omega}_{H}&=\underset{\omega\in\mathbb{R}^{d_D+d_X}}{\text{argmin}}\sum_{i=1}^n \big(H_i-\omega (X_i',D_i')'\big)^2+\eta_{H,n,\omega}||\omega||_1
\end{align*}

We can then use these estimates in place of the linear regression estimates in the algorithms detailed in this section.

A number of methods exist for choosing the penalty parameters in Lasso regression, for example cross-validation.

\section{Inference and Consistency}

The methods in the previous section estimate a low-dimensional parameter of interest $\beta_0$ (and perhaps $d_W$) in the presence of possibly high-dimensional nuisance parameters. We take the standard approach to asymptotic analysis in such settings which is to find conditions under which the estimates are root-$n$ consistent and admit an asymptotic Gaussian approximation. We focus on the doubly-robust estimator with sample-splitting $\hat{\beta}_{DR}$ detailed in the previous section.

The doubly-robust estimator is a Double-Machine Learning 2 (DML2) estimator of the kind analyzed in section 3.2 in \citet{Chernozhukov2018}. DML2 estimators (along with the DML1 estimators in \citet{Chernozhukov2018}) have the advantage that they are root-$n$ consistent and asymptotically normal centered at the true parameter under relatively weak conditions on the rates at which the nuisance parameters converge.

\citet{Chernozhukov2018} suggest a variance estimator for DML2 estimators.  In the case of $\hat{\beta}_{DR}$ this simplifies to:
\[
\hat{\sigma}^2=\frac{1}{n}\sum_{j=1}^J\sum_{i\in \mathcal{I}_j}\hat{\Sigma}_X^{+}\hat{g}_i\hat{g}_i '\hat{\Sigma}_X^{+}
\]
For each $j\in1,...,J$ and $i\in\mathcal{I}_j$, $\hat{g}_i$ is defined as follows:
\[
\hat{g}_i= g_i(\hat{\beta}_{DR};\hat{\xi}_j,\hat{\mu}_j,\hat{\gamma}_{Y,j},\hat{\gamma}_{V,j},\hat{\gamma}_{X,j},\hat{\gamma}_{X,j},\hat{\omega}_{Z,j},\hat{\omega}_{Y,j},\hat{\omega}_{V,j})
\]
$\hat{g}_i$ is a sample analogue of $g_i$, which is defined by:
\[
{g}_i= g_i(\beta_0;\xi_0,\mu_0,\gamma_{Y,0},\gamma_{V,0},\gamma_{X,0},\gamma_{X,0},\omega_{Z,0},\omega_{Y,0},\omega_{V,0})
\]
If the variance estimator is consistent and $\hat{\beta}_{DR}$ is asymptotically  Gaussian centered at $\beta_0$, then a confidence interval for $l'\beta_0$ (where $l$ is some vector) can be obtained as follows: 
\[
CI=\big[l'\hat{\beta}\pm \Phi^{-1}(1-\alpha /2)\sqrt{l'\hat{\sigma}^2 l/n}\big]
\]
The formula above is suggested in \citet{Chernozhukov2018}. $\Phi$ is the cumulative distribution function of a standard Gaussian random variable.

We now present high-level assumptions for root-$n$ consistency and asymptotic normality of the doubly-robust estimator with sample splitting as defined in (\ref{drdef1}). Note that our results apply for any choice of estimators for the nuisance parameters not just those specified in Section 3. 

  Our asymptotic analysis is based on Theorems 3.1 and 3.2 in  \citet{Chernozhukov2018}. The Assumptions 1.1-1.4 and  4.1-4.3 (stated below) act as primitive conditions for the assumptions in that paper.

In order to derive results that are uniform over some parameter space, we suppose that for each sample size $n$, the data generating process, denoted by $P$, belongs to some set $\mathcal{P}_n$. The Assumptions below then restrict $\mathcal{P}_n$. 

For notational convenience, for any random column vector $H_i$ we let $\Sigma_H=E[H_i H_i']$, however in the case of $H_i=(X_i',D_i')'$ we write $\Sigma_{XD}$. As in previous sections, if $b$ is a vector then $||b||$ is the Euclidean norm of $b$. If $A$ is a matrix then $||A||$ is the Euclidean matrix norm of $A$, so if $A$ has $d$ rows then $||A||=\sup_{b\in\mathbb{R}^d: ||b||=1}||Ab||$. For a positive semi-definite matrix $A$, $A^{1/2}$ is the unique positive semi-definite matrix $B$ so that $BB=A$ and if $A$ is strictly positive definite then $A^{-1/2}$ is the inverse of $A^{1/2}$.

\theoremstyle{definition}
\newtheorem*{A4.1}{Assumption 4.1 (Restrictions on the DGP)}
\begin{A4.1}
If $P\in\mathcal{P}_n$ the following hold with all scalars on the right-hand sides finite. i. $\Sigma_X$, $\Sigma_{\tilde{X}}$ ,$\Sigma_{\tilde{V}}$, $\Sigma_{\bar{V}}$, $\Sigma_{\bar{Z}}$, $\Sigma_{D}$, and $\Sigma_{XD}$, are finite, and strictly positive definite. $E[\tilde{Y}_{i}^{2}]>0$ and $E[\bar{Y}_{i}^{2}]>0$. ii. $E[{Y}_{i}^{2}]\leq\sigma_{Y}^2$,
$||\Sigma_X^{1/2} \beta_{0}||\leq\bar{\beta}$,
$||\Sigma_{X}||\leq \sigma_{X}^2$, $||\Sigma_{\tilde{V}}^{1/2} \xi_{0}||\leq\bar{\xi}$,
$||\mu_{0}\Sigma_{\bar{Z}}^{1/2}||\leq\bar{\mu}$.  iii. With probability 1 for $H=X,V,Y$   $||\Sigma_{\tilde{H}}^{-1/2}E[\tilde{H}_{i}\tilde{H}_{i}'|{D}_{i}]^{1/2}||\leq\bar{\sigma}_{\tilde{H}|D}$, for $H=Z,V,Y$ $
 ||\Sigma_{\bar{H}}^{-1/2}E[\bar{H}_i \bar{H}_i '|X_{i},D_i]^{1/2}||\leq\bar{\sigma}_{\bar{H}|XD}
 $, 
  $||\Sigma_{\tilde{V}}^{-1/2}E[\tilde{V}_{i}\tilde{V}_{i}'|\tilde{X}_{i}]^{1/2}||\leq \bar{\sigma}_{\tilde{V}|\tilde{X}}$,   $||\Sigma_{\bar{V}}^{-1/2}E[\bar{V}_{i}\bar{V}_{i}'|\bar{Z}_{i}]^{1/2}||\leq \bar{\sigma}_{\bar{V}|\bar{Z}}$, and $E[\bar{Y}_i^2|\bar{Z}_i]/E[\bar{Y}_i^2]\leq \bar{\sigma}_{\bar{Y}|\bar{Z}}^2$. 
iv. $E[||\Sigma_D^{-1/2} D_i||^4]\leq d_D s_D^2$ and $E[||\Sigma_{XD}^{-1/2} (X_i',D_i')'||^4]\leq (d_D+d_X) s_{XD}^2$. 

\end{A4.1}

\theoremstyle{definition}
\newtheorem*{A4.2}{Assumption 4.2 (Convergence rates of the nuisance parameter estimates)}
\begin{A4.2}
There is a sequence $\alpha_n$ with $\alpha_n\to 0$ so that if $P\in\mathcal{P}_n$ then  
with probability at least $1-\alpha_n$ the following hold for $j=1,...,J$. 
i. $||(\hat{\mu}_j-\mu_{0})\Sigma_{\bar{Z}}^{1/2}||\leq\delta_{\mu,n}$.
ii. $||\Sigma_V^{1/2}(\hat{\xi}_j-\xi_{0})||\leq\delta_{\xi,n}$.
iii. For $H=X,V,Y$, $||\Sigma_{\tilde{H}}^{-1/2}(\hat{\gamma}_{H,j}-\gamma_{H,0})\Sigma_D^{1/2}||\leq\delta_{\gamma,H,n}$.
iv. For $H=Z,V,Y$, $
||\Sigma_{\bar{H}}^{-1/2}(\hat{\omega}_{H,j}-\omega_{H,0})\Sigma_{XD}^{1/2}||\leq\delta_{\omega,H,n}$.
\end{A4.2}

\theoremstyle{definition}
\newtheorem*{A4.3}{Assumption 4.3 (Higher order moments)}
\begin{A4.3}
For some $q>2$, if $P\in\mathcal{P}_n$ then the following inequalities hold with the scalars on the right-hand sides finite.  
i.  $E\big[||\tilde{X}_{i}\tilde{V}_{i}'\xi_{0}||^{q}\big]^{1/q}\leq\bar{S}_{q}$, $E\big[||\mu_{0}\bar{Z}_{i}\bar{Y}_{i}||^{q}\big]^{1/q}\leq\bar{S}_{q}$, and $E\big[||\mu_{0}\bar{Z}_{i}\bar{V}_{i}'\xi_{0}||^{q}\big]^{1/q}\leq\bar{S}_{q}$ ii. $E\big[||\Sigma_{D}^{-1/2}D_{i}D_{i}\Sigma_{D}^{-1/2}||^{q}\big]\leq d_{D}^{2}S_{q,D,D}$, for $H=X,Y,V$, $E\big[||\Sigma_{\tilde{X}}^{-1/2}\tilde{X}_{i}\tilde{H}_{i}\Sigma_{\tilde{H}}^{-1/2}||^{q}\big]\leq d_{X}d_{H}S_{q,\tilde{X},\tilde{H}}$ and for $H=X,Y,V$  $E\big[||\Sigma_{D}^{-1/2}D_{i}\tilde{H}_{i}\Sigma_{\tilde{H}}^{-1/2}||^{q}\big]\leq d_{D}d_{H}S_{q,D,\tilde{H}}$. iii. For $H=Z,Y,V$,  $E\big[||\Sigma_{\bar{Z}}^{-1/2}\bar{Z}_{i}\bar{H}_{i}\Sigma_{\bar{H}}^{-1/2}||^{q}\big]\leq d_{Z}d_{H}S_{q,\bar{Z},\bar{Z}}$ and $E\big[||\Sigma_{XD}^{-1/2}(X_{i}',D_{i}')'\bar{H}_{i}\Sigma_{\bar{H}}^{-1/2}||^{q}\big]\leq(d_{X}+d_{D})d_{H}S_{q,XD,\bar{Z}}$.

\end{A4.3}

Assumption 4.1 imposes bounds on the magnitudes of some population objects. The bounds must apply for all sample sizes $n$. The bounds given in the assumption do not change with the sample size apart from in the case of 4.1.iv. The upper bound $d_D s_D^2$ in 4.1.iv may grow with the sample size at because $d_D$, the number of additional controls, may grow with $n$, likewise for $(d_D + d_X ) s_{XD}^2$. This flexibility is important because $E[||\Sigma_D^{1/2} D_i||^4]$ is equal to the sum of fourth moments of each of the $d_D$ components of $\Sigma_D^{1/2} D_i$. Suppose the bound on $E[||\Sigma_D^{1/2} D_i||^4]$ were fixed, this would imply that as $d_D$ grows, the average fourth moment of each component of $\Sigma_D^{1/2} D_i$ would have to go to zero. Note that the bounds in 4.1.iii are trivially satisfied under conditional homoskedasticity. In that case each of the upper bounds in 4.1.iii simply equals $1$.

Assumption 4.2 simply imposes convergence rates for each of the nuisance parameter estimates. Note that the convergence rates are required to hold uniformly over sequences of DGPs in $\{\mathcal{P}_n\}_{n=1}^\infty$.

Assumption 4.3 imposes some bounds on higher-order moments. This condition is required to hold for consistency of the variance estimate $\hat{\sigma}$ and thus for asymptotic validity of the confidence intervals.

\theoremstyle{plain}
\newtheorem*{Th2}{Theorem 2}
\begin{Th2}
	
	Suppose that for each $n$, $P\in\mathcal{P}_{n}$ so that Assumptions
	1.1-1.4, Assumptions 4.1, 4.2, and 4.3 all hold, $d_{X}$ is fixed,
	and all of the eigenvalues of $E[g_{i}g_{i}']$ are bounded below
	away from zero. Suppose that the nuisance parameter estimates are
	all consistent. That is, $\delta_{\mu,n}\prec1$, $\delta_{\xi,n}\prec1$,
	for $H=X,Y,V$ we have $\delta_{\gamma,H,n}\prec1$, and for $H=Z,Y,V$
	$\delta_{\omega,H,n}\prec1$. Moreover, suppose that:
	\begin{align}
		(\sqrt{d_{D}}+\sqrt{n})\delta_{\gamma,X,n}(\delta_{\gamma,Y,n}+\delta_{\gamma,V,n}+\delta_{\gamma,X,n}) & \prec1 \label{asy1}\\
		(\sqrt{d_{D}}+\sqrt{n})\delta_{\omega,Z,n}(\delta_{\omega,Y,n}+\delta_{\omega,V,n}) & \prec1 \label{asy2}
	\end{align}
	
	Finally suppose that there exits a constant $c$ so that:
	\begin{align}
		c\geq & d_{V}^{1/q}\delta_{\xi,n}+d_{D}^{2/q}(\delta_{\gamma,Y,n}+\delta_{\gamma,V,n}+\delta_{\gamma,X,n})+(\delta_{\mu,n}+\delta_{\xi,n})d_{Z}^{1/q}d_{V}^{1/q} 
		\nonumber \\
		+ & d_{D}^{1/q}(d_{Z}^{1/q}+d_{D}^{1/q}\delta_{\omega,Z,n})\big(\delta_{\omega,Y,n}+\delta_{\omega,V,n}\big) \label{asy3}
	\end{align}
	
	Where $q>2$ is the constant in Assumption 4.3. Then uniformly over
	all $P\in\mathcal{P}_{n}$, $\hat{\beta}$ is root-$n$ consistent
	and asymptotically normal:
	\[
	\sqrt{n}\sigma^{-1}(\beta_{0}-\hat{\beta})\rightsquigarrow N(0,I)
	\]
	
	Where the asymptotic variance $\sigma$ is given by: 
	$
	\sigma=\Sigma_{X}^{-1}E[g_{i}g_{i}']\Sigma_{X}^{-1}
	$. 	Moreover, the variance estimator $\hat{\sigma}$ is consistent for
	$\sigma$ and the confidence described earlier in this section have
	asymptotically correct coverage.

\end{Th2}

Theorem 2 establishes uniform root-$n$ consistency of the estimator and asymptotic validity of the confidence intervals.  The theorem requires conditions (\ref{asy1}), (\ref{asy2}), and (\ref{asy3}). One set of primitive conditions for (\ref{asy1}) and (\ref{asy2}) is that $d_D$ grows more slowly than $n$ and all of the nuisance parameter estimates go to zero at a strictly faster rate than $n^{-1/4}$.

The condition (\ref{asy3}) directly restricts the rate at which $d_D$, $d_V$, and $d_Z$ may grow with the sample size. The condition is weaker when $q$ is large. Indeed, if the nuisance parameter estimates are consistent then (\ref{asy3}) must hold for a sufficiently large value of $q$. However when $q$ is larger then Assumption 4.3  requires the existence of even higher order moments.

\section{Simulation Study}

In order to assess the efficacy of the methods we present in Section 3 we carry out a Monte Carlo simulation. We implement our methods on a number of simulated datasets. For each simulation, we draw observations independently and identically from the following model:

\begin{align*}
	V_i&=B_0 W_i+ \upsilon_i\\
	X_i&=T_0 W_i+ \epsilon_i\\
	Z_i&=C_0 W_i+ G_0 X_i + \eta_i\\
	Y_i&=X_i'\beta_0+F_0 W_i + \chi_0 V_i + e_i
\end{align*}

The residuals $\upsilon_i$, $\epsilon_i$, $\eta_i$, and $e_i$ are drawn independently of each other from zero mean Gaussian distributions: $W_i\sim N(0,I)$, $\upsilon_i\sim N(0,\Sigma_V)$, $\epsilon_i\sim N(0,\Sigma_X)$, $\eta_i\sim N(0,\Sigma_Z)$, and $e_i\sim N(0,\Sigma_Y)$. Note that we do not include additional controls $D_i$ in our simulations.

In each simulation we must choose parameters $\beta_0$, $B_0$, $C_0$, $G_0$, $T_0$, $F_0$, $\chi_0$, $\Sigma_Y$, $\Sigma_V$, $\Sigma_X$, and $\Sigma_Z$. Rather than use a fixed value of each parameter in all of our simulations, we draw the parameters at random in each simulation. Thus our simulation results show the weighted average performance of our estimators over a parameter space.

We draw the parameters as follows. The elements of the coefficient matrices $\beta_0$, $B_0$, $C_0$, $G_0$, $T_0$, $F_0$, and $\chi_0$ are all independently mean-zero normal with variance equal to the square root of the number of columns of the matrix. For example, the elements of $F_0$ are all independent with distribution $N(0,1/\sqrt{d_W})$. This choice of the variances of the normal distributions ensures that the ratio of the variance in each variable to the residual variance remains roughly constant as the dimension changes.

The covariance matrices have a re-scaled inverse Wishart distribution, for example $d_V p\Sigma_V^{-1} \sim W_{d_V}(I,d_V p)$. The natural number $p$ is a hyper-parameter that determines the degrees of freedom of the Wishart distribution. 

We are left with hyperparameters $p$, $d_W$, $d_X$, $d_V$, $d_Z$, and the sample size $n$. In all of our simulations we let $d_X=1$ so that there is a single treatment of interest. We set $p=2$ which means the covariance matrices are concentrated around the identity. In all of our simulations $d_Z=d_V$ so there are the same number of proxies in $Z_i$ as in $V_i$. We carry out simulations for a range of choices for the remaining hyperparameters $d_W$, $d_V$, and $n$.

\begin{figure}[h]
	\subfloat[$d_W=10$]{
		\includegraphics[scale=0.20]{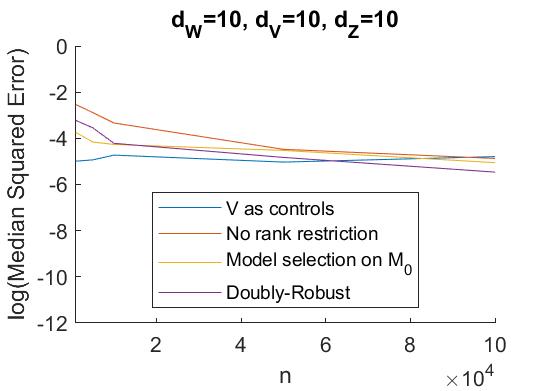}
		\includegraphics[scale=0.20]{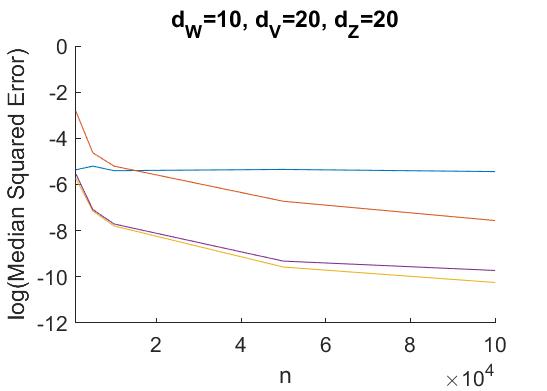}
		\includegraphics[scale=0.20]{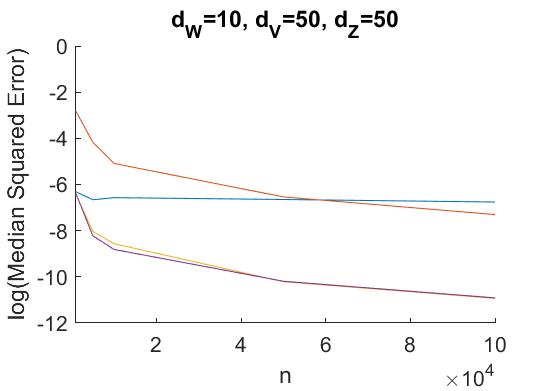}}\\
	\subfloat[$d_W=20$]{
		\includegraphics[scale=0.20]{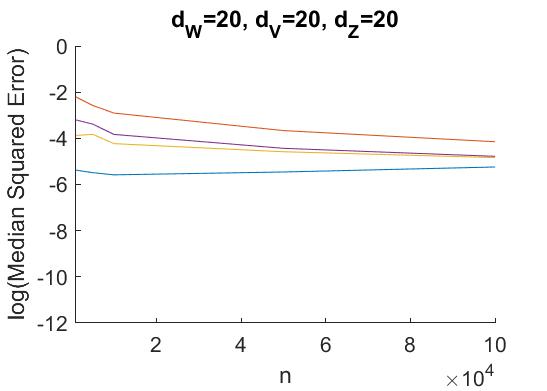}
		\includegraphics[scale=0.20]{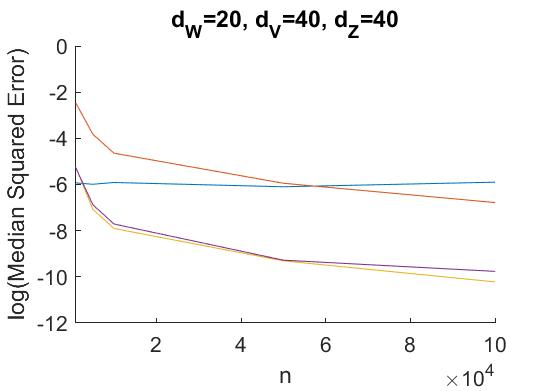}
		\includegraphics[scale=0.20]{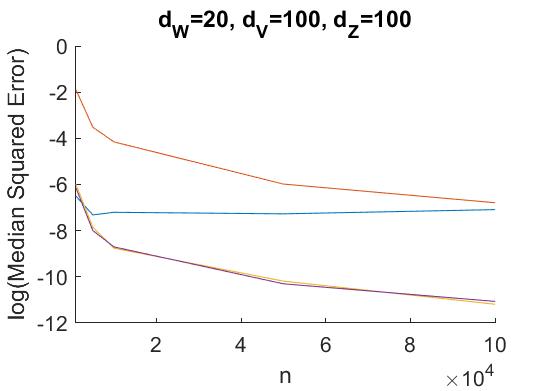}}\\
	\caption{Simulated Median Squared Errors}
	{\scriptsize{}}%
	\noindent\begin{minipage}[t]{1\columnwidth}%
		{\scriptsize{}Median Squared Errors on the y-axes are the medians  of $||\hat{\beta}-\beta_0||^2$ over $500$ simulated datasets for various estimators $\hat{\beta}$. The different figures correspond to different choices for the number of confounding factors $d_W$, and the numbers of proxies $d_V$ and $d_Z$.}%
	\end{minipage}{\scriptsize\par}
\end{figure}

Figure 5.1 shows the mean-squared errors of alternative estimators for a variety of different hyperparameters. The estimators that are compared are: a naive least-squares estimator that simply treats $V_i$ as a set of controls, and three different proxy control estimators. These are the proxy control estimator in Algorithm 1 with $r=d_V$ (i.e, no rank restriction), the computationally expedient adaptive estimator in Algorithm 2, and the doubly-robust estimator with sample-splitting in Algorithm 3.

The proxy control estimator with no rank restriction is equivalent to the two-stage least squares strategy of \citet{Griliches1977} in which ${V}_i$ is a vector of endogenous regressors, ${X}_i$ is a vector of exogenous regressors, and ${Z}_i$ is a vector of instruments. In all cases shows in Figure 5.1 this estimator performs worse (in terms of median squared error) than the other two proxy control estimators. The difference is particularly stark when the ratio of the number of proxies to the number of confounders is large.

In the left-most sub-figures in Figure 5.1, the number of proxies in each group is the same as the number of unobserved confounders. Therefore there is no reduced-rank structure or sparsity in the true parameters. Nonetheless, the model selection estimators out-perform the unrestricted estimator. This may be due to the regularizing effect of the reduced-rank regression and $\ell_1$ penalization.

The rank-selection estimator in Algorithm 2 and doubly-robust estimator perform similarly in all simulations, particularly in large samples. The double-robustness of the estimator in Algorithm 3 is intended to reduce the bias, but may come at the cost of increasing the variance.

The naive estimator that treats $V_i$ as a set of controls is necessarily inconsistent, and indeed the mean squared error of this estimator is stable for all but the smallest sample sizes shown in the figures. Nonetheless, the estimator regularly out-performs the unrestricted proxy control estimator in small samples, and it outperforms this estimator even in the largest samples in the most high dimensional case where $d_V=d_Z=100$. The naive method involves estimation of a $d_V$-dimensional nuisance parameter, whereas in the unrestricted proxy control method the nuisance parameters have dimension $d_V(1+d_Z + d_X)$. The larger dimension of the nuisance parameters may lead to high variance of the unrestricted proxy control method, which may dominate the bias in the naive estimator in finite samples.

The naive estimator out-performs all the other methods when $d_W=d_V=d_Z=20$, even in large samples. When $d_W=d_V=d_Z=10$ the naive estimator out-performs the other methods for sample-sizes in the bottom half of the range we examine, but performs worse that all three proxy methods in the largest samples. When the number of proxies in each group exceeds the number of confounders, the adaptive proxy control methods substantially outperformed the naive estimator in all but the very smallest sample sizes we examine.
\begin{figure}[h]
	\subfloat[$d_W=10$]{
		\includegraphics[scale=0.20]{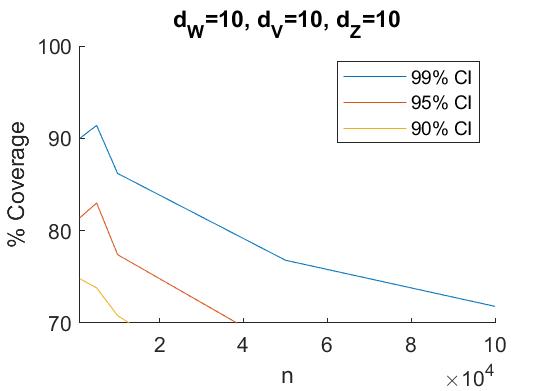}
		\includegraphics[scale=0.20]{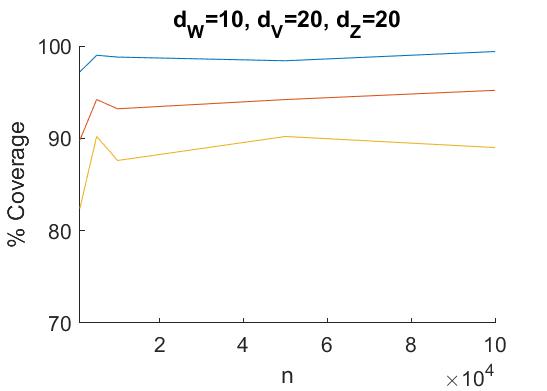}
		\includegraphics[scale=0.20]{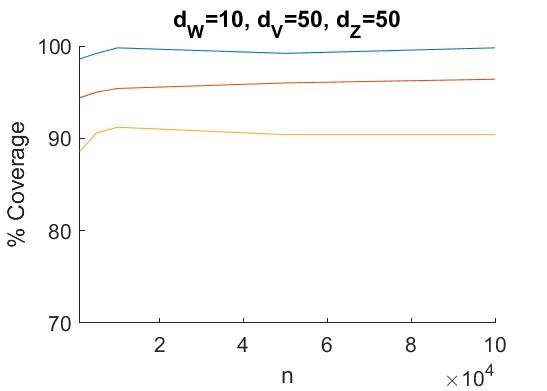}}\\
	\subfloat[$d_W=20$]{
		\includegraphics[scale=0.20]{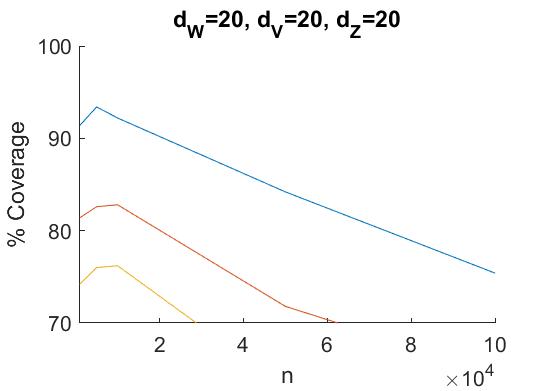}
		\includegraphics[scale=0.20]{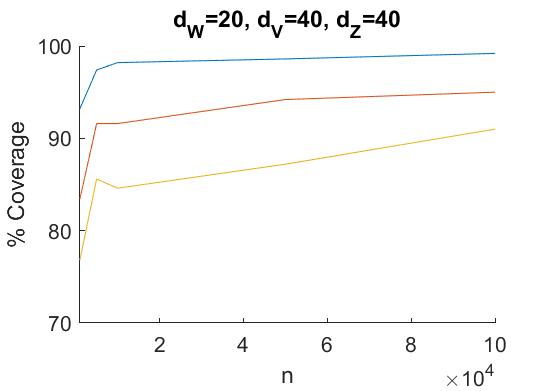}
		\includegraphics[scale=0.20]{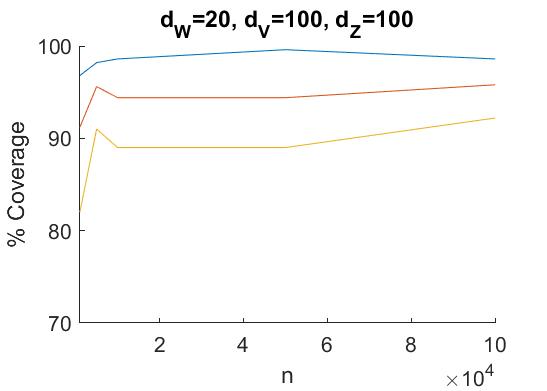}}\\
	\caption{Simulated Confidence Interval Coverage}
	{\scriptsize{}}%
	\noindent\begin{minipage}[t]{1\columnwidth}%
		{\scriptsize{}Confidence interval coverage of the treatment parameter. on the y-axes are percentages of $500$ simulated datasets in which confidence intervals contain $\beta_0$. The different figures correspond to different choices for the number of confounding factors $d_W$, and the numbers of proxies $d_V$ and $d_Z$.}%
	\end{minipage}{\scriptsize\par}
\end{figure}

Figure 5.2 shows the percentage of simulations in which $99\%$, $95\%$, and $90\%$ confidence intervals cover the true parameter $\beta_0$ (recall $\beta_0$ is drawn at random in each simulation). The confidence intervals are those based on a Gaussian approximation for the doubly-robust estimator as described in Section 4.

We see from the left-most sub-figures that the confidence intervals greatly under-cover when the number of proxies in $V_i$ and $Z_i$ equal the number of confounding factors. When there are more proxies than confounders the coverage is close to the desired levels, particularly in large samples.

In Table 1 we give the proportion of simulations in which the rank $r$ selected according to Algorithm 2, is equal to the number of confounders $d_W$ (which is the rank of the matrix $M_0$). In each row we see that the probability of correctly selecting the rank increases with the sample size. For all choices of $d_W$, $d_V$, and $d_Z$ other than those in the first two rows of the table, the algorithm selects the correct rank in at least 87\% of simulations when the sample size is $10,000$ or higher, and at least 69\% of the time when the sample size is $5000$ or higher.

However, when $d_V=d_Z=d_W=10$ the algorithm selects the incorrect rank in over 80\% of simulations, even with $100,000$ observations. When $d_V=d_Z=d_W=20$ the incorrect rank is chosen over 90\% of the time. This may explain the under-coverage when $d_V=d_Z=d_W$ as seen in Figure 5.2. In these cases the algorithm regularly selects a rank that is smaller than the true rank (the algorithm cannot select a rank smaller than the minimum of $d_V$ and $d_Z$), and this leads to bias in the resulting coefficient estimates. This bias then leads to under-coverage of the confidence intervals, and because the performance of the rank selection only gradually increases with the sample size, this bias decreases more slowly than the variance, leading to increasingly poor coverage as the sample size grows.

%\begin{landscape}
\begin{table}[h]
	\caption{Frequency of Correct Rank Selection}
	\centering
	\begin{tabular}{c|c||c|c|c|c|c}
		\multicolumn{1}{c}{} & \multicolumn{1}{c}{} & \multicolumn{5}{c}{\underline{Sample Size}}\tabularnewline
		\multicolumn{1}{c}{$d_{W}$} & \multicolumn{1}{c}{$d_{V}=d_Z$} & \multicolumn{1}{c}{$1000$} & \multicolumn{1}{c}{$5000$} & \multicolumn{1}{c}{$10000$} & \multicolumn{1}{c}{$50000$} &
		\multicolumn{1}{c}{$100000$}
		\tabularnewline
		\specialrule{.2em}{.1em}{.1em} 
		10  & 10  &  0.048  &  0.062  &  0.078	& 0.108   & 0.138 \tabularnewline
		20  & 20  &  0.068  &  0.066  &  0.072 & 0.076   & 0.088\tabularnewline
		10  & 20  &  0.478  &  0.922  &  0.984 & 0.992   & 0.986\tabularnewline
		20  & 40  &  0.204  &  0.696  &  0.876  & 1.000   & 1.000\tabularnewline
		10  & 50  &  0.700  &  0.998  &  1.000 &  1.000  &  1.000\tabularnewline
		20  & 100 &  0.062  &  0.940  &  1.000 &  1.000  &  1.000\tabularnewline
	\end{tabular}
	{\scriptsize{}}%
	\noindent\begin{minipage}[t]{1\columnwidth}%
		{\scriptsize{}Figures are the proportion of the $500$ simulated datasets in which the $r$ selected by Algorithm 2 is equal to $d_W$. Rows corresponds to different choices of $d_W$, $d_V$, and $d_Z$, columns correspond to different choices of the sample size $n$.}%
	\end{minipage}{\scriptsize\par}
	
\end{table}
%\end{landscape}

\section{Conclusion}

We present novel identification results for the linear model with proxy controls. Our identification results suggest method of moments estimators that can take advantage of the dimension reduction when the number of unobserved confounding factors is smaller than the number of proxies. We present model selection methods that adapt to the unknown number of confounding factors. We provide conditions for uniform root-$n$ consistency of our estimates and asymptotic validity of an inference procedure. Our simulation results suggest that our estimators are more effective than proxy control methods that do not exploit the dimension reduction, particularly when the the number of proxies substantially exceeds the number of unobserved confounders. In the latter case inference based on our doubly-robust adaptive proxy control method performs well.

\bibliographystyle{authordate1}
\bibliography{manyproxiesrefs}

\appendix

\section{Algorithm 3}

Below is a full step-by-step description of the doubly robust estimation procedure with sample splitting.

\begin{algorithm}[h]
	\caption{Doubly-robust estimation with sample-splitting}
	\label{alg3}
	
	Returns estimates $\hat{\beta}_{DR}$  of $\beta_0$.
	\begin{algorithmic}[1]
		\FOR{$j=1,2,...,J$}
		\FOR{$H=W,\, V,\, Z,\, Y,\, X$}
		\STATE $\hat{\gamma}_{H,j}\leftarrow (\sum_{i\in\mathcal{I}_{-j}}D_i D_i')^{+} \sum_{i\in\mathcal{I}_{-j}}D_i H_i$
		\STATE $\hat{\omega}_{H,j}\leftarrow \big(\sum_{i\in\mathcal{I}_{-j}}(X_i',D_i')'(X_i',D_i')\big)^{+} \sum_{i\in\mathcal{I}_{-j}}(X_i',D_i')' H_i$
		\FOR{$i\in \mathcal{I}_j$}
		\STATE $\hat{H}_{i}\leftarrow H_{i}-\hat{\gamma}_{H,j}'D_i $
		\STATE $\check{H}_{i}\leftarrow H_{i}-\hat{\omega}_{H,j}'(X_i',D_i')'$
		\ENDFOR
		\ENDFOR
		\STATE $\hat{\Omega}_j\leftarrow\sum_{i\in\mathcal{I}_{-j}}(\hat{Z}_{j,i}',\hat{X}_{j,i}' )'(\hat{Z}_{j,i}',\hat{X}_{j,i}' )$
		\STATE $\hat{Q}_j \leftarrow \sum_{i\in\mathcal{I}_{-j}}\hat{V}_{j,i}(\hat{Z}_{j,i}',\hat{X}_{j,i}')
		\hat{\Omega}_j^{+}
		\sum_{i\in\mathcal{I}_{-j}} \hat{V}_{j,i}(\hat{Z}_{j,i}',\hat{X}_{j,i}' )'
		$
		\STATE $\hat{E}_j\leftarrow eigen(\hat{Q}_j)$
		\STATE $\hat{r}_j\leftarrow \#\{\text{eigenvalues of $\hat{Q}$ $\geq$ $\lambda_n$}\}$
		\STATE $\hat{M}_j\leftarrow\hat{\Omega}_j^{+} \sum_{k\in\mathcal{I}_{-j}}(\hat{Z}_{j,i}',\hat{X}_{j,i}')'\hat{V}_{j,i}'\hat{E}_{j,[:,1:\hat{r}_j]}\hat{E}_{j,[:,1:\hat{r}_j]}'$
		\STATE $\hat{\mu}_j\leftarrow\sum_{i\in\mathcal{I}_{-j}}\hat{X}_{j,i}(\hat{Z}_{j,i}',\hat{X}_{j,i}')\hat{M}_j'\big(\hat{M}_{j,[:,1:d_Z]}\sum_{i\in\mathcal{I}_{-j}}\check{Z}_{j,i}\check{Z}_{j,i}'\hat{M}_{j,[:,1:d_Z]}'\big)^{+}\hat{M}_{j,[:,1:d_Z]}$
		\STATE $\hat{M}_{j,d_V}\leftarrow\hat{\Omega}_j^{+} \sum_{i\in\mathcal{I}_{-j}}(\hat{Z}_{j,i}',\hat{X}_{j,i}')'\hat{V}_{j,i}'$
		\STATE 	$\hat{\xi}_j\leftarrow\underset{\xi\in\mathbb{R}^{d_V}}{\text{argmin}}\sum_{i\in\mathcal{I}_{-j}}\big(\check{Y}_i-\check{Z}_{j,i}\hat{M}_{j,d_V,[:,1:d_Z]}'\xi\big)^2+\delta_n ||\xi||_1$
		\STATE $\hat{\Sigma}_X \leftarrow \sum_{j=1}^J\sum_{i\in\mathcal{I}_j}\hat{X}_{j,i}\hat{X}_{j,i}'/n$
		\ENDFOR
		\STATE $\hat{\beta}_{DR}\to \tilde{\Sigma}^{+}\frac{1}{n}\sum_{j=1}^J\sum_{i\in \mathcal{I}_j}\big(\hat{X}_{i,j} \big(\hat{Y}_{i,j} - \hat{V}_{i,j}' \hat{\xi}_j \big)
		-\hat{\mu}_j \check{Z}_{i,j} (\check{Y}_{i,j} -\check{V}_{i,j}' \hat{\xi}_j)\big)$

	\end{algorithmic}
\end{algorithm}

\section{Proofs}

\begin{proof}[Proof Theorem 1] 
	
	Let $C_{0}'=E[\bar{Z}_{i}\bar{Z}_{i}']^{-1}E[\bar{Z}_{i}\tilde{W}_{i}']$
	and $G_{0}'=E[\tilde{X}_{i}\tilde{X}_{i}']^{-1}E[\tilde{X}_{i}\tilde{W}_{i}]$
	(the inverses are well defined under Assumption 1.4), then we can
	write: 
	\begin{equation}
		\tilde{W}_{i}=C_{0}\bar{Z}_{i}+G_{0}\tilde{X}_{i}+a_{i}\label{eq:W-1}
	\end{equation}
	
	Where $E[a_{i}]=0$ and $E[\bar{Z}_{i}a_{i}']=E[\tilde{X}_{i}a_{i}']=0$.
	Note that by the properties of partialling out $E[\bar{Z}\tilde{W}']=E[\bar{Z}{W}']$, and so under Assumption 1.3 $C_{0}$ has full row rank.
	
	Partialling out $D_{i}$ from both sides of (\ref{V}) and using $E[D_{i}\upsilon_{i}']=0$
	we get: 
	\begin{equation}
		\tilde{V}_{i}=B_{0}\tilde{W}_{i}+\upsilon_{i}\label{eq:V2}
	\end{equation}
	
	Under Assumptions 1.2.iii and 1.3 $B_{0}$ has full column rank. To
	see this note that $E[X_{i}\upsilon_{i}']=0$ and $E[D_{i}\upsilon_{i}']=0$,
	which implies $E[\tilde{X}_{i}\upsilon_{i}']=0$ and so $\tilde{V}_{i}=\bar{V}_{i}$. Moreover, by properties of partialling out, $E[\tilde{V}{W}']=E[\tilde{V}\tilde{W}']$. So $E[\tilde{V}\tilde{W}']=E[\bar{V}{W}']$, 
	and thus $B_{0}=E[\bar{V}{W}']E[\tilde{W}\tilde{W}']^{-1}$.
	By Assumption 1.2 $E[\bar{V}{W}']$ has full column rank, and
	thus $B_{0}$ has full column rank.
	
	Using (\ref{eq:U}) to substitute for $U_{i}$ in equation (\ref{eq:1}),
	we get: 
	\[
	Y_{i}=X_{i}'\beta_{0}+A_{0}W_{i}+L_{0}D_{i}+\varepsilon_{i}
	\]
	
	Partialling out $D_{i}$ from both sides of the above and using that
	$E[D_{i}\varepsilon_{i}]=0$ we get: 
	\begin{equation}
		\tilde{Y}_{i}=\tilde{X}_{i}'\beta_{0}+A_{0}\tilde{W}_{i}+\varepsilon_{i}\label{eq:Y1}
	\end{equation}
	
	We will use equations (\ref{eq:W-1}), (\ref{eq:V2}), and (\ref{eq:Y1}),
	and Assumption 1.2 to derive a set of moment conditions.
	
	Substituting (\ref{eq:W-1}) for $\tilde{W}_{i}$ in (\ref{eq:V2})
	we get: 
	\[
	\tilde{V}_{i}=(B_{0}C_{0})\bar{Z}_{i}+(B_{0}G_{0})\tilde{X}_{i}+(B_{0}a_{i}+\upsilon_{i})
	\]
	
	And so, since $E[\tilde{X}_{i}\bar{Z}_{i}']=0$:
	
	\begin{align}
		E[\tilde{X}_{i}\tilde{V}_{i}'] & =E[\tilde{X}_{i}\tilde{X}_{i}']G_{0}'B_{0}'+E[\tilde{X}_{i}\upsilon_{i}']\nonumber \\
		& =E[\tilde{X}_{i}\tilde{X}_{i}']G_{0}'B_{0}'\label{eq:momen1}
	\end{align}
	
	Where the second line follows from Assunmption 1.1.iii and $E[D_{i}\upsilon_{i}']=0$.
	Furthermore we see: 
	\begin{align}
		E[\bar{Z}_{i}\tilde{V}_{i}'] & =E[\bar{Z}_{i}\bar{Z}_{i}']C_{0}'B_{0}'+E[\bar{Z}_{i}\upsilon_{i}']\nonumber \\
		& =E[\bar{Z}_{i}\bar{Z}_{i}']C_{0}'B_{0}'\label{eq:momen2}
	\end{align}
	
	Where the second line again follows from Assunmption 1.1.iii and $E[D_{i}\upsilon_{i}']=0$.
	
	Substituting (\ref{eq:W-1}) for $\tilde{W}_{i}$ in (\ref{eq:Y1})
	we get: 
	\begin{align*}
		\tilde{Y}_{i} & =\tilde{X}_{i}'\beta_{0}+A_{0}C_{0}\bar{Z}_{i}+A_{0}G_{0}\tilde{X}_{i}+A_{0}a_{i}+\varepsilon_{i}
	\end{align*}
	
	And so: 
	\begin{align}
		E[\tilde{X}_{i}\tilde{Y}_{i}] & =E[\tilde{X}_{i}\tilde{X}_{i}'](\beta_{0}+G_{0}'A_{0}')+E[\tilde{X}_{i}a_{i}']A_{0}'+E[\tilde{X}_{i}\varepsilon_{i}]\nonumber \\
		& =E[\tilde{X}_{i}\tilde{X}_{i}'](\beta_{0}+G_{0}'A_{0}')\label{eq:momen3}
	\end{align}
	
	Where the second line follows from Assumption 1.1.ii, Assumption 1.1.i which states $E[D_{i}\varepsilon_{i}]=0$,
	and $E[\tilde{X}_{i}a_{i}']=0$. Furthermore:
	
	\begin{align}
		E[\bar{Z}_{i}\tilde{Y}_{i}] & =E[\bar{Z}_{i}\bar{Z}_{i}']C_{0}'A_{0}'+E[\bar{Z}a_{i}']A_{0}'+E[\bar{Z}_{i}\varepsilon_{i}]\nonumber \\
		& =E[\bar{Z}_{i}\bar{Z}_{i}']C_{0}'A_{0}'\label{eq:momen4}
	\end{align}
	
	Where the second line follows from Assumption 1.1.iii, $E[D_{i}\varepsilon_{i}]=0$
	and $E[\bar{Z}_{i}a_{i}']=0$.
	
	Because $E[\tilde{X}\bar{Z}_{i}']=0$ we can combine (\ref{eq:momen1}),
	(\ref{eq:momen2}), (\ref{eq:momen3}), and (\ref{eq:momen4}) into
	one moment condition as follows: 
	
	\[
	E\bigg[\bigg(\begin{pmatrix}\tilde{V}_{i}\\
		\tilde{Y}_{i}
	\end{pmatrix}-\begin{pmatrix}B_{0}C_{0} & B_{0}G_{0}\\
		A_{0}C_{0} & A_{0}G_{0}+\beta_{0}'
	\end{pmatrix}\begin{pmatrix}\bar{Z}_{i}\\
		\tilde{X}_{i}
	\end{pmatrix}\bigg)(\bar{Z}_{i}',\tilde{X}_{i}')\bigg]=0
	\]
	
	Note that $E[\bar{Z}_{i}\bar{Z}_{i}']=E[\tilde{Z}_{i}\bar{Z}_{i}']$,
	and $E[\bar{Z}_{i}\tilde{X}_{i}']=0$, so the above is equivalent
	to:
	
	\begin{equation}
		E\bigg[\bigg(\begin{pmatrix}\tilde{V}_{i}\\
			\tilde{Y}_{i}
		\end{pmatrix}-\begin{pmatrix}B_{0}C_{0} & B_{0}G_{0}\\
			A_{0}C_{0} & A_{0}G_{0}+\beta_{0}'
		\end{pmatrix}\begin{pmatrix}\tilde{Z}_{i}\\
			\tilde{X}_{i}
		\end{pmatrix}\bigg)(\bar{Z}_{i}',\tilde{X}_{i}')\bigg]=0\label{eq:firstmm}
	\end{equation}
	
	Moreover, $(\bar{Z}_{i}',\tilde{X}_{i}')=(\tilde{Z}_{i}',\tilde{X}_{i}')O$
	where $O$ is the matrix defined by: 
	\[
	O=\begin{pmatrix}I_{d_{Z}} & 0\\
		-E[\tilde{X}_{i}\tilde{X}_{i}']^{+}E[\tilde{X}_{i}\tilde{Z}_{i}'] & I_{d_{X}}
	\end{pmatrix}
	\]
	
	Under Assumption 1.4 this matrix is non-singular and so (\ref{eq:firstmm})
	is equivalent to
	\begin{equation}
		E\bigg[\bigg(\begin{pmatrix}\tilde{V}_{i}\\
			\tilde{Y}_{i}
		\end{pmatrix}-\begin{pmatrix}B_{0}C_{0} & B_{0}G_{0}\\
			A_{0}C_{0} & A_{0}G_{0}+\beta_{0}'
		\end{pmatrix}\begin{pmatrix}\tilde{Z}_{i}\\
			\tilde{X}_{i}
		\end{pmatrix}\bigg)(\tilde{Z}_{i}',\tilde{X}_{i}')\bigg]=0\label{eq:num1}
	\end{equation}
	
	So the moment condition (\ref{sparemom}) is satisfied by $r=d_{W}$,
	$A=A_{0}$, $B=B_{0}$, $C=C_{0}$, and $G=G_{0}$, and $\beta=\beta_{0}$.
	
	We also need to show there exists a solution to (\ref{sparemom})
	whenever $d_{W}<r\leq d_{V}$. We can construct a solution as follows.
	Firstly, let $\beta=\beta_{0}$. Let the first $d_{W}$ rows of $C$
	and $G$ equal $C_{0}$ and $G_{0}$ respectively, let the final $r-d_{W}$
	rows of $C$ and $G$ have all entries equal to $0$. Let the first
	$d_{W}$ entries of $A$ equal $A_{0}$ and the final $r-d_{W}$ entries
	be zeros. Finally, let the first $d_{W}$ columns of $B$ equal $B_{0}$,
	and the final $r-d_{W}$ columns be any linearly independent vectors
	in the orthogonal complement of the null space of $B_{0}$ (such a
	collection must exist because $B$ has $d_{V}$ rows and $r\leq d_{V}$).
	Then $B$ has full column rank, the matrices are of the required dimensions
	and we have $BC=B_{0}C_{0}$, $BG=B_{0}G_{0}$, $AC=A_{0}C_{0}$,
	and $AG=A_{0}G_{0}$ so by (\ref{eq:num1}) the solution satisfies
	the moment condition.
	
	Now, consider a vector $\beta$, natural number $r\leq\min\{d_{V},d_{Z}\}$,
	and matrices $A$, $B$, $C$, and $G$ of respective dimensions $1\times r$,
	$d_{V}\times r$, $r\times d_{Z}$ and $r\times d_{X}$, with $B$
	and $C$ of rank $r$ so that:
	
	\[
	E\bigg[\bigg(\begin{pmatrix}\tilde{V}_{i}\\
		\tilde{Y}_{i}
	\end{pmatrix}-\begin{pmatrix}BC & BG\\
		AC & AG+\beta'
	\end{pmatrix}\begin{pmatrix}\tilde{Z}_{i}\\
		\tilde{X}_{i}
	\end{pmatrix}\bigg)(\tilde{Z}_{i}',\tilde{X}_{i}')\bigg]=0
	\]
	
	We will prove the first statement of the theorem by showing that $\beta=\beta_{0}$.
	We will then prove the second statement of the theorem by showing
	that $r\geq d_{W}$, since we have already shown there exists a solution
	with $r=d_{W}$, we then have that the smallest $r$ such that a solution
	exists is $d_{W}$.
	
	Under Assumption 1.4 $E\big[(\tilde{Z}_{i}',\tilde{X}_{i}')'(\tilde{Z}_{i}',\tilde{X}_{i}')\big]$
	is non-singular, so we can combine the above with (\ref{eq:num1})
	to get the following four equalities:
	
	\begin{align}
		B_{0}C_{0} & =BC\label{eq:last1}\\
		B_{0}G_{0} & =BG\label{eq:last2}\\
		A_{0}C_{0} & =AC\label{eq:last3}\\
		A_{0}G_{0}-\beta_{0}' & =AG-\beta'\label{eq:last4}
	\end{align}
	
	Recall that $C_{0}$ has full row rank and thus $C_{0}C_{0}'$ is
	non-singular. Define $Q=C_{0}'(C_{0}C_{0}')^{-1}G_{0}$. Post-multiplying
	both sides of (\ref{eq:last3}) by $Q$ we get $A_{0}G_{0}=ACQ$ and
	substituting this into (\ref{eq:last4}) gives: 
	\begin{equation}
		ACQ-\beta_{0}'=AG-\beta'\label{eq:finnear}
	\end{equation}
	
	Now, post-multiplying both sides of (\ref{eq:last1}) by $Q$ we get
	$B_{0}G_{0}=BCQ$. Substituting into (\ref{eq:last2}) $BG=BCQ$.
	Premultiplying both sides by $A(B'B)^{-1}B'$ (recall $B$ has full
	column rank and so $B'B$ is non-singular) we get $AG=ACQ$. Substituting
	into (\ref{eq:finnear}) we get $\beta=\beta_{0}$, as required.
	
	Next we show $r\geq d_{W}$, which establishes identification of $d_{W}$.
	Post-multiplying both sides of (\ref{eq:last1}) by $C_{0}'(C_{0}C_{0}')^{-1}$
	(recall $C_{0}C_{0}'$ is non-singular), we get $B_{0}=BCC_{0}'(C_{0}C_{0}')^{-1}$.
	Recall $B_{0}$ is of full column rank, that is $rank(B_{0})=d_{W}$,
	and so $BCC_{0}'(C_{0}C_{0}')^{-1}$ is of rank $d_{W}$. For any
	two matrices $M_{1}$ and $M_{2}$, $rank(M_{1}M_{2})\leq\min\{rank(M_{1}),rank(M_{2})\}$,
	and so: 
	\[
	d_{W}=rank\big(BCC_{0}'(C_{0}C_{0}')^{-1}\big)\leq rank(B)
	\]
	Since $B$ has $r$ columns, $rank(B)\leq r$, and so $r\geq d_{W}$.
\end{proof}

	\theoremstyle{plain} \newtheorem*{P1}{Lemma 1} \begin{P1}
		Consider a vector $\beta$ and natural number $r\leq d_{V}$. There
		exist matrices $A$, $B$, $C$, and $G$ of dimensions $1\times r$,
		$d_{V}\times r$, $r\times d_{Z}$, and $r\times d_{X}$, with $rank(B)=r$
		that satisfy (\ref{sparemom}) if and only if there exists a $d_{V}\times(d_{Z}+d_{X})$
		matrix $M$ with $rank(M)=r$ and a vector $\xi\in\mathbb{R}^{d_{V}}$
		with at most $r$ non-zero entries so that: 
		\begin{align*}
			E\big[\big(\tilde{V}_{i}-M(\tilde{Z}_{i}',\tilde{X}_{i}')'\big)(\tilde{Z}_{i}',\tilde{X}_{i}')\big] & =0\\
			E\big[\big(\tilde{Y}_{i}-\beta'\tilde{X}_{i}-\xi'M(\tilde{Z}_{i}',\tilde{X}_{i}')'\big)(\tilde{Z}_{i}',\tilde{X}_{i}')\big] & =0
		\end{align*}
	\end{P1} \begin{proof}[Proof of Lemma 1] 
		
		First let us show the `only if'. Fix $r\leq d_{V}$ and $\beta$.
		Suppose $A$, $B$, $C$, and $G$ have dimensions $1\times r$, $d_{V}\times r$,
		$r\times d_{Z}$ and $r\times d_{X}$, with $B$ and $C$ of rank
		$r$ and satisfy (\ref{sparemom}). 
		
		$B$ is $d_{V}\times r$ and has rank $r$, thus $B$ is of full row
		rank and so $B'QB$ is non-singular for any non-singular matrix $Q$
		so letting $\xi=QB(B'QB)^{-1}A'$ we have: 
		\[
		E\bigg[\bigg(\begin{pmatrix}\tilde{V}_{i}\\
			\tilde{Y}_{i}-\beta'\tilde{X}_{i}
		\end{pmatrix}-\begin{pmatrix}B(C,G)\\
			\xi'B(C,G)
		\end{pmatrix}\begin{pmatrix}\tilde{Z}_{i}\\
			\tilde{X}_{i}
		\end{pmatrix}\bigg)(\tilde{Z}_{i}',\tilde{X}_{i}')\bigg]=0
		\]
		
		Let $M=B(C,G)$, then $M$ has rank $r$. Now, since $B'$ is $r\times d_{V}$
		with rank $r$, it has an $r$-dimensional range and $d_{V}-r$ dimensional
		null space. So we can choose $Q$ so that it has $r$ linearly independent
		rows in the range of $B$ and $d_{V}-r$ linearly independent rows
		in the null-space of $B$, in which case $Q$ is non-singular and
		$\xi=QB(B'QB)^{-1}A'$ has at most $r$ non-zero entries (the entries
		corresponding to the rows of $Q$ in the null space are all zero).
		So the `if' holds.
		
		Now the `only if'. Let $M$ and $\xi$ satisfy (\ref{mom1st}) and
		(\ref{mom2nd}) with $rank(M)=r$, $M$ of dimension $d_{V}\times(d_{Z}+d_{X})$.
		By elementary linear algebra, any $d_{V}\times(d_{Z}+d_{X})$ matrix
		$M$ of rank $r$ can be written as the product $M=M_{1}M_{2}$ where
		$M_{1}$ is a $d_{V}\times r$ matrix and $M_{2}$ is a $r\times(d_{Z}+d_{X})$
		matrix and $M_{1}$ and $M_{2}$ have rank $r$. So let $B=M_{1}$,
		$(C,G)=M_{2}$, and $A=\xi'B$ and we are done. \end{proof}
	
	\begin{proof}[Proof of Corollary 1] Immediately follows from Theorem
		1 and Lemma 1. 
	\end{proof}

\begin{proof}[Proof that the moment condition holds under Assumptions 1.1-1.4]

Recall that the moment condition is defined as follows:
\begin{align*}
	& E\big[g_{i}(\beta_{0};\xi_{0},\mu_{0},\gamma_{Y,0},\gamma_{V,0},\gamma_{X,0},\gamma_{X,0},\omega_{Z,0},\omega_{Y,0},\omega_{V,0})\big]\\
	= & E\big[\tilde{X}_{i}\big(\tilde{Y}_{i}-\tilde{V}_{i}'\xi_{0}-\tilde{X}_{i}'\beta_{0}\big)\big]-\mu_{0}E\big[\bar{Z}_{i}(\bar{Y}_{i}-\xi_{0}'\bar{V}_{i})\big]=0
\end{align*}

Using the first moment equality in Corollary 1 to substitute for
$M$ in the second moment equality we get:
\begin{equation}
	E\big[(\tilde{Z}_{i}',\tilde{X}_{i}')'\big(\tilde{Y}_{i}-\tilde{X}_{i}'\beta_{0}-\tilde{V}_{i}'\xi_{0}\big)\big]=0\label{eq:2bpart}
\end{equation}

Where we have set $\beta=\beta_{0}$ and $\xi=\xi_{0}$ which satisfy
the moment conditions in Corollary 1. The expression above implies:
\[
E\big[\tilde{X}_{i}\big(\tilde{Y}_{i}-\beta_{0}'\tilde{X}_{i}'-\xi_{0}'\tilde{V}_{i}\big)\big]=0
\]

Next, premultiplying both sides of (\ref{eq:2bpart}) by $\big(I_{d_{Z}},-E[\tilde{Z}_{i}\tilde{X}_{i}']E[\tilde{Z}_{i}\tilde{Z}_{i}']^{+}\big)$
we get:
\[
E\big[\bar{Z}_{i}\big(\tilde{Y}_{i}-\tilde{X}_{i}'\beta_{0}-\tilde{V}_{i}'\xi_{0}\big)\big]=0
\]

By elementary properties of partialling out we get $E[\bar{Z}_{i}\tilde{Y}_{i}]=E[\bar{Z}_{i}\bar{Y}_{i}]$,
$E[\bar{Z}_{i}\tilde{V}_{i}]=E[\bar{Z}_{i}\bar{V}_{i}]$, and $E[\bar{Z}_{i}\tilde{X}_{i}]=E[\bar{Z}_{i}\bar{X}_{i}]=0$.
And so the moment condition above implies:
\[
E\big[\bar{Z}_{i}(\bar{Y}_{i}-\xi_{0}'\bar{V}_{i})\big]=0
\]

Combining gives the result.
\end{proof}

\begin{proof}[Proof that the doubly-robust moment condition is doubly-robust under Assumptions 1.1-1.4]

We will go thorugh each argument of $g_{i}$ in turn and show that the
moment condition is robust to that argument.

\textbf{Step 1: Prove $E\big[g_{i}(\beta_{0};\xi,\mu_{0},\gamma_{Y,0},\gamma_{V,0},\gamma_{X,0},\gamma_{X,0},\omega_{Z,0},\omega_{Y,0},\omega_{V,0})\big]=0$
	for all $\xi$}

Note that for any $\xi$:
\begin{align*}
	& E\big[g_{i}(\beta_{0};\xi,\mu_{0},\Sigma_{0},\gamma_{Y,0},\gamma_{V,0},\gamma_{X,0},\gamma_{X,0},\omega_{Z,0},\omega_{Y,0},\omega_{V,0})\big]\\
	= & E\big[\tilde{X}_{i}(\tilde{Y}_{i}-\tilde{V}_{i}'\xi-\tilde{X}_{i}'\beta_{0})\big]-\mu_{0}E\big[\bar{Z}_{i}(\bar{Y}_{i}-\xi'\bar{V}_{i})\big]\\
	= & E[\tilde{X}_{i}(\tilde{Y}_{i}-\tilde{X}_{i}'\beta_{0})]-\mu_{0}E[\bar{Z}_{i}\bar{Y}_{i}]+\big(\mu_{0}E[\bar{Z}_{i}\bar{V}_{i}']-E[\tilde{X}_{i}\tilde{V}_{i}']\big)\xi
\end{align*}

We will show that $E[\tilde{X}_{i}\tilde{V}_{i}']=\mu_{0}E[\bar{Z}_{i}\bar{V}_{i}']$
and so the expression above is independent of $\xi$. Since the moment
condition holds when $\xi=\xi_{0}$, this proves the result. Substituting the definition of $\mu_{0}$ we have:
\begin{align}
	& \mu_{0}E[\bar{Z}_{i}\bar{V}_{i}']\nonumber \\
	= & E[\tilde{X}_{i}(\tilde{Z}_{i}',\tilde{X}_{i}')]M_{0}'\big(M_{0,[:,1:d_{Z}]}E[\bar{Z}_{i}\bar{Z}_{i}']M_{0,[:,1:d_{Z}]}'\big)^{+}M_{0,[:,1:d_{Z}]}E[\bar{Z}_{i}\bar{V}_{i}']\label{eq:myusub}
\end{align}

By the definition of $M_{0}$:
\[
E\bigg[\big(\tilde{V}_{i}-M_{0}(\tilde{Z}_{i}',\tilde{X}_{i}')'\big)(\tilde{Z}_{i}',\tilde{X}_{i}')\bigg]=0
\]

Post-multiplying by $\big(I_{d_{Z}},-E[\tilde{Z}_{i}\tilde{X}_{i}']E[\tilde{Z}_{i}\tilde{Z}_{i}']^{+}\big)'$
we get:
\[
E\bigg[\big(\tilde{V}_{i}-M_{0}(\tilde{Z}_{i}',\tilde{X}_{i}')'\big)\bar{Z}_{i}'\bigg]=0
\]

And then by the properties of partialling out:
\begin{align*}
	0=E\bigg[\big(\tilde{V}_{i}-M_{0}(\tilde{Z}_{i}',\tilde{X}_{i}')'\big)\bar{Z}_{i}'\bigg] & =E\bigg[\big(\bar{V}_{i}-M_{0}(\bar{Z}_{i}',\bar{X}_{i}')'\big)\bar{Z}_{i}'\bigg]\\
	& =E\bigg[\big(\bar{V}_{i}-M_{0}(\bar{Z}_{i}',0_{1\times d_{X}})'\big)\bar{Z}_{i}'\bigg]\\
	& =E\big[(\bar{V}_{i}-M_{0,[:,1:d_{Z}]}\bar{Z}_{i})\bar{Z}_{i}'\big]
\end{align*}

And so we have $M_{0,[:,1:d_{Z}]}E[\bar{Z}_{i}\bar{Z}_{i}']=E[\bar{V}_{i}\bar{Z}_{i}']$.
Using this to substitute for $E[\bar{V}_{i}\bar{Z}_{i}']$ in (\ref{eq:myusub})
and simplifying we get:
\[
	\mu_{0}E[\bar{Z}_{i}\bar{V}_{i}']= E\big[\tilde{X}_{i}(\tilde{Z}_{i}',\tilde{X}_{i}')M_{0}'\big]
\]

Recall that $E\bigg[\big(\tilde{V}_{i}-M_{0}(\tilde{Z}_{i}',\tilde{X}_{i}')'\big)(\tilde{Z}_{i}',\tilde{X}_{i}')\bigg]=0$,
which implies that $E[\tilde{V}_{i}\tilde{X}_{i}']=E\big[M_{0}(\tilde{Z}_{i}',\tilde{X}_{i}')'\tilde{X}_{i}'\big]$.
Substituting this into the above we get:
\[
\mu_{0}E[\bar{Z}_{i}\bar{V}_{i}']=E[\tilde{X}_{i}\tilde{V}_{i}]
\]
As desired.

\textbf{Step 2: Prove $E\big[g_{i}(\beta_{0};\xi_{0},\mu,\Sigma_{0},\gamma_{Y,0},\gamma_{V,0},\gamma_{X,0},\gamma_{X,0},\omega_{Z,0},\omega_{Y,0},\omega_{V,0})\big]=0$
	for all $\mu$}

In the proof that the moment condition holds we showed that:
\[
E\big[\bar{Z}_{i}(\bar{Y}_{i}-\xi_{0}'\bar{V}_{i})\big]=0
\]

And so for any $\mu$:
\begin{align*}
	& E\big[g_{i}(\beta_{0};\xi_{0},\mu,\gamma_{Y,0},\gamma_{V,0},\gamma_{X,0},\gamma_{X,0},\omega_{Z,0},\omega_{Y,0},\omega_{V,0})\big]\\
	= & E\big[\tilde{X}_{i}\big(\tilde{Y}_{i}-\tilde{V}_{i}'\xi_{0}-\tilde{X}_{i}'\beta_{0}\big)\big]-\mu E\big[\bar{Z}_{i}(\bar{Y}_{i}-\xi_{0}'\bar{V}_{i})\big]\\
	= & E\big[\tilde{X}_{i}\big(\tilde{Y}_{i}-\tilde{V}_{i}'\xi_{0}-\tilde{X}_{i}'\beta_{0}\big)\big]
\end{align*}

The final expression above does not depend on $\mu$. The moment condition
holds for $\mu=\mu_{0}$ and so it must hold for all $\mu$.

\textbf{Step 3: Prove robustness for the remaining arguments}

By the properties of partialling out, for any $\gamma_{Y}$:
\begin{align*}
	E\big[\tilde{X}_{i}\tilde{Y}_{i}(\gamma_{Y})\big] & =E[\tilde{X}_{i}Y_{i}]-E[\tilde{X}_{i}D_{i}']\gamma_{Y}\\
	& =E[\tilde{X}_{i}\tilde{Y}_{i}]
\end{align*}

$\gamma_{Y}$ only enters $E\big[g_{i}(\beta_{0};\xi_{0},\mu_0,\gamma_{Y},\gamma_{V,0},\gamma_{X,0},\gamma_{X,0},\omega_{Z,0},\omega_{Y,0},\omega_{V,0})\big]$,
through the expression above, so we are robust to $\gamma_{Y}$. Similarly
for any $\gamma_{V}$, $E\big[\tilde{X}_{i}\tilde{V}_{i}(\gamma_{V})'\big]=E[\tilde{X}_{i}\tilde{V}_{i}']$,
which shows we are robust to $\gamma_{V}$. For any $\gamma_{X,2}$,
$E\big[\tilde{X}_{i}\tilde{X}_{i}'(\gamma_{X,2})\big]=E[\tilde{X}_{i}\tilde{X}_{i}']$
so we are robust to $\gamma_{X,2}$. For any $\gamma_{X,1}$:
\[
E\big[\tilde{X}_{i}(\gamma_{X,1})\big(\tilde{Y}_{i}-\tilde{V}_{i}'\xi_{0}-\tilde{X}_{i}'\beta_{0}\big)\big]=E\big[\tilde{X}_{i}\big(\tilde{Y}_{i}-\tilde{V}_{i}'\xi_{0}-\tilde{X}_{i}'\beta_{0}\big)\big]
\]

Which shows we are robust to $\gamma_{X,1}$. For any $\omega_{Y}$, $E\big[\bar{Z}_{i}\bar{Y}_{i}(\omega_{Y})\big]=E[\bar{Z}_{i}\bar{Y}_{i}]$,
for any $\omega_{V}$ we have  $E\big[\bar{Z}_{i}\bar{V}_{i}(\omega_{V})'\big]=E[\bar{Z}_{i}\bar{V}_{i}']$,
and for any $\omega_{Z}$:
\[
E\big[\bar{Z}_{i}(\omega_{Z})(\bar{Y}_{i}-\bar{V}_{i}'\xi_{0})\big]=E\big[\bar{Z}_{i}(\bar{Y}_{i}-\bar{V}_{i}'\xi_{0})\big]
\]

So the moment condition is also robust to $\omega_{Y}$, $\omega_{V}$,
and $\omega_{Z}$.
\end{proof}

\begin{proof}[Proof of Theorem 2]
	
	To prove the result we confirm that the conditions for Theorems 3.1
	and 3.2 in \citet{Chernozhukov2018} hold. The result follows immediately from those theorems. 
	
	Theorems 3.1 and 3.2 in \citet{Chernozhukov2018} require Assumptions 3.1 and 3.2 in that
	paper. Let us begin with Assumption 3.1. This states that a) the true
	parameter ($\beta_{0}$ in our case) satisfies the moment condition.
	b) That the moment condition is linear in this parameter. c) That
	the map from the parameters to the moment is twice continuously Gateux
	differentiable, and d) that the score is Neyman orthogonal (or `near
	Neyman orthogonal'). By Lemma 2 the moment condition is valid so a)
	hold. By Lemma 3 the score is doubly-robust and therefore Neyman-orthogonal
	so d) holds. The score is linear in $\beta_{0}$ and it is linear
	in each of its arguments and is thus continuously twice Gauteax differentiable,
	so b) and c) hold. Thus Assumption 3.1 of \citet{Chernozhukov2018} is satisfied. 
	
	We now show that Assumption 3.2 of \citet{Chernozhukov2018} holds. this constitutes the
	bulk of the proof. Below we restate this assumption as it applies
	in our setting. It will be convenient to collect all the nuisance
	parameters into one single parameter. In particular, let $\eta_{0}$
	contain the true values of all the nuisance parameters so that:
	\[
	\eta_{0}=(\mu_{0},\xi_{0},\gamma_{X,0},\gamma_{Y,0},\gamma_{V,0},\omega_{Z,0},\omega_{Y,0},\omega_{V,0})
	\]
	In the above, the parentheses indicate an ordered set rather than
	horizontal concatenation of matrices. Similarly, let $\hat{\eta}_{j}$
	be the collection of all the nuisance parameter estimates for the
	$j^{th}$ subsample:
	\[
	\hat{\eta}_{j}=(\hat{\mu}_{j},\hat{\xi}_{j},\hat{\gamma}_{X,j},\hat{\gamma}_{Y,j},\hat{\gamma}_{V,j},\hat{\omega}_{Z,j},\hat{\omega}_{Y,j},\hat{\omega}_{V,j})
	\]
	Moreover, for some $\eta=(\mu,\xi,\gamma_{X},\gamma_{Y},\gamma_{V},\omega_{Z},\omega_{Y},\omega_{V})$
	we define $g_{i}(\eta)$ as follows:
	\[
	g_{i}(\beta,\eta)=g_{i}(\beta,\mu,\xi,\gamma_{X},\gamma_{X},\gamma_{Y},\gamma_{V},\omega_{Z},\omega_{Y},\omega_{V})
	\]
	Assumption 3.2 of \citet{Chernozhukov2018} states $E[g_{i}g_{i}']$ has eigenvalues all
	bounded below away from zero and that for each $n$ there is a set
	$\mathcal{T}_{n}$, a sequence $\alpha_{n}\to0$, a constant $c_{1}$,
	and a sequence $\delta_{n}\to0$, so that if $P\in\mathcal{P}_{n}$
	the conditions below all hold.
	\begin{enumerate}
		\item With probability at least $1-\alpha_{n}$, $\hat{\eta}_{j}\in\mathcal{T}_{n}$
		for all $j=1,...,J$.
		\item $\sup_{\eta\in\mathcal{T}_{n}}E\big[||g_{i}(\beta_{0},\eta)||^{q}\big]^{1/q}\leq c_{1}$
		\item $\sup_{\eta\in\mathcal{T}_{n}}E\big[||\tilde{X}_{i}(\gamma_{X})\tilde{X}_{i}(\gamma_{X})'||^{q}\big]^{1/q}\leq c_{1}$
		\item $\sup_{\eta\in\mathcal{T}_{n}}||\Sigma_{\tilde{X}}-E\big[\tilde{X}_{i}(\gamma_{X})\tilde{X}_{i}(\gamma_{X})'\big]||\leq\delta_{n}$
		\item $\sup_{\eta\in\mathcal{T}_{n}}E\big[||g_{i}(\beta_{0},\eta_{0})-g_{i}(\beta_{0},\eta)||^{2}\big]^{1/2}\leq\delta_{n}$
		\item $\sup_{r\in(0,1),\eta\in\mathcal{T}_{n}}||\frac{\partial^{2}}{\partial r^{2}}E\bigg[g_{i}\big(\beta_{0},\eta_{0}+r(\eta-\eta_{0})\big)\bigg]||\leq\delta_{n}/\sqrt{n}$
	\end{enumerate}
	We will show that the conditions above are satisfied with:
	\begin{align*}
		\delta_{n} & \leq c_{3,n}+c_{4,n}+\sqrt{n}c_{5,n}
	\end{align*}
	
	Where the sequences $c_{3,n}$, $c_{4,n}$, and $c_{5,n}$ have rates
	$c_{3,n}\precsim\delta_{\gamma,X,n}^{2}$,
	
	\begin{align*}
		c_{4,n} & \precsim\delta_{\gamma,Y,n}+\delta_{\gamma,X,n}+\delta_{\gamma,V,n}+\delta_{\omega,Y,n}+\delta_{\omega,Z,n}+\delta_{\omega,V,n}+\delta_{\xi,n}+\delta_{\mu,n}\\
		+ & \sqrt{d_{D}}\delta_{\gamma,X,n}(\delta_{\gamma,Y,n}+\delta_{\gamma,V,n}+\delta_{\gamma,X,n})\\
		+ & \sqrt{d_{D}}\delta_{\omega,Z,n}(\delta_{\omega,Y,n}+\delta_{\omega,V,n})
	\end{align*}
	
	and:
	\begin{align*}
		c_{5,n} & \precsim\delta_{\gamma,X,n}(\delta_{\gamma,Y,n}+\delta_{\gamma,V,n}+\delta_{\gamma,X,n})+\delta_{\omega,Z,n}(\delta_{\omega,Y,n}+\delta_{\omega,V,n})
	\end{align*}
	
	Then under the premises of the Theorem $\delta_{n}\prec1$.
	
	The following notation will prove useful. For any $\eta=(\mu,\xi,\gamma_{X},\gamma_{Y},\gamma_{V},\omega_{Z},\omega_{Y},\omega_{V})$,
	let $\delta_{\xi,n}=\xi-\xi_{0}$, $\delta_{\mu,n}=\mu-\mu_{0}$,
	for $H=X,Y,V$ let $\Delta_{\gamma,H}=\gamma_{H}-\gamma_{H,0}$, and
	for $H=Z,Y,V$ let $\Delta_{\omega,H}=\omega_{H}-\omega_{H,0}$.
	
	Some facts we use repeatedly in our arguments are as follows. For
	any random vector $Q_{i}$ and a non-random matrix $A$ with number
	of columns equal to the length of $Q_{i}$, then letting $\Sigma_{Q}=E[Q_{i}Q_{i}']$
	and $r_{A}$ the number of rows of $A$:
	\begin{equation}
		||A\Sigma_{Q}^{1/2}||^{2}\leq E[||AQ_{i}||^{2}]\leq r_{A}||A\Sigma_{Q}^{1/2}||^{2}\label{eq:basic1}
	\end{equation}
	As an immediate corollary, if $b$ is a column vector then $E[||b'Q_{i}||^{2}]=||\Sigma_{Q}^{1/2}b||^{2}$.
	In addition, if $R_{i}$ is a vector of residuals from population
	least squares regression of a variable $Q_{i}$ on some regressors,
	then $||\Sigma_{R}^{1/2}b||^{2}\leq||\Sigma_{Q}^{1/2}b||^{2}$. Thus
	we have $||\Sigma_{\bar{V}}\xi_{0}||\leq||\Sigma_{\tilde{V}}\xi_{0}||\leq\bar{\xi}$
	and similarly, for any $\xi$ we have $||\Sigma_{\bar{V}}\Delta_{\xi}||\leq||\Sigma_{\tilde{V}}\Delta_{\xi}||\leq\delta_{\xi,n}$.
	Finally, $E[\bar{Y}_{i}^{2}]\leq E[\tilde{Y}_{i}^{2}]\leq\sigma_{Y}^{2}$,
	and $||\Sigma_{\tilde{X}}||\leq\sigma_{X}^{2}$.
	
	\textbf{Condition 1}
	
	Under Assumption 4.2, the set $\mathcal{T}_{n}$ defined as follows
	satisfies Condition 1 for some $\alpha_{n}\to0$. $\eta\in\mathcal{T}_{n}$
	if and only if $||\Delta_{\mu}\Sigma_{\bar{Z}}^{1/2}||\leq\delta_{\mu,n}$,
	$||\Sigma_{\tilde{V}}^{1/2}\Delta_{\xi}||\leq\delta_{\xi,n}$, for
	$H=Y,V,X$ $||\Sigma_{\tilde{H}}^{-1/2}\Delta_{\gamma,H}\Sigma_{D}^{1/2}||\leq\delta_{\gamma,H}$,
	and for $H=Z,V,Y$ we have $||\Sigma_{\tilde{H}}^{-1/2}\Delta_{\gamma,H}\Sigma_{XD}^{1/2}||\leq\delta_{\omega,H}$.
	In our discussion of the remaining conditions we take $\mathcal{T}_{n}$
	to be this set.
	
	We now proceed by bounding in turn $m_{n}$, $m_{n}'$, $r_{n}$,
	$r_{n}'$, and $\lambda_{n}'$, the scalars on the left-hand sides
	of the equations in Conditions 2-6.
	
	\textbf{Condition 2}
	
	We will show that there is a sequence so that if $P\in\mathcal{P}_{n}$
	and $\eta\in\mathcal{T}_{n}$ then $E\big[||g_{i}(\beta_{0},\eta)||^{q}\big]^{1/q}\leq c_{1,n}$
	where: 
	\begin{align*}
		c_{1,n}\precsim & 1+d_{V}^{1/q}\delta_{\xi,n}+d_{D}^{2/q}(\delta_{\gamma,Y,n}+\delta_{\gamma,V,n}+\delta_{\gamma,X,n})\\
		+ & (\delta_{\mu,n}+\delta_{\xi,n})d_{Z}^{1/q}d_{V}^{1/q}\\
		+ & d_{D}^{1/q}(d_{Z}^{1/q}+d_{D}^{1/q}\delta_{\omega,Z,n})\big(\delta_{\omega,Y,n}+\delta_{\omega,V,n}\big)
	\end{align*}
	
	Under the premises of the Theorem the expression on the right-hand
	side above is bounded by a constant $c$ given in the theorem and
	so $c_{1,n}\precsim1$.
	
	Let $\eta\in\mathcal{T}_{n}$. In our setting $E\big[||g_{i}(\beta_{0},\eta)||^{q}\big]^{1/q}$
	is equal to:
	\[
	E\bigg[||\tilde{X}_{i}(\gamma_{X})\big(\tilde{Y}_{i}(\gamma_{Y})-\tilde{V}_{i}(\gamma_{V})'\xi-\tilde{X}_{i}(\gamma_{X})'\beta_{0}\big)-\mu\bar{Z}_{i}(\omega_{Z})\big(\bar{Y}_{i}(\omega_{Y})-\bar{V}_{i}(\omega_{V})'\xi\big)||^{q}\bigg]^{1/q}
	\]
	
	By the triangle inequality:
	
	\begin{align}
		& E\bigg[||\tilde{X}_{i}(\gamma_{X})\big(\tilde{Y}_{i}(\gamma_{Y})-\tilde{V}_{i}(\gamma_{V})'\xi-\tilde{X}_{i}(\gamma_{X})'\beta_{0}\big)-\mu\bar{Z}_{i}(\omega_{Z})\big(\bar{Y}_{i}(\omega_{Y})-\bar{V}_{i}(\omega_{V})'\xi\big)||^{q}\bigg]^{1/q}\nonumber \\
		\leq & E\bigg[||\tilde{X}_{i}(\gamma_{X})\big(\tilde{Y}_{i}(\gamma_{Y})-\tilde{V}_{i}(\gamma_{V})'\xi-\tilde{X}_{i}(\gamma_{X})'\beta_{0}\big)||^{q}\bigg]^{1/q}\nonumber \\
		+ & E\bigg[||\mu\bar{Z}_{i}(\omega_{Z})\big(\bar{Y}_{i}(\omega_{Y})-\bar{V}_{i}(\omega_{V})'\xi\big)||^{q}\bigg]^{1/q}\label{eq:bigb}
	\end{align}
	
	Let us upper-bound the first term on the right-hand above. Using the
	definitions of $\tilde{X}_{i}(\cdot),$ $\tilde{Y}_{i}(\cdot)$, and
	$\tilde{V}_{i}(\cdot)$ and the triangle inequality we get:
	
	\begin{align*}
		& E\bigg[||\tilde{X}_{i}(\gamma_{X})\big(\tilde{Y}_{i}(\gamma_{Y})-\tilde{V}_{i}(\gamma_{V})'\xi-\tilde{X}_{i}(\gamma_{X})'\beta_{0}\big)||^{q}\bigg]^{1/q}\\
		\leq & E\big[||\tilde{X}_{i}(\tilde{Y}_{i}+\tilde{X}_{i}'\beta_{0}+\tilde{V}_{i}'\Delta_{\xi})||^{q}\big]^{1/q}\\
		+ & E\big[||\tilde{X}_{i}\tilde{V}_{i}'\xi_{0}||^{q}\big]^{1/q}\\
		+ & E\big[||(\tilde{X}_{i}+\Delta_{\gamma,X}D_{i})D_{i}'(\Delta_{\gamma,Y}+\Delta_{\gamma,V}'\xi+\Delta_{\gamma,X}'\beta_{0})||^{q}\big]^{1/q}
	\end{align*}
	
	We will bound each term on the right-hand side of the inequality above
	in turn. Again by the triangle inequality:
	
	\begin{align*}
		& E\big[||\tilde{X}_{i}(\tilde{Y}_{i}+\tilde{X}_{i}'\beta_{0}+\tilde{V}_{i}'\Delta_{\xi})||^{q}\big]^{1/q}\\
		\leq & E\big[||\tilde{X}_{i}\tilde{Y}_{i}||^{q}\big]^{1/q}+E\big[||\tilde{X}_{i}\tilde{X}_{i}'\beta_{0}||^{q}\big]^{1/q}\\
		+ & E\big[||\tilde{X}_{i}\tilde{V}_{i}'\Delta_{\xi}||^{q}\big]^{1/q}
	\end{align*}
	
	Applying the properties of the matrix norm and then using Assumptions
	4.1 and 4.3 we get:
	\begin{align*}
		E\big[||\tilde{X}_{i}\tilde{Y}_{i}||^{q}\big]^{1/q} & \leq||\Sigma_{\tilde{X}}^{1/2}||E\big[||\Sigma_{\tilde{X}}^{-1/2}\tilde{X}_{i}\tilde{Y}_{i}/E[\tilde{Y}_{i}^{2}]^{1/2}||^{q}\big]^{1/q}E[\tilde{Y}_{i}^{2}]^{1/2}\\
		& \leq\sigma_{X}d_{X}^{1/q}S_{q,\tilde{X},\tilde{Y}}^{1/q}\sigma_{Y}
	\end{align*}
	
	Again, using the properties of the matrix norm and Assumptions 4.1
	and 4.3:
	\begin{align*}
		E\big[||\tilde{X}_{i}\tilde{X}_{i}'\beta_{0}||^{q}\big]^{1/q} & \leq||\Sigma_{\tilde{X}}^{1/2}||E\big[||\Sigma_{\tilde{X}}^{-1/2}\tilde{X}_{i}\tilde{X}_{i}'\Sigma_{\tilde{X}}^{-1/2}||^{q}\big]^{1/q}||\Sigma_{X}^{-1/2}\beta_{0}||\\
		& \leq\sigma_{X}d_{X}^{2/q}S_{q,\tilde{X},\tilde{X}}^{1/q}\bar{\beta}
	\end{align*}
	
	And following similar steps but also using the properties of $\mathcal{T}_{n}$
	to get $||\Sigma_{\tilde{V}}^{1/2}\Delta_{\xi}||\leq\delta_{\xi}$:
	
	\begin{align*}
		E\big[||\tilde{X}_{i}\tilde{V}_{i}'\Delta_{\xi}||^{q}\big]^{1/q} & \leq||\Sigma_{\tilde{X}}^{1/2}||E\big[||\Sigma_{\tilde{X}}^{-1/2}\tilde{X}_{i}\tilde{V}_{i}'\Sigma_{\tilde{V}}^{-1/2}||^{q}\big]^{1/q}||\Sigma_{\tilde{V}}^{1/2}\Delta_{\xi}||\\
		& \leq\sigma_{X}d_{X}^{1/q}d_{V}^{1/q}S_{q,\tilde{X},\tilde{V}}^{1/q}\delta_{\xi,n}
	\end{align*}
	
	And so:
	\begin{align*}
		& E\big[||\tilde{X}_{i}(\tilde{Y}_{i}+\tilde{X}_{i}'\beta_{0}+\tilde{V}_{i}'\Delta_{\xi})||^{q}\big]^{1/q}\\
		\leq & \sigma_{X}d_{X}^{1/q}\big(S_{q,\tilde{X},\tilde{Y}}^{1/q}\sigma_{Y}+d_{X}^{1/q}S_{q,\tilde{X},\tilde{X}}^{1/q}\bar{\beta}+d_{V}^{1/q}S_{q,\tilde{X},\tilde{V}}^{1/q}\delta_{\xi,n})
	\end{align*}
	
	By Assumption 4.3 $E\big[||\tilde{X}_{i}\tilde{V}_{i}'\xi_{0}||^{q}\big]^{1/q}\leq\bar{S}_{q}$.
	Next note that, by the properties of the matrix norm:
	
	\begin{align*}
		& E\big[||(\tilde{X}_{i}+\Delta_{\gamma,X}D_{i})D_{i}'(\Delta_{\gamma,Y}+'\Delta_{\gamma,V}'\xi+\Delta_{\gamma,X}'\beta_{0})||^{q}\big]^{1/q}\\
		\leq & E\big[||(\tilde{X}_{i}+\Delta_{\gamma,X}D_{i})D_{i}'\Sigma_{D}^{-1/2}||^{q}\big]^{1/q}||\Sigma_{D}^{-1/2}(\Delta_{\gamma,Y}+\Delta_{\gamma,V}'\xi+\Delta_{\gamma,X}'\beta_{0})||
	\end{align*}
	
	Consider the first term in the product on the right-hand side above.
	Applying the triangle inequality and the properties of the matrix
	norm:
	\begin{align*}
		& E\big[||(\tilde{X}_{i}+\Delta_{\gamma,X}D_{i})D_{i}'\Sigma_{D}^{-1/2}||^{q}\big]^{1/q}\\
		\leq & E\big[||\tilde{X}_{i}D_{i}'\Sigma_{D}^{-1/2}||^{q}\big]^{1/q}+E\big[||\Delta_{\gamma,X}D_{i}D_{i}'\Sigma_{D}^{-1/2}||^{q}\big]^{1/q}\\
		\leq & ||\Sigma_{\tilde{X}}^{1/2}||E\big[||\Sigma_{\tilde{X}}^{-1/2}\tilde{X}_{i}D_{i}'\Sigma_{D}^{-1/2}||^{q}\big]^{1/q}\\
		+ & ||\Sigma_{\tilde{X}}^{1/2}||\cdot||\Sigma_{\tilde{X}}^{-1/2}\Delta_{\gamma,X}\Sigma_{D}^{1/2}||E\big[||\Sigma_{D}^{-1/2}D_{i}D_{i}'\Sigma_{D}^{-1/2}||^{q}\big]^{1/q}\\
		\leq & \sigma_{X}(d_{X}^{1/q}S_{q,\tilde{X},D}^{1/q}+d_{D}^{1/q}S_{q,D,D}^{1/q})d_{D}^{1/q}
	\end{align*}
	
	Where the last line follows by Assumptions 4.1 and 4.3. Next note
	that again applying the triangle inequality and properties of the
	matrix norm:
	\begin{align*}
		& ||\Sigma_{D}^{-1/2}(\Delta_{\gamma,Y}+'\Delta_{\gamma,V}'\xi+\Delta_{\gamma,X}'\beta_{0})||\\
		\leq & ||\Sigma_{D}^{-1/2}\Delta_{\gamma,Y}E[\tilde{Y}_{i}^{2}]^{-1/2}||E[\tilde{Y}_{i}^{2}]^{1/2}+||\Sigma_{D}^{-1/2}\Delta_{\gamma,V}'\Sigma_{\tilde{V}}^{-1/2}||\cdot||\xi_{0}+\Delta_{\xi}||\\
		+ & ||\Sigma_{D}^{-1/2}\Delta_{\gamma,X}'\Sigma_{\tilde{X}}^{-1/2}||\cdot||\Sigma_{\tilde{X}}^{1/2}\beta_{0}||\\
		\leq & \delta_{\gamma,Y,n}\sigma_{Y}+\delta_{\gamma,V,n}(\bar{\xi}+\delta_{\xi,n})+\delta_{\gamma,X,n}\bar{\beta}
	\end{align*}
	
	Where the final line follows from Assumptions 4.1 and properties of
	$\mathcal{T}_{n}$. Combining we get:
	\begin{align*}
		& E\big[||(\tilde{X}_{i}+\Delta_{\gamma,X}D_{i})D_{i}'(\Delta_{\gamma,Y}+'\Delta_{\gamma,V}'\xi+\Delta_{\gamma,X}'\beta_{0})||^{q}\big]^{1/q}\\
		\leq & \sigma_{X}(d_{X}^{1/q}S_{q,\tilde{X},D}^{1/q}+d_{D}^{1/q}S_{q,D,D}^{1/q})d_{D}^{1/q}(\delta_{\gamma,Y,n}\sigma_{Y}+\delta_{\gamma,V,n}(\bar{\xi}+\delta_{\xi,n})+\delta_{\gamma,X,n}\bar{\beta})
	\end{align*}
	
	In all we get:
	\begin{align*}
		& E\bigg[||\tilde{X}_{i}(\gamma_{X})\big(\tilde{Y}_{i}(\gamma_{Y})-\tilde{V}_{i}(\gamma_{V})'\xi-\tilde{X}_{i}(\gamma_{X})'\beta_{0}\big)||^{q}\bigg]^{1/q}\\
		\leq & \sigma_{X}d_{X}^{1/q}\big(\sigma_{Y}S_{q,\tilde{X},\tilde{Y}}^{1/q}+d_{X}^{1/q}\bar{\beta}S_{q,\tilde{X},\tilde{X}}^{1/q}+d_{V}^{1/q}S_{q,\tilde{X},\tilde{V}}^{1/q}\delta_{\xi,n}\big)+\bar{S}_{q}\\
		+ & \sigma_{X}(d_{X}^{1/q}S_{q,\tilde{X},D}^{1/q}+d_{D}^{1/q}S_{q,D,D}^{1/q})d_{D}^{1/q}\big(\delta_{\gamma,Y,n}\sigma_{Y}+\delta_{\gamma,V,n}(\bar{\xi}+\delta_{\xi,n})+\delta_{\gamma,X,n}\bar{\beta}\big)\\
		\precsim & 1+d_{V}^{1/q}\delta_{\xi,n}+d_{D}^{2/q}(\delta_{\gamma,Y,n}+\delta_{\gamma,V,n}+\delta_{\gamma,X,n}).
	\end{align*}
	
	It remains to bound the second term on the right-hand side of \ref{eq:bigb}.
	Using the definitions of $\bar{Z}_{i}(\cdot),$ $\bar{Y}_{i}(\cdot)$,
	and $\bar{V}_{i}(\cdot)$ and the triangle inequality
	\begin{align}
		& E\bigg[||\mu\bar{Z}_{i}(\omega_{Z})\big(\bar{Y}_{i}(\omega_{Y})-\bar{V}_{i}(\omega_{V})'\xi\big)||^{q}\bigg]^{1/q}\nonumber \\
		\leq & E\big[||\Delta_{\mu}\bar{Z}_{i}(\bar{Y}_{i}+\bar{V}_{i}'\xi_{0})||^{q}\big]^{1/q}\nonumber \\
		+ & E\big[||\mu_{0}\bar{Z}_{i}\bar{Y}_{i}||^{q}\big]^{1/q}+E\big[||\mu_{0}\bar{Z}_{i}\bar{V}_{i}'\xi_{0}||^{q}\big]^{1/q}\nonumber \\
		+ & E\big[||\mu\bar{Z}_{i}\bar{V}_{i}'\Delta_{\xi}||^{q}\big]^{1/q}\nonumber \\
		+ & E\bigg[||\mu\big(\bar{Z}_{i}+\Delta_{\omega,Z}(X_{i}',D_{i}')'\big)(X_{i}',D_{i}')(\Delta_{\omega,Y}-\Delta_{\omega,V}'\xi)||^{q}\bigg]^{1/q}\label{eq:finb}
	\end{align}
	
	We will bound each of the terms on the right-hand side of the inequality
	above. First note that by the triangle inequality and properties of
	the matrix norm:
	\begin{align*}
		& E\big[||\Delta_{\mu}\bar{Z}_{i}(\bar{Y}_{i}+\bar{V}_{i}'\xi_{0})||^{q}\big]^{1/q}\\
		\leq & ||\Delta_{\mu}\Sigma_{\bar{Z}}^{1/2}||\big(E\big[||\Sigma_{\bar{Z}}^{-1/2}\bar{Z}_{i}\bar{Y}_{i}E[\tilde{Y}_{i}^{2}]^{-1/2}||^{q}\big]^{1/q}E[\tilde{Y}_{i}^{2}]^{1/2}+E\big[||\Sigma_{\bar{Z}}^{-1/2}\bar{Z}_{i}\bar{V}_{i}'\Sigma_{\bar{V}}^{-1/2}||^{q}\big]^{1/q}||\Sigma_{\bar{V}}^{1/2}\xi_{0}||\big)\\
		\leq & \delta_{\mu,n}\big(d_{Z}^{1/q}S_{q,\bar{Z},\bar{Y}}^{1/q}\sigma_{Y}+d_{Z}^{1/q}d_{V}^{1/q}S_{q,\bar{Z},\bar{V}}^{1/q}\bar{\xi}\big)
	\end{align*}
	
	Where the final line follows from Assumptions 4.1, 4.2, and the definition
	of $\mathcal{T}_{n}$. Next, by Assumption 4.3:
	\[
	E\big[||\mu_{0}\bar{Z}_{i}\bar{Y}_{i}||^{q}\big]^{1/q}+E\big[||\mu_{0}\bar{Z}_{i}\bar{V}_{i}'\xi_{0}||^{q}\big]^{1/q}\leq2\bar{S}_{q}
	\]
	
	Next applying using the properties of the matrix norm and the triangle
	inequality and then Assumptions 4.2 and 4.3:
	
	\begin{align*}
		& E\big[||\mu\bar{Z}_{i}\bar{V}_{i}'\Delta_{\xi}||^{q}\big]^{1/q}\\
		\leq & (||\mu\Sigma_{\bar{Z}}^{1/2}||+||\Delta_{\mu}\Sigma_{\bar{Z}}^{1/2}||)E\big[||\Sigma_{\bar{Z}}^{-1/2}\bar{Z}_{i}\bar{V}_{i}'\Sigma_{\bar{V}}^{-1/2}||^{q}\big]^{1/q}||\Sigma_{\bar{V}}^{1/2}\Delta_{\xi}||\\
		\leq & (\bar{\mu}+\delta_{\mu,n})\big(d_{Z}^{1/q}d_{V}^{1/q}S_{q,\bar{Z},\bar{V}}^{1/q}\delta_{\xi,n}\big)
	\end{align*}
	
	Now we bound the final term in \ref{eq:finb}. Using basic properties
	of the matrix norm:
	\begin{align*}
		& E\bigg[||\mu\big(\bar{Z}_{i}+\Delta_{\omega,Z}(X_{i}',D_{i}')'\big)(X_{i}',D_{i}')(\Delta_{\omega,Y}-\Delta_{\omega,V}'\xi)||^{q}\bigg]^{1/q}\\
		\leq & ||\mu\Sigma_{\bar{Z}}^{1/2}||E\bigg[||\Sigma_{\bar{Z}}^{-1/2}\big(\bar{Z}_{i}+\Delta_{\omega,Z}(X_{i}',D_{i}')'\big)(X_{i}',D_{i}')\Sigma_{XD}^{-1/2}||^{q}\bigg]^{1/q}||\Sigma_{XD}^{1/2}(\Delta_{\omega,Y}-\Delta_{\omega,V}'\xi)||
	\end{align*}
	
	By the triangle inequality and Assumptions 4.1 and the definition
	of $\mathcal{T}_{n}$ we have $||\mu\Sigma_{\bar{Z}}^{1/2}||\leq\bar{\mu}+\delta_{\mu,n}$.
	Next applying properties of the matrix norm and the triangle inequality
	and we get:
	\begin{align*}
		& E\bigg[||\Sigma_{\bar{Z}}^{-1/2}\big(\bar{Z}_{i}+\Delta_{\omega,Z}(X_{i}',D_{i}')'\big)(X_{i}',D_{i}')\Sigma_{XD}^{-1/2}||^{q}\bigg]^{1/q}\\
		\leq & E\big[||\Sigma_{\bar{Z}}^{-1/2}\bar{Z}_{i}(X_{i}',D_{i}')\Sigma_{XD}^{-1/2}||^{q}\big]^{1/q}\\
		+ & ||\Sigma_{\bar{Z}}^{-1/2}\Delta_{\omega,Z}\Sigma_{XD}^{1/2}||E\big[||\Sigma_{XD}^{-1/2}(X_{i}',D_{i}')'(X_{i}',D_{i}')\Sigma_{XD}^{-1/2}||^{q}\big]^{1/q}\\
		\leq & d_{Z}^{1/q}(d_{X}+d_{D})^{1/q}S_{q,\bar{Z},XD}^{1/q}+\delta_{\omega,Z,n}(d_{X}+d_{D})^{2/q}S_{q,XD,XD}^{1/q}
	\end{align*}
	
	Where the final line follows from the properties of $\mathcal{T}_{n}$
	and Assumption 4.3. Finally, using the triangle inequality, the properties
	of the matrix norm along with Assumptions 4.1 and 4.2:
	\begin{align*}
		& ||\Sigma_{XD}^{1/2}(\Delta_{\omega,Y}-\Delta_{\omega,V}'\xi)||\\
		\leq & ||\Sigma_{XD}^{1/2}\Delta_{\omega,Y}E[\bar{Y}_{i}^{2}]^{-1/2}||E[\bar{Y}_{i}^{2}]^{1/2}+||\Sigma_{XD}^{1/2}\Delta_{\omega,V}'\Sigma_{\bar{V}}^{-1/2}||\big(||\Sigma_{\bar{V}}^{1/2}\xi_{0}||+||\Sigma_{\bar{V}}^{1/2}\Delta_{\xi}||\big)\\
		\leq & \delta_{\omega,Y,n}\sigma_{Y}+\delta_{\omega,V,n}(\bar{\xi}+\delta_{\xi,n})
	\end{align*}
	
	Combining we get:
	\begin{align*}
		& E\bigg[||\mu\big(\bar{Z}_{i}+\Delta_{\omega,Z}(X_{i}',D_{i}')'\big)(X_{i}',D_{i}')(\Delta_{\omega,Y}-\Delta_{\omega,V}'\xi)||^{q}\bigg]^{1/q}\\
		\leq & (\bar{\mu}+\delta_{\mu,n})\big(d_{Z}^{1/q}(d_{X}+d_{D})^{1/q}S_{q,\bar{Z},XD}^{1/q}+\delta_{\omega,Z,n}(d_{X}+d_{D})^{2/q}S_{q,XD,XD}^{1/q}\big)\\
		& \times\big(\delta_{\omega,Y,n}\sigma_{Y}+\delta_{\omega,V,n}(\bar{\xi}+\delta_{\xi,n})\big)
	\end{align*}
	
	And so:
	\begin{align*}
		& E\bigg[||\mu\bar{Z}_{i}(\omega_{Z})\big(\bar{Y}_{i}(\omega_{Y})-\bar{V}_{i}(\omega_{V})'\xi\big)||^{q}\bigg]^{1/q}\\
		\leq & \delta_{\mu,n}\big(d_{Z}^{1/q}S_{q,\bar{Z},\bar{Y}}^{1/q}\sigma_{Y}+d_{Z}^{1/q}d_{V}^{1/q}S_{q,\bar{Z},\bar{V}}^{1/q}\bar{\xi}\big)\\
		+ & 2\bar{S}_{q}\\
		+ & (\bar{\mu}+\delta_{\mu,n})\big(d_{Z}^{1/q}d_{V}^{1/q}S_{q,\bar{Z},\bar{V}}^{1/q}\delta_{\xi,n}\big)\\
		+ & \bigg((\bar{\mu}+\delta_{\mu,n})\big(d_{Z}^{1/q}(d_{X}+d_{D})^{1/q}S_{q,\bar{Z},XD}^{1/q}+\delta_{\omega,Z,n}(d_{X}+d_{D})^{2/q}S_{q,XD,XD}^{1/q}\big)\\
		& \times\big(\delta_{\omega,Y,n}\sigma_{Y}+\delta_{\omega,V,n}(\bar{\xi}+\delta_{\xi,n})\big)\bigg)\\
		\precsim & 1+(\delta_{\mu,n}+\delta_{\xi,n})d_{Z}^{1/q}d_{V}^{1/q}\\
		+ & d_{D}^{1/q}(d_{Z}^{1/q}+d_{D}^{1/q}\delta_{\omega,Z,n})\big(\delta_{\omega,Y,n}+\delta_{\omega,V,n}\big)
	\end{align*}
	
	Combining everything we get:
	\begin{align*}
		c_{1,n}= & \sigma_{X}d_{X}^{1/q}\big(\sigma_{Y}S_{q,\tilde{X},\tilde{Y}}^{1/q}+d_{X}^{1/q}\bar{\beta}S_{q,\tilde{X},\tilde{X}}^{1/q}+d_{V}^{1/q}S_{q,\tilde{X},\tilde{V}}^{1/q}\delta_{\xi,n}\big)+\bar{S}_{q}\\
		+ & \sigma_{X}(d_{X}^{1/q}S_{q,\tilde{X},D}^{1/q}+d_{D}^{1/q}S_{q,D,D}^{1/q})d_{D}^{1/q}\big(\delta_{\gamma,Y,n}\sigma_{Y}+\delta_{\gamma,V,n}(\bar{\xi}+\delta_{\xi,n})+\delta_{\gamma,X,n}\bar{\beta}\big)\\
		+ & \delta_{\mu,n}\big(d_{Z}^{1/q}S_{q,\bar{Z},\bar{Y}}^{1/q}\sigma_{Y}+d_{Z}^{1/q}d_{V}^{1/q}S_{q,\bar{Z},\bar{V}}^{1/q}\bar{\xi}\big)\\
		+ & 2\bar{S}_{q}\\
		+ & (\bar{\mu}+\delta_{\mu,n})\big(d_{Z}^{1/q}d_{V}^{1/q}S_{q,\bar{Z},\bar{V}}^{1/q}\delta_{\xi,n}\big)\\
		+ & \bigg((\bar{\mu}+\delta_{\mu,n})\big(d_{Z}^{1/q}(d_{X}+d_{D})^{1/q}S_{q,\bar{Z},XD}^{1/q}+\delta_{\omega,Z,n}(d_{X}+d_{D})^{2/q}S_{q,XD,XD}^{1/q}\big)\\
		& \times\big(\delta_{\omega,Y,n}\sigma_{Y}+\delta_{\omega,V,n}(\bar{\xi}+\delta_{\xi,n})\big)\bigg)\\
		\precsim & 1+d_{V}^{1/q}\delta_{\xi,n}+d_{D}^{2/q}(\delta_{\gamma,Y,n}+\delta_{\gamma,V,n}+\delta_{\gamma,X,n})\\
		+ & (\delta_{\mu,n}+\delta_{\xi,n})d_{Z}^{1/q}d_{V}^{1/q}\\
		+ & d_{D}^{1/q}(d_{Z}^{1/q}+d_{D}^{1/q}\delta_{\omega,Z,n})\big(\delta_{\omega,Y,n}+\delta_{\omega,V,n}\big)
	\end{align*}
	
	\textbf{Condition 3}
	
	We now consider $\sup_{\eta\in\mathcal{T}_{n}}E\big[||\tilde{X}_{i}(\gamma_{X})\tilde{X}_{i}(\gamma_{X})'||^{q}\big]^{1/q}$.
	We will show that if $P\in\mathcal{P}_{n}$ and $\eta\in\mathcal{T}_{n}$
	then $E\big[||\tilde{X}_{i}(\gamma_{X})\tilde{X}_{i}(\gamma_{X})'||^{q}\big]^{1/q}\leq c_{2,n}$
	where: 
	
	\begin{align*}
		c_{2,n} & =\sigma_{X}d_{X}^{2/q}S_{q,\tilde{X},\tilde{X}}^{1/q}\\
		& +\delta_{\gamma,X,n}\sigma_{X}^{2}d_{D}^{1/q}(d_{X}^{1/q}S_{q,\tilde{X},D}^{1/q}+\delta_{\gamma,X,n}d_{D}^{1/q}S_{q,D,D}^{1/q})\\
		& \precsim1+\delta_{\gamma,X,n}d_{D}^{1/q}(1+\delta_{\gamma,X,n}d_{D}^{1/q})
	\end{align*}
	
	Under the premises of the Theorem the expression on the right-hand
	side above is bounded by a constant and so $c_{2,n}\precsim1$. Since
	$c_{1,n}\precsim1$ and $c_{2,n}\precsim1$ there must exist a sufficiently
	large constant $c_{1}$ so that Conditions 1 and 2 are satisfied.
	
	Let $\eta\in\mathcal{T}_{n}$. By the triangle inequality and the
	definition of $\tilde{X}_{i}(\cdot)$ and $\tilde{X}_{i}$:
	
	\begin{align*}
		& E\big[||\tilde{X}_{i}(\gamma_{X})\tilde{X}_{i}(\gamma_{X})'||^{q}\big]^{1/q}\\
		\leq & E\big[||\tilde{X}_{i}\tilde{X}_{i}'||^{q}\big]^{1/q}\\
		+ & E\big[||(\tilde{X}_{i}+\Delta_{\gamma,X}D_{i})D_{i}'\Delta_{\gamma,X}'||^{q}\big]^{1/q}
	\end{align*}
	
	By the properties of the matrix norm and Assumptions 4.1 and 4.3:
	\begin{align*}
		E\big[||\tilde{X}_{i}\tilde{X}_{i}'||^{q}\big]^{1/q} & \leq||\Sigma_{\tilde{X}}||E\big[||\Sigma_{\tilde{X}}^{-1/2}\tilde{X}_{i}\tilde{X}_{i}'\Sigma_{\tilde{X}}^{-1/2}||^{q}\big]^{1/q}\\
		& \leq\sigma_{X}d_{X}^{2/q}S_{q,\tilde{X},\tilde{X}}^{1/q}
	\end{align*}
	
	Again y the properties of the matrix norm and Assumption 4.1 and 4.2:
	
	\begin{align*}
		& E\big[||(\tilde{X}_{i}+\Delta_{\gamma,X}D_{i})D_{i}'\Delta_{\gamma,X}'||^{q}\big]^{1/q}\\
		\leq & E\big[||(\tilde{X}_{i}+\Delta_{\gamma,X}D_{i})D_{i}'\Sigma_{D}^{-1/2}||^{q}\big]^{1/q}||\Sigma_{D}^{1/2}\Delta_{\gamma,X}'\Sigma_{\tilde{X}}^{-1/2}||\cdot||\Sigma_{\tilde{X}}||^{1/2}\\
		\leq & E\big[||(\tilde{X}_{i}+\Delta_{\gamma,X}D_{i})D_{i}'\Sigma_{D}^{-1/2}||^{q}\big]^{1/q}\delta_{\gamma,X,n}\sigma_{X}
	\end{align*}
	
	Finally, using the triangle inequality and properties of the matrix
	norm and then Assumptions 4.1, 4.2, and 4.3:
	
	\begin{align*}
		E\big[||(\tilde{X}_{i}+\Delta_{\gamma,X}D_{i})D_{i}'\Sigma_{D}^{-1/2}||^{q}\big]^{1/q} & \leq||\Sigma_{\tilde{X}}||^{1/2}E\big[||\Sigma_{\tilde{X}}^{-1/2}\tilde{X}_{i}D_{i}'\Sigma_{D}^{-1/2}||^{q}\big]^{1/q}\\
		& +||\Sigma_{\tilde{X}}||^{1/2}||\Sigma_{\tilde{X}}^{-1/2}\Delta_{\gamma,X}\Sigma_{D}^{1/2}||E\big[||\Sigma_{D}^{-1/2}D_{i}D_{i}'\Sigma_{D}^{-1/2}||^{q}\big]^{1/q}\\
		& \leq\sigma_{X}d_{X}^{1/q}d_{D}^{1/q}S_{q,\tilde{X},D}^{1/q}+\sigma_{X}\delta_{\gamma,X,n}d_{D}^{2/q}S_{q,D,D}^{1/q}
	\end{align*}
	
	Combining gives the result.
	
	\textbf{Condition 4}
	
	We now consider $\sup_{\eta\in\mathcal{T}_{n}}||\Sigma_{\tilde{X}}-E\big[\tilde{X}_{i}(\gamma_{X})\tilde{X}_{i}(\gamma_{X})'\big]||$.
	We will show that if $\eta\in\mathcal{T}_{n}$ and $P\in\mathcal{P}_{n}$
	then letting the sequence $c_{3,n}=d_{X}\delta_{\gamma,X,n}^{2}\sigma_{X}^{2}$
	we have
	\begin{align*}
		||\Sigma_{\tilde{X}}-E\big[\tilde{X}_{i}(\gamma_{X})\tilde{X}_{i}(\gamma_{X})'\big]|| & \leq c_{3,n}\\
		& \precsim\delta_{\gamma,X,n}^{2}
	\end{align*}
	
	Using the properties of partialling out:
	\begin{align*}
		\Sigma_{\tilde{X}}-E\big[\tilde{X}_{i}(\gamma_{X})\tilde{X}_{i}(\gamma_{X})'\big] & =E\bigg[\big(\tilde{X}_{i}-\tilde{X}_{i}(\gamma_{X})\big)\big(\tilde{X}_{i}-\tilde{X}_{i}(\gamma_{X})\big)'\bigg]\\
		& =(\gamma_{X,0}-\gamma_{X})\Sigma_{D}(\gamma_{X,0}-\gamma_{X})'
	\end{align*}
	
	And so:
	\begin{align*}
		||\Sigma_{\tilde{X}}-E\big[\tilde{X}_{i}(\gamma_{X})\tilde{X}_{i}(\gamma_{X})'\big]|| & =||(\gamma_{X,0}-\gamma_{X})\Sigma_{D}(\gamma_{X,0}-\gamma_{X})'||\\
		& \leq E\big[||D_{i}(\gamma_{X,0}-\gamma_{X})'||^{2}\big]\\
		& \leq d_{X}||\Sigma_{D}^{1/2}(\gamma_{X,0}-\gamma_{X})'||^{2}\\
		& \leq d_{X}||\Sigma_{D}^{1/2}(\gamma_{X,0}-\gamma_{X})'\Sigma_{\tilde{X}}^{-1/2}||^{2}||\Sigma_{\tilde{X}}||
	\end{align*}
	
	Where the second line follows by \ref{eq:basic1} and the third line
	follows from the properties of the matrix norm. By the properties
	of $\mathcal{T}_{n}$, $||\Sigma_{D}^{1/2}(\gamma_{X,0}-\gamma_{X})'\Sigma_{\tilde{X}}^{-1/2}||\leq\delta_{\gamma,X,n}$,
	and Assumption 4.1 states that $||\Sigma_{\tilde{X}}||\leq\sigma_{X}^{2}$,
	combining with the above gives the result.
	
	\textbf{Condition 5}
	
	We will show that if $\eta\in\mathcal{T}_{n}$ and $P\in\mathcal{P}_{n}$
	then $E\big[||g_{i}(\beta_{0},\eta_{0})-g_{i}(\beta_{0},\eta)||^{2}\big]^{1/2}\leq c_{4,n}$
	where $c_{4,n}$ has rate below: 
	\begin{align*}
		c_{4,n} & \precsim\delta_{\gamma,Y,n}+\delta_{\gamma,X,n}+\delta_{\gamma,V,n}+\delta_{\omega,Y,n}+\delta_{\omega,Z,n}+\delta_{\omega,V,n}+\delta_{\xi,n}+\delta_{\mu,n}\\
		+ & \sqrt{d_{D}}\delta_{\gamma,X,n}(\delta_{\gamma,Y,n}+\delta_{\gamma,V,n}+\delta_{\gamma,X,n})\\
		+ & \sqrt{d_{D}}\delta_{\omega,Z,n}(\delta_{\omega,Y,n}+\delta_{\omega,V,n})
	\end{align*}
	
	Let $\eta\in\mathcal{T}_{n}$. In our setting $E\big[||g_{i}(\beta_{0},\eta_{0})-g_{i}(\beta_{0},\eta)||^{2}\big]^{1/2}$
	is equal to the following:
	\begin{align*}
		& E\bigg[||\tilde{X}_{i}(\gamma_{X})\big(\tilde{Y}_{i}(\gamma_{Y})-\tilde{V}_{i}(\gamma_{V})'\xi-\tilde{X}_{i}(\gamma_{X})'\beta_{0}\big)-\mu\bar{Z}_{i}(\omega_{Z})\big(\bar{Y}_{i}(\omega_{Y})-\bar{V}_{i}(\omega_{V})'\xi\big)\\
		& -\tilde{X}_{i}(\tilde{Y}_{i}-\tilde{V}_{i}'\xi_{0}-\tilde{X}_{i}'\beta_{0})+\mu_{0}\bar{Z}_{i}(\bar{Y}_{i}-\bar{V}_{i}'\xi_{0})||^{2}\bigg]^{1/2}
	\end{align*}
	
	By the triangle inequality the above is bounded by:
	\begin{align}
		& E\big[||\tilde{X}_{i}(\gamma_{X})\tilde{Y}_{i}(\gamma_{Y})-\tilde{X}_{i}\tilde{Y}_{i}||^{2}\big]^{1/2}\nonumber \\
		+ & E\big[||\tilde{X}_{i}(\gamma_{X})\tilde{V}_{i}(\gamma_{V})'\xi-\tilde{X}_{i}\tilde{V}_{i}'\xi_{0}||^{2}\big]^{1/2}\nonumber \\
		+ & E\big[||\big(\tilde{X}_{i}(\gamma_{X})\tilde{X}_{i}(\gamma_{X})'-\tilde{X}_{i}\tilde{X}_{i}'\big)\beta_{0}||^{2}\big]^{1/2}\nonumber \\
		+ & E\big[||\mu\bar{Z}_{i}(\omega_{Z})\bar{Y}_{i}(\omega_{Y})-\mu_{0}\bar{Z}_{i}\bar{Y}_{i}||^{2}\big]^{1/2}\nonumber \\
		+ & E\big[||\mu\bar{Z}_{i}(\omega_{Z})\bar{V}_{i}(\omega_{V})'\xi-\mu_{0}\bar{Z}_{i}\bar{V}_{i}'\xi_{0}||^{2}\big]^{1/2}\label{eq:decomp}
	\end{align}
	
	Before we begin bound the terms in the expression above, we prove
	an inequality given below. Let $R_{i}$, $H_{i}$, and $Q_{i}$ be
	random vectors and let $R_{i}(\chi)=R_{i}-\chi Q_{i}$ for $\chi=\chi_{R}$
	and $\chi_{R,0}$, and let $H_{i}(\chi)=H_{i}-\chi Q_{i}$ for $\chi=\chi_{H}$
	and $\chi_{H,0}$. Define $\Sigma_{R}=E\big[R_{i}(\chi_{R,0})R_{i}(\chi_{R,0})'\big]$,
	$\Sigma_{H}=E\big[H_{i}(\chi_{H,0})H_{i}(\chi_{H,0})'\big]$, and
	$\Sigma_{Q}=E[Q_{i}Q_{i}']$. Let $A$ be a matrix with $r_{A}$ rows
	and the number of columns equal to the length of $R_{i}$. Let $\zeta$
	be a vector of the same length as $H_{i}$. Define $||A\Sigma_{R}^{1/2}||=\bar{A}$,
	$||\Sigma_{H}^{1/2}\zeta||=\bar{\zeta}$, $||\Sigma_{R}||=\sigma_{R}^{2}$,
	and $||\Sigma_{H}||=\sigma_{H}^{2}$, and $E[||\Sigma_{Q}^{-1/2}Q_{i}||^{4}]=s_{Q}^{2}$.
	Let $\bar{\sigma}_{Q|R}$ and $\bar{\sigma}_{Q|H}$ be the smallest
	almost sure bounds on $||\Sigma_{R}^{-1/2}E\big[R_{i}(\chi_{R,0})R_{i}(\chi_{R,0})'|Q_{i}\big]^{1/2}||$
	and $||\Sigma_{H}^{-1/2}E\big[H_{i}(\chi_{H,0})H_{i}(\chi_{H,0})'|Q_{i}\big]^{1/2}||$
	respectively. We will show that:
	
	\begin{align}
		& E\bigg[||A\big(R_{i}(\chi_{R})H_{i}(\chi_{H})'-R_{i}(\chi_{R,0})H_{i}(\chi_{H,0})'\big)\zeta||^{2}\bigg]^{1/2}\nonumber \\
		\leq & \sqrt{r_{A}}\bar{A}\bar{\zeta}\bar{\sigma}_{R|Q}||\Sigma_{H}^{-1/2}(\chi_{H,0}-\chi_{H})\Sigma_{Q}^{1/2}||\nonumber \\
		+ & \sqrt{r_{A}}\bar{A}\bar{\zeta}\bar{\sigma}_{H|Q}||\Sigma_{R}^{-1/2}(\chi_{R,0}-\chi_{R})\Sigma_{Q}^{1/2}||\nonumber \\
		+ & s_{Q}\sqrt{d_{Q}}\bar{A}\bar{\zeta}||\Sigma_{H}^{-1/2}(\chi_{H,0}-\chi_{H})\Sigma_{Q}^{1/2}||\cdot||\Sigma_{R}^{-1/2}(\chi_{R,0}-\chi_{R})\Sigma_{Q}^{1/2}||\label{eq:firstver}
	\end{align}
	
	To see this, we first apply the triangle inequality and the properties
	of the matrix norm:
	
	\begin{align}
		& E\bigg[||A\big(R_{i}(\chi_{R})H_{i}(\chi_{H})'-R_{i}(\chi_{R,0})H_{i}(\chi_{H,0})'\big)\zeta||^{2}\bigg]^{1/2}\nonumber \\
		\leq & E\big[||AR_{i}(\chi_{R,0})Q_{i}'(\chi_{R,0}-\chi_{R})'\zeta||^{2}\big]^{1/2}\nonumber \\
		+ & E\big[||A(\chi_{R,0}-\chi_{R})Q_{i}H_{i}(\chi_{H,0})'\zeta||^{2}\big]^{1/2}\nonumber \\
		+ & E\big[||A(\chi_{R,0}-\chi_{R})Q_{i}Q_{i}'(\chi_{H,0}-\chi_{H})'\zeta||^{2}\big]^{1/2}\label{eq:intbound}
	\end{align}
	
	To help bound the above we will show that $E\big[||AR_{i}(\chi_{R,0})||^{2}|Q_{i}\big]\leq r_{A}\bar{\sigma}_{R|Q}^{2}\bar{A}^{2}$
	and $E\big[||\zeta'H_{i}(\chi_{H,0})||^{2}|Q_{i}\big]\leq\bar{\sigma}_{H|Q}^{2}\bar{\zeta}^{2}$
	each with probability $1$. To see this, we first apply \ref{eq:basic1}
	to get:
	\[
	E\big[||AR_{i}(\chi_{R,0})||^{2}|Q_{i}\big]\leq r_{A}||AE\big[R_{i}(\chi_{R,0})R_{i}(\chi_{R,0})'|Q_{i}\big]^{1/2}||^{2}
	\]
	
	Using the properties of the matrix norm we then get:
	\begin{align*}
		E\big[||AR_{i}(\chi_{R,0})||^{2}|Q_{i}\big] & \leq r_{A}||A\Sigma_{R}^{1/2}||^{2}||\Sigma_{R}^{-1/2}E\big[R_{i}(\chi_{R,0})R_{i}(\chi_{R,0})'|Q_{i}\big]^{1/2}||^{2}\\
		& \leq r_{A}\bar{\sigma}_{R|Q}^{2}\bar{A}^{2}
	\end{align*}
	
	The inequality $E\big[||\zeta'H_{i}(\chi_{H,0})||^{2}|Q_{i}\big]\leq\bar{\sigma}_{H|Q}^{2}\bar{\zeta}^{2}$
	follows from the same steps (note that $\zeta$ is a column vector
	and so $\zeta'$ has one row, hence there is no term analogous to
	$r_{A}$ in the upper bound).
	
	Now we bound the first term on the right-hand side of \ref{eq:intbound}: 
	
	\begin{align*}
		& E\big[||AR_{i}(\chi_{R,0})Q_{i}'(\chi_{H,0}-\chi_{H})'\zeta||^{2}\big]\\
		= & E\bigg[E\big[||AR_{i}(\chi_{R,0})||^{2}|Q_{i}\big]||\zeta'(\chi_{H,0}-\chi_{H})Q_{i}||^{2}\bigg]\\
		\leq & r_{A}\bar{\sigma}_{R|Q}^{2}\bar{A}^{2}E\big[||\zeta'(\chi_{H,0}-\chi_{H})Q_{i}||^{2}\big]\\
		= & r_{A}\bar{\sigma}_{R|Q}^{2}\bar{A}^{2}||\zeta'(\chi_{H,0}-\chi_{H})\Sigma_{Q}^{1/2}||^{2}\\
		\leq & r_{A}\bar{\sigma}_{R|Q}^{2}\bar{A}^{2}\bar{\zeta}^{2}||\Sigma_{H}^{-1/2}(\chi_{H,0}-\chi_{H})\Sigma_{Q}^{1/2}||^{2}
	\end{align*}
	
	Following similar steps we get:
	
	\begin{align*}
		& E\big[||A(\chi_{R,0}-\chi_{R})Q_{i}H_{i}(\chi_{H,0})'\zeta||^{2}\big]\\
		= & E\bigg[||A(\chi_{R,0}-\chi_{R})Q_{i}||^{2}E\big[||H_{i}(\chi_{H,0})'\zeta||^{2}|Q_{i}\big]\bigg]\\
		\leq & \bar{\sigma}_{H|Q}^{2}\bar{\zeta}^{2}E\big[||A(\chi_{R,0}-\chi_{R})Q_{i}||^{2}\big]\\
		\leq & r_{A}\bar{\sigma}_{H|Q}^{2}\bar{\zeta}^{2}||A(\chi_{R,0}-\chi_{R})\Sigma_{Q}^{1/2}||^{2}\\
		\leq & r_{A}\bar{A}^{2}\bar{\sigma}_{H|Q}^{2}\bar{\zeta}^{2}||\Sigma_{R}^{-1/2}(\chi_{R,0}-\chi_{R})\Sigma_{Q}^{1/2}||^{2}
	\end{align*}
	
	Where the penultimate line uses \ref{eq:basic1}. Finally, with repeated
	application of the properties of the matrix norm:
	
	\begin{align*}
		& E\big[||A(\chi_{R,0}-\chi_{R})Q_{i}Q_{i}'(\chi_{H,0}-\chi_{H})'\zeta||^{2}\big]\\
		= & E\big[||A(\chi_{R,0}-\chi_{R})\Sigma_{Q}^{1/2}\Sigma_{Q}^{-1/2}Q_{i}Q_{i}'\Sigma_{Q}^{-1/2}\Sigma_{Q}^{1/2}(\chi_{H,0}-\chi_{H})'\zeta||^{2}\big]\\
		\leq & E\big[||A(\chi_{R,0}-\chi_{R})\Sigma_{Q}^{1/2}||^{2}\cdot||\Sigma_{Q}^{-1/2}Q_{i}Q_{i}'\Sigma_{Q}^{-1/2}||^{2}\cdot||\Sigma_{Q}^{1/2}(\chi_{H,0}-\chi_{H})'\zeta||^{2}\big]\\
		= & ||A(\chi_{R,0}-\chi_{R})\Sigma_{Q}^{1/2}||^{2}E[||\Sigma_{Q}^{-1/2}Q_{i}Q_{i}'\Sigma_{Q}^{-1/2}||^{2}]||\Sigma_{Q}^{1/2}(\chi_{H,0}-\chi_{H})'\zeta||^{2}\\
		= & E[||\Sigma_{Q}^{-1/2}Q_{i}||^{4}]||A(\chi_{R,0}-\chi_{R})\Sigma_{Q}^{1/2}||^{2}||\Sigma_{Q}^{1/2}(\chi_{H,0}-\chi_{H})'\zeta||^{2}\\
		\leq & s_{Q}^{2}d_{Q}||\Sigma_{Q}^{1/2}(\chi_{H,0}-\chi_{H})'\zeta||^{2}||A(\chi_{R,0}-\chi_{R})\Sigma_{Q}^{1/2}||\\
		\leq & s_{Q}^{2}d_{Q}\bar{A}^{2}\bar{\zeta}^{2}\\
		\times & ||\Sigma_{H}^{-1/2}(\chi_{H,0}-\chi_{H})\Sigma_{Q}^{1/2}||^{2}\\
		\times & ||\Sigma_{R}^{-1/2}(\chi_{R,0}-\chi_{R})\Sigma_{Q}^{1/2}||^{2}
	\end{align*}
	
	Combining we get \ref{eq:firstver}.
	
	Consider the first term in \ref{eq:decomp}. We apply \ref{eq:firstver}
	with $A$ the $d_{X}\times d_{X}$ identity matrix, $\zeta=1$, $R_{i}(\cdot)=\tilde{X}_{i}(\cdot)$,
	$H_{i}(\cdot)=\tilde{Y}_{i}(\cdot)$, $\chi_{R}=\gamma_{X}$, $\chi_{R,0}=\gamma_{X,0}$,
	$\chi_{H}=\gamma_{Y}$, and $\chi_{H,0}=\gamma_{Y,0}$. Note that
	in this case $r_{A}=d_{X}$, $\bar{A}\leq\sigma_{X}$, $\bar{\zeta}\leq\sigma_{Y}$,
	$\bar{\sigma}_{H|Q}\leq\bar{\sigma}_{\tilde{Y}|D}$, $\bar{\sigma}_{R|Q}\leq\bar{\sigma}_{\tilde{X}|D}$,
	and $s_{Q}\leq s_{D}$. We get:
	
	\begin{align*}
		& E\big[||\tilde{X}_{i}(\gamma_{X})\tilde{Y}_{i}(\gamma_{Y})-\tilde{X}_{i}\tilde{Y}_{i}||^{2}\big]^{1/2}\\
		\leq & \sigma_{X}\sigma_{Y}(\sqrt{d_{X}}\bar{\sigma}_{\tilde{X}|D}\delta_{\gamma,Y,n}+\sqrt{d_{X}}\bar{\sigma}_{\tilde{Y}|D}\delta_{\gamma,X,n}+\sqrt{d_{D}}s_{D}\delta_{\gamma,X,n}\delta_{\gamma,Y,n})
	\end{align*}
	
	Where we have used the properties of $\mathcal{T}_{n}$ for the bounds
	$\delta_{\gamma,Y,n}$ and $\delta_{\gamma,X,n}$. 
	
	Next we bound the second term in \ref{eq:decomp}. By the triangle
	inequality:
	
	\begin{align*}
		& E\big[||\tilde{X}_{i}(\gamma_{X})\tilde{V}_{i}(\gamma_{V})'\xi-\tilde{X}_{i}\tilde{V}_{i}'\xi_{0}||^{2}\big]^{1/2}\\
		\leq & E\big[||\big(\tilde{X}_{i}(\gamma_{X})\tilde{V}_{i}(\gamma_{V})'-\tilde{X}_{i}\tilde{V}_{i}'\big)\xi_{0}||^{2}\big]^{1/2}\\
		+ & E\bigg[||\big(\tilde{X}_{i}(\gamma_{X})\tilde{V}_{i}(\gamma_{V})'-\tilde{X}_{i}\tilde{V}_{i}'\big)\Delta_{\xi}||^{2}\bigg]^{1/2}\\
		+ & E\bigg[||\tilde{X}_{i}\tilde{V}_{i}'\Delta_{\xi}||^{2}\bigg]^{1/2}
	\end{align*}
	
	Let us apply \ref{eq:firstver} with $A$ the identity, $\zeta=\xi_{0}$,
	$R_{i}(\cdot)=\tilde{X}_{i}(\cdot)$, $H_{i}(\cdot)=\tilde{V}_{i}(\cdot)$,
	$\chi_{R}=\gamma_{X}$, $\chi_{R,0}=\gamma_{X,0}$, $\chi_{H}=\gamma_{V}$,
	and $\chi_{H,0}=\gamma_{V,0}$. We get:
	
	\begin{align*}
		& E\big[||\big(\tilde{X}_{i}(\gamma_{X})\tilde{V}_{i}(\gamma_{V})'-\tilde{X}_{i}\tilde{V}_{i}'\big)\xi_{0}||^{2}\big]^{1/2}\\
		\leq & \sigma_{X}\bar{\xi}(\sqrt{d_{X}}\bar{\sigma}_{\tilde{X}|D}\delta_{\gamma,V,n}+\sqrt{d_{X}}\bar{\sigma}_{\tilde{V}|D}\delta_{\gamma,X,n}+\sqrt{d_{D}}s_{D}\delta_{\gamma,X,n}\delta_{\gamma,V,n})
	\end{align*}
	
	Where we have again used the properties of $\mathcal{T}_{n}$ for
	the bounds $\delta_{\gamma,V,n}$, $\delta_{\gamma,X,n}$, and $\delta_{\gamma,XV}$.
	Doing just as above but with $\zeta=\xi_{0}-\xi$ we get:
	\begin{align*}
		& E\big[||\big(\tilde{X}_{i}(\gamma_{X})\tilde{V}_{i}(\gamma_{V})'-\tilde{X}_{i}\tilde{V}_{i}'\big)\Delta_{\xi}||^{2}\big]^{1/2}\\
		\leq & \sigma_{X}\delta_{\xi,n}(\sqrt{d_{X}}\bar{\sigma}_{\tilde{X}|D}\delta_{\gamma,V,n}+\sqrt{d_{X}}\bar{\sigma}_{\tilde{V}|D}\delta_{\gamma,X,n}+\sqrt{d_{D}}s_{D}\delta_{\gamma,X,n}\delta_{\gamma,V,n})
	\end{align*}
	
	Where we have used that if $\eta\in\mathcal{T}_{n}$ $||\Sigma_{\tilde{V}}^{1/2}(\xi-\xi)||^{2}\leq\delta_{\xi,n}^{2}$.
	Finally we note that:
	
	\begin{align*}
		E\big[||\tilde{X}_{i}\tilde{V}_{i}'\Delta_{\xi}||^{2}\big]^{1/2} & =E\big[||\tilde{X}_{i}||^{2}||E[\Sigma_{\tilde{V}}^{-1/2}\tilde{V}_{i}\tilde{V}_{i}'\Sigma_{\tilde{V}}^{-1/2}|\tilde{X}_{i}]^{1/2}\Sigma_{\tilde{V}}^{1/2}\Delta_{\xi}||^{2}\big]^{1/2}\\
		& \leq E\big[||\tilde{X}_{i}||^{2}||E[\Sigma_{\tilde{V}}^{-1/2}\tilde{V}_{i}\tilde{V}_{i}'\Sigma_{\tilde{V}}^{-1/2}|\tilde{X}_{i}]||\big]^{1/2}||\Sigma_{\tilde{V}}^{1/2}\Delta_{\xi}||\\
		& =\leq E\big[||\tilde{X}_{i}||^{2}||\Sigma_{\tilde{V}}^{-1/2}E[\tilde{V}_{i}\tilde{V}_{i}'|\tilde{X}_{i}]^{1/2}||^{2}\big]^{1/2}||\Sigma_{\tilde{V}}^{1/2}\Delta_{\xi}||\\
		& \leq E[||\tilde{X}_{i}||^{2}]^{1/2}\delta_{\xi,n}\bar{\sigma}_{\tilde{V}|\tilde{X}}\\
		& \leq\sqrt{d_{X}}\sigma_{X}\delta_{\xi,n}\bar{\sigma}_{\tilde{V}|\tilde{X}}
	\end{align*}
	
	Where we have used that $E[||\tilde{X}_{i}||^{2}]=trace\big(E[\tilde{X}_{i}\tilde{X}_{i}']\big)=d_{X}\sigma_{X}^{2}$.
	
	So in all, the second term in \ref{eq:decomp} is bounded by:
	\begin{align*}
		& E\big[||\tilde{X}_{i}(\gamma_{X})\tilde{V}_{i}(\gamma_{V})'\xi-\tilde{X}_{i}\tilde{V}_{i}'\xi_{0}||^{2}\big]^{1/2}\\
		\leq & \sigma_{X}(\bar{\xi}+\delta_{\xi,n})(\sqrt{d_{X}}\bar{\sigma}_{\tilde{X}|D}\delta_{\gamma,V,n}+\sqrt{d_{X}}\bar{\sigma}_{\tilde{V}|D}\delta_{\gamma,X,n}+\sqrt{d_{D}}s_{D}\delta_{\gamma,X,n}\delta_{\gamma,V,n})\\
		+ & \sqrt{d_{X}}\sigma_{X}\delta_{\xi,n}\bar{\sigma}_{\tilde{V}|\tilde{X}}
	\end{align*}
	
	Next we bound the third term in \ref{eq:decomp}. Once again let us
	apply \ref{eq:firstver} with $A$ the identity, $\zeta=\beta_{0}$,
	$R_{i}(\cdot)=H(\cdot)=\tilde{X}_{i}(\cdot)$, $\chi_{R}=\chi_{H}=\gamma_{X}$,
	and $\chi_{R,0}=\chi_{H,0}=\gamma_{X,0}$. We get, using the properties
	of $\mathcal{T}_{n}$:
	\begin{align*}
		E\big[||\tilde{X}_{i}(\gamma_{X})\tilde{X}_{i}(\gamma_{Y})'\beta_{0}-\tilde{X}_{i}\tilde{X}_{i}'\beta_{0}||^{2}\big]^{1/2} & \leq\sigma_{X}\bar{\beta}(2\sqrt{d_{X}}\bar{\sigma}_{\tilde{X}|D}\delta_{\gamma,X,n}+\sqrt{d_{D}}s_{D}\delta_{\gamma,X,n}^{2})
	\end{align*}
	
	Next we bound the fourth term in \ref{eq:decomp}. By the triangle
	inequality:
	\begin{align*}
		& E\big[||\mu\bar{Z}_{i}(\omega_{Z})\bar{Y}_{i}(\omega_{Y})-\mu_{0}\bar{Z}_{i}\bar{Y}_{i}||^{2}\big]^{1/2}\\
		\leq & E\big[||\mu_{0}\bar{Z}_{i}(\omega_{Z})\bar{Y}_{i}(\omega_{Y})-\mu_{0}\bar{Z}_{i}\bar{Y}_{i}||^{2}\big]^{1/2}\\
		+ & E\big[||\Delta_{\mu}\big(\bar{Z}_{i}(\omega_{Z})\bar{Y}_{i}(\omega_{Y})-\bar{Z}_{i}\bar{Y}_{i}\big)||^{2}\big]^{1/2}\\
		+ & E\big[||\Delta_{\mu}\bar{Z}_{i}\bar{Y}_{i}||^{2}\big]^{1/2}
	\end{align*}
	
	Let us apply \ref{eq:firstver} with $A=\mu_{0}$ the identity, $\zeta=1$,
	$R_{i}(\cdot)=\bar{Z}_{i}(\cdot)$, $H_{i}(\cdot)=\bar{Y}_{i}(\cdot)$,
	$\chi_{R}=\omega_{Z}$, $\chi_{R,0}=\omega_{Z,0}$, $\chi_{H}=\omega_{Y}$,
	and $\chi_{H,0}=\omega_{Y,0}$. Note that $\mu_{0}$ has $d_{X}$
	rows so $r_{A}=d_{X}$. We get:
	
	\begin{align*}
		& E\big[||\mu_{0}\bar{Z}_{i}(\omega_{Z})\bar{Y}_{i}(\omega_{Y})-\mu_{0}\bar{Z}_{i}\bar{Y}_{i}||^{2}\big]^{1/2}\\
		\leq & \bar{\mu}\sigma_{Y}(\sqrt{d_{X}}\bar{\sigma}_{\bar{Z}|XD}\delta_{\omega,Y,n}+\sqrt{d_{X}}\bar{\sigma}_{\bar{Y}|XD}\delta_{\omega,Z,n}+\sqrt{d_{X}+d_{D}}s_{XD}\delta_{\omega,Z,n}\delta_{\omega,Y,n})
	\end{align*}
	
	Repeating the above but with $A=\mu_{0}-\mu$ we get:
	\begin{align*}
		& E\big[||\mu_{0}\bar{Z}_{i}(\omega_{Z})\bar{Y}_{i}(\omega_{Y})-\mu_{0}\bar{Z}_{i}\bar{Y}_{i}||^{2}\big]^{1/2}\\
		\leq & \delta_{\mu,n}\sigma_{Y}(\sqrt{d_{X}}\bar{\sigma}_{\bar{Z}|XD}\delta_{\omega,Y,n}+\sqrt{d_{X}}\bar{\sigma}_{\bar{Y}|XD}\delta_{\omega,Z,n}+\sqrt{d_{X}+d_{D}}s_{XD}\delta_{\omega,Z,n}\delta_{\omega,Y,n})
	\end{align*}
	
	And now note that: 
	
	\begin{align*}
		E\big[||\Delta_{\mu}\bar{Z}_{i}\bar{Y}_{i}||^{2}\big]^{1/2} & =E\big[||\Delta_{\mu}\bar{Z}_{i}||^{2}E[\bar{Y}_{i}^{2}|\bar{Z}_{i}]\big]^{1/2}\\
		& \leq\sigma_{\bar{Y}|\bar{Z}}\sigma_{Y}E\big[||\Delta_{\mu}\bar{Z}_{i}||^{2}\big]^{1/2}\\
		& \leq\sigma_{\bar{Y}|\bar{Z}}\sigma_{Y}\sqrt{d_{X}}||\Delta_{\mu}\Sigma_{\bar{Z}}^{1/2}||\\
		& \leq\sqrt{d_{X}}\sigma_{\bar{Y}|\bar{Z}}\sigma_{Y}\delta_{\mu,n}
	\end{align*}
	
	Where the final inequality uses \ref{eq:basic1}. And so in all:
	\begin{align*}
		& E\big[||\mu\bar{Z}_{i}(\omega_{Z})\bar{Y}_{i}(\omega_{Y})-\mu_{0}\bar{Z}_{i}\bar{Y}_{i}||^{2}\big]^{1/2}\\
		\leq & \sigma_{Y}(\bar{\mu}+\delta_{\mu,n})(\sqrt{d_{X}}\bar{\sigma}_{\bar{Z}|XD}\delta_{\omega,Y,n}+\sqrt{d_{X}}\bar{\sigma}_{\bar{Y}|XD}\delta_{\omega,Z,n}+\sqrt{d_{X}+d_{D}}s_{XD}\delta_{\omega,Z,n}\delta_{\omega,Y,n})\\
		+ & \sqrt{d_{X}}\sigma_{\bar{Y}|\bar{Z}}\sigma_{Y}\delta_{\mu,n}
	\end{align*}
	
	Next we bound the final term in \ref{eq:decomp}. By the triangle
	inequality:
	
	\begin{align*}
		& E\big[||\mu\bar{Z}_{i}(\omega_{Z})\bar{V}_{i}(\omega_{V})'\xi-\mu_{0}\bar{Z}_{i}\bar{V}_{i}'\xi_{0}||^{2}\big]^{1/2}\\
		\leq & E\big[||\mu_{0}\bar{Z}_{i}(\omega_{Z})\bar{V}_{i}(\omega_{V})'\xi_{0}-\mu_{0}\bar{Z}_{i}\bar{V}_{i}'\xi_{0}||^{2}\big]^{1/2}\\
		+ & E\bigg[||\Delta_{\mu}\big(\bar{Z}_{i}(\omega_{Z})\bar{V}_{i}(\omega_{V})'\xi_{0}-\mu_{0}\bar{Z}_{i}\bar{V}_{i}'\xi_{0}\big)||^{2}\bigg]^{1/2}\\
		+ & E\bigg[||\Delta_{\mu}\big(\bar{Z}_{i}(\omega_{Z})\bar{V}_{i}(\omega_{V})'-\bar{Z}_{i}\bar{V}_{i}'\big)\Delta_{\xi}||^{2}\bigg]^{1/2}\bigg]^{1/2}\\
		+ & E\bigg[||\mu_{0}\big(\bar{Z}_{i}(\omega_{Z})\bar{V}_{i}(\omega_{V})'-\bar{Z}_{i}\bar{V}_{i}'\big)\Delta_{\xi}||^{2}\\
		+ & E\big[||\Delta_{\mu}\bar{Z}_{i}\bar{V}_{i}'\xi_{0}||^{2}\big]^{1/2}\\
		+ & E\big[||\Delta_{\mu}\bar{Z}_{i}\bar{V}_{i}'\Delta_{\xi}||^{2}\big]^{1/2}\\
		+ & E\big[||\mu_{0}\bar{Z}_{i}\bar{V}_{i}'\Delta_{\xi}||^{2}\big]^{1/2}
	\end{align*}
	
	Let us apply \ref{eq:firstver} with $A=\mu_{0}$ the identity, $\zeta=\xi_{0}$,
	$R_{i}(\cdot)=\bar{Z}_{i}(\cdot)$, $H_{i}(\cdot)=\bar{V}_{i}(\cdot)$,
	$\chi_{R}=\omega_{Z}$, $\chi_{R,0}=\omega_{Z,0}$, $\chi_{H}=\omega_{V}$,
	and $\chi_{H,0}=\omega_{V,0}$. We get:
	
	\begin{align*}
		& E\big[||\mu_{0}\bar{Z}_{i}(\omega_{Z})\bar{V}_{i}(\omega_{V})'\xi_{0}-\mu_{0}\bar{Z}_{i}\bar{V}_{i}'\xi_{0}||^{2}\big]^{1/2}\\
		\leq & \bar{\mu}\bar{\xi}(\sqrt{d_{X}}\bar{\sigma}_{\bar{Z}|XD}\delta_{\omega,V,n}+\sqrt{d_{X}}\bar{\sigma}_{\bar{V}|XD}\delta_{\omega,Z,n}+\sqrt{d_{X}+d_{D}}s_{XD}\delta_{\omega,Z,n}\delta_{\omega,V,n})
	\end{align*}
	
	Repeating the above but with $A=\mu_{0}-\mu$ instead of $A=\mu_{0}$
	we get:
	\begin{align*}
		& E\bigg[||\Delta_{\mu}\big(\bar{Z}_{i}(\omega_{Z})\bar{V}_{i}(\omega_{V})'\xi_{0}-\mu_{0}\bar{Z}_{i}\bar{V}_{i}'\xi_{0}\big)||^{2}\bigg]^{1/2}\\
		\leq & \delta_{\mu,n}\bar{\xi}(\sqrt{d_{X}}\bar{\sigma}_{\bar{Z}|XD}\delta_{\omega,V,n}+\sqrt{d_{X}}\bar{\sigma}_{\bar{V}|XD}\delta_{\omega,Z,n}+\sqrt{d_{X}+d_{D}}s_{XD}\delta_{\omega,Z,n}\delta_{\omega,V,n})
	\end{align*}
	
	Using the steps above but with $\zeta=\xi_{0}-\xi$ instead of $\zeta=\xi_{0}$:
	\begin{align*}
		& E\bigg[||\Delta_{\mu}\big(\bar{Z}_{i}(\omega_{Z})\bar{V}_{i}(\omega_{V})'-\bar{Z}_{i}\bar{V}_{i}'\big)\Delta_{\xi}||^{2}\bigg]^{1/2}\\
		\leq & \delta_{\mu,n}\delta_{\xi,n}(\sqrt{d_{X}}\bar{\sigma}_{\bar{Z}|XD}\delta_{\omega,V,n}+\sqrt{d_{X}}\bar{\sigma}_{\bar{V}|XD}\delta_{\omega,Z,n}+\sqrt{d_{X}+d_{D}}s_{XD}\delta_{\omega,Z,n}\delta_{\omega,V,n})
	\end{align*}
	
	And applying the steps above but with $A=\mu_{0}$ instead of $A=\mu_{0}-\mu$:
	\begin{align*}
		& E\bigg[||\Delta_{\mu}\big(\bar{Z}_{i}(\omega_{Z})\bar{V}_{i}(\omega_{V})'\xi_{0}-\mu_{0}\bar{Z}_{i}\bar{V}_{i}'\xi_{0}\big)||^{2}\bigg]^{1/2}\\
		\leq & \bar{\mu}\delta_{\xi,n}(\sqrt{d_{X}}\bar{\sigma}_{\bar{Z}|XD}\delta_{\omega,V,n}+\sqrt{d_{X}}\bar{\sigma}_{\bar{V}|XD}\delta_{\omega,Z,n}+\sqrt{d_{X}+d_{D}}s_{XD}\delta_{\omega,Z,n}\delta_{\omega,V,n})
	\end{align*}
	
	Next note that:
	
	\begin{align*}
		E[||\bar{V}_{i}'\xi_{0}||^{2}|\bar{Z}_{i}] & =||E[\bar{V}_{i}'\bar{V}_{i}|\bar{Z}_{i}]^{1/2}\Sigma_{\bar{V}}^{-1/2}\Sigma_{\bar{V}}^{1/2}\xi_{0}||^{2}\\
		& \leq||E[\bar{V}_{i}'\bar{V}_{i}|\bar{Z}_{i}]^{1/2}\Sigma_{\bar{V}}^{-1/2}||^{2}\cdot||\Sigma_{\bar{V}}^{1/2}\xi_{0}||^{2}\\
		& \leq\bar{\sigma}_{\bar{V}|\bar{Z}}^{2}\bar{\xi}^{2}
	\end{align*}
	
	Applying the above we get:
	
	\begin{align*}
		E\big[||\Delta_{\mu}\bar{Z}_{i}\bar{V}_{i}'\xi_{0}||^{2}\big] & =E\big[||\Delta_{\mu}\bar{Z}_{i}||^{2}\cdot E[||\bar{V}_{i}'\xi_{0}||^{2}|\bar{Z}_{i}]\big]\\
		& \leq\bar{\sigma}_{\bar{V}|\bar{Z}}^{2}\bar{\xi}^{2}E\big[||\Delta_{\mu}\bar{Z}_{i}||^{2}\big]\\
		& \leq\bar{\sigma}_{\bar{V}|\bar{Z}}^{2}\bar{\xi}^{2}d_{X}||\Delta_{\mu}\Sigma_{Z}^{1/2}||^{2}\\
		& \leq d_{X}\delta_{\mu,n}^{2}\bar{\xi}^{2}\bar{\sigma}_{\bar{V}|\bar{Z}}^{2}
	\end{align*}
	
	Where the penultimate line follows from \ref{eq:basic1}. Applying
	the same steps with $\xi_{0}-\xi$ in place of $\xi_{0}$:
	\[
	E\big[||\Delta_{\mu}\bar{Z}_{i}\bar{V}_{i}'\Delta_{\xi}||^{2}\big]^{1/2}\leq\sqrt{d_{X}}\delta_{\mu,n}\delta_{\xi,n}\bar{\sigma}_{\bar{V}|\bar{Z}}
	\]
	
	Applying the same steps with $\mu_{0}$ in place of $\mu_{0}-\mu$:
	\[
	E\big[||\mu_{0}\bar{Z}_{i}\bar{V}_{i}'\Delta_{\xi}||^{2}\big]^{1/2}\leq\sqrt{d_{X}}\bar{\mu}\delta_{\xi,n}\bar{\sigma}_{\bar{V}|\bar{Z}}
	\]
	
	So the final term in \ref{eq:decomp} is bounded by:
	\begin{align*}
		& E\big[||\mu\bar{Z}_{i}(\omega_{Z})\bar{V}_{i}(\omega_{V})'\xi-\mu_{0}\bar{Z}_{i}\bar{V}_{i}'\xi_{0}||^{2}\big]^{1/2}\\
		\leq & (\bar{\mu}+\delta_{\mu,n})(\bar{\xi}+\delta_{\xi,n})(\sqrt{d_{X}}\bar{\sigma}_{\bar{Z}|XD}\delta_{\omega,V,n}+\sqrt{d_{X}}\bar{\sigma}_{\bar{V}|XD}\delta_{\omega,Z,n}+\sqrt{d_{X}+d_{D}}s_{XD}\delta_{\omega,Z,n}\delta_{\omega,V,n})\\
		+ & (\delta_{\mu,n}\bar{\xi}+\bar{\mu}\delta_{\xi,n}+\delta_{\mu,n}\delta_{\xi,n})\sqrt{d_{X}}\bar{\sigma}_{\bar{V}|\bar{Z}}
	\end{align*}
	
	Combining everything, we see that \ref{eq:decomp} and therefore $E\big[||g_{i}(\beta_{0},\eta_{0})-g_{i}(\beta_{0},\eta)||^{2}\big]^{1/2}$
	is bounded by $c_{4,n}$ given by:
	\begin{align*}
		c_{4,n}= & \sigma_{X}\sigma_{Y}(\sqrt{d_{X}}\bar{\sigma}_{\tilde{X}|D}\delta_{\gamma,Y,n}+\sqrt{d_{X}}\bar{\sigma}_{\tilde{Y}|D}\delta_{\gamma,X,n}+\sqrt{d_{D}}s_{D}\delta_{\gamma,X,n}\delta_{\gamma,Y,n})\\
		+ & \sigma_{X}(\bar{\xi}+\delta_{\xi,n})(\sqrt{d_{X}}\bar{\sigma}_{\tilde{X}|D}\delta_{\gamma,V,n}+\sqrt{d_{X}}\bar{\sigma}_{\tilde{V}|D}\delta_{\gamma,X,n}+\sqrt{d_{D}}s_{D}\delta_{\gamma,X,n}\delta_{\gamma,V,n})\\
		+ & \sqrt{d_{X}}\sigma_{X}\delta_{\xi,n}\bar{\sigma}_{\tilde{V}|\tilde{X}}\\
		+ & \sigma_{X}\bar{\beta}(2\sqrt{d_{X}}\bar{\sigma}_{\tilde{X}|D}\delta_{\gamma,X,n}+\sqrt{d_{D}}s_{D}\delta_{\gamma,X,n}^{2})\\
		+ & \sigma_{Y}(\bar{\mu}+\delta_{\mu,n})(\sqrt{d_{X}}\bar{\sigma}_{\bar{Z}|XD}\delta_{\omega,Y,n}+\sqrt{d_{X}}\bar{\sigma}_{\bar{Y}|XD}\delta_{\omega,Z,n}+\sqrt{d_{X}+d_{D}}s_{XD}\delta_{\omega,Z,n}\delta_{\omega,Y,n})\\
		+ & \sqrt{d_{X}}\sigma_{\bar{Y}|\bar{Z}}\sigma_{Y}\delta_{\mu,n}\\
		+ & (\bar{\mu}+\delta_{\mu,n})(\bar{\xi}+\delta_{\xi,n})(\sqrt{d_{X}}\bar{\sigma}_{\bar{Z}|XD}\delta_{\omega,V,n}+\sqrt{d_{X}}\bar{\sigma}_{\bar{V}|XD}\delta_{\omega,Z,n}+\sqrt{d_{X}+d_{D}}s_{XD}\delta_{\omega,Z,n}\delta_{\omega,V,n})\\
		+ & (\delta_{\mu,n}\bar{\xi}+\bar{\mu}\delta_{\xi,n}+\delta_{\mu,n}\delta_{\xi,n})\sqrt{d_{X}}\bar{\sigma}_{\bar{V}|\bar{Z}}
	\end{align*}
	
	The terms on the right hand side without $n$ subscripts are independent
	of $n$ with the sole exception of $d_{D}$. As such we get:
	\begin{align*}
		c_{4,n} & \precsim\delta_{\gamma,Y,n}+\delta_{\gamma,X,n}+\delta_{\gamma,V,n}+\delta_{\omega,Y,n}+\delta_{\omega,Z,n}+\delta_{\omega,V,n}+\delta_{\xi,n}+\delta_{\mu,n}\\
		+ & \sqrt{d_{D}}\delta_{\gamma,X,n}(\delta_{\gamma,Y,n}+\delta_{\gamma,V,n}+\delta_{\gamma,X,n})\\
		+ & \sqrt{d_{D}}\delta_{\omega,Z,n}(\delta_{\omega,Y,n}+\delta_{\omega,V,n})
	\end{align*}
	
	\textbf{Condition 6 }
	
	We now consider $\sup_{r\in(0,1),\eta\in\mathcal{T}_{n}}||\frac{\partial^{2}}{\partial r^{2}}E\bigg[g_{i}\big(\beta_{0},\eta_{0}+r(\eta-\eta_{0})\big)\bigg]||$.
	We will show that if $r\in(0,1)$, $\eta\in\mathcal{T}_{n}$ and $P\in\mathcal{P}_{n}$
	then $||\frac{\partial^{2}}{\partial r^{2}}E\bigg[g_{i}\big(\beta_{0},\eta_{0}+r(\eta-\eta_{0})\big)\bigg]||$
	is bounded by $c_{5,n}$ given below:
	\begin{align*}
		c_{5,n} & =2\sigma_{X}\delta_{\gamma,X,n}(\delta_{\gamma,Y,n}\sigma_{Y}+\delta_{\gamma,V,n}\bar{\xi}+2\delta_{\gamma,V,n}\delta_{\xi,n}+\delta_{\gamma,X,n}\bar{\beta})\\
		& +6(\bar{\mu}+\delta_{\mu,n})\delta_{\omega,Z,n}(\delta_{\omega,Y,n}\sigma_{Y}+\delta_{\omega,V,n}\bar{\xi}+3\delta_{\omega,V,n}\delta_{\xi,n})\\
		& \precsim\delta_{\gamma,X,n}(\delta_{\gamma,Y,n}+\delta_{\gamma,V,n}+\delta_{\gamma,X,n})+\delta_{\omega,Z,n}(\delta_{\omega,Y,n}+\delta_{\omega,V,n})
	\end{align*}
	
	From the definition of $g_{i}$ we have the following: 
	
	\begin{align*}
		& E\bigg[g_{i}\big(\beta_{0},\eta_{0}+r(\eta-\eta_{0})\big)\bigg]\\
		= & E\bigg[\tilde{X}_{i}(\gamma_{X,0}+r\Delta_{\gamma,X})\big(\tilde{Y}_{i}(\gamma_{Y,0}+r\Delta_{\gamma,Y})-\tilde{V}_{i}(\gamma_{V,0}+r\Delta_{\gamma,V})'(\xi_{0}+r\Delta_{\xi})-\tilde{X}_{i}(\gamma_{X,0}+r\Delta_{\gamma,X})'\beta_{0}\big)\bigg]\\
		- & (\mu_{0}+r\Delta_{\mu})\bigg[\bar{Z}_{i}(\omega_{Z,0}+r\Delta_{\omega,Z})\big(\bar{Y}_{i}(\omega_{Y,0}+r\Delta_{\omega,Y})-\bar{V}_{i}(\omega_{V,0}+r\Delta_{\omega,V})'(\xi_{0}+r\Delta_{\xi})\big)\bigg]
	\end{align*}
	
	Substituting definitions into the right-hand side and then using out
	assumptions to simplify the expression this yields the following:
	
	\begin{align*}
		& E\bigg[g_{i}\big(\beta_{0},\eta_{0}+r(\eta-\eta_{0})\big)\bigg]\\
		= & r^{2}\Delta_{\gamma,X}\Sigma_{D}\big(\Delta_{\gamma,Y}'-\Delta_{\gamma,V}'(\xi_{0}+r\Delta_{\xi})-\Delta_{\gamma,X}'\beta_{0}\big)\\
		- & r^{2}\mu_{0}\Delta_{\omega,Z}'\Sigma_{XD}\big(\Delta_{\omega,Y}'-\Delta_{\omega,V}'(\xi_{0}+r\Delta_{\xi})\big)\\
		- & r^{3}\Delta_{\mu}\Delta_{\omega,Z}'\Sigma_{XD}\big(\Delta_{\omega,Y}'-\Delta_{\omega,V}'(\xi_{0}+r\Delta_{\xi})\big)
	\end{align*}
	
	Taking second derivatives we get the equality below:
	\begin{align*}
		& \frac{\partial^{2}}{\partial r^{2}}E\bigg[g_{i}\big(\beta_{0},\eta_{0}+r(\eta-\eta_{0})\big)\bigg]\\
		= & 2\Delta_{\gamma,X}\Sigma_{D}\big(\Delta_{\gamma,Y}'-\Delta_{\gamma,V}'(\xi_{0}+3r\Delta_{\xi})-\Delta_{\gamma,X}'\beta_{0}\big)\\
		- & 2\mu_{0}\Delta_{\omega,Z}'\Sigma_{XD}\big(\Delta_{\omega,Y}'-\Delta_{\omega,V}'(\xi_{0}+3r\Delta_{\xi})\big)\\
		- & 6r\Delta_{\mu}\Delta_{\omega,Z}'\Sigma_{XD}\big(\Delta_{\omega,Y}'-\Delta_{\omega,V}'(\xi_{0}+2r\Delta_{\xi})\big)
	\end{align*}
	
	With repeated application of the triangle inequality, Cauchy-Schwartz,
	and the properties of the matrix norm, and using the properties of
	$\mathcal{T}_{n}$ we get:
	
	\begin{align*}
		& ||\frac{\partial^{2}}{\partial r^{2}}E\bigg[g_{i}\big(\beta_{0},\eta_{0}+r(\eta-\eta_{0})\big)\bigg]||\\
		= & 2||\Sigma_{\tilde{X}}^{1/2}||\delta_{\gamma,X,n}\big(\delta_{\gamma,Y,n}E[\tilde{Y}_{i}^{2}]^{1/2}+\delta_{\omega,V,n}(||\Sigma_{\tilde{V}}^{1/2}\xi_{0}||+2r||\Sigma_{\tilde{V}}^{1/2}\Delta_{\xi}||)+\delta_{\gamma,X}||\Sigma_{\tilde{X}}^{1/2}\beta_{0}||\big)\\
		+ & 2||\mu_{0}\Sigma_{\bar{Z}}^{1/2}||\delta_{\omega,Z,n}\big(\delta_{\omega,Y,n}E[\bar{Y}_{i}^{2}]^{1/2}+\delta_{\omega,V,n}(||\Sigma_{\bar{V}}^{1/2}\xi_{0}||+3r||\Sigma_{\bar{V}}^{1/2}\Delta_{\xi}||)\big)\\
		+ & 6r||\Delta_{\mu}\Sigma_{\bar{Z}}^{1/2}||\delta_{\omega,Z,n}\big(\delta_{\omega,Y,n}E[\bar{Y}_{i}^{2}]^{1/2}+\delta_{\omega,V,n}(||\Sigma_{\bar{V}}^{1/2}\xi_{0}||+2r||\Sigma_{\bar{V}}^{1/2}\Delta_{\xi}||)\big)
	\end{align*}
	
	The expression above is maximized over $r\in[0,1]$ by setting $r=1$.
	Simplifying and employing properties of $\mathcal{T}_{n}$ and Assumption
	4.1 we get the bound below:
	
	\begin{align*}
		c_{5,n} & =2\sigma_{X}\delta_{\gamma,X,n}(\delta_{\gamma,Y,n}\sigma_{Y}+\delta_{\gamma,V,n}\bar{\xi}+2\delta_{\gamma,V,n}\delta_{\xi,n}+\delta_{\gamma,X,n}\bar{\beta})\\
		& +6(\bar{\mu}+\delta_{\mu,n})\delta_{\omega,Z,n}(\delta_{\omega,Y,n}\sigma_{Y}+\delta_{\omega,V,n}\bar{\xi}+3\delta_{\omega,V,n}\delta_{\xi,n})\\
		& \precsim\delta_{\gamma,X,n}(\delta_{\gamma,Y,n}+\delta_{\gamma,V,n}+\delta_{\gamma,X,n})+\delta_{\omega,Z,n}(\delta_{\omega,Y,n}+\delta_{\omega,V,n})
	\end{align*}

\end{proof}

\end{document}